\documentclass[
 reprint,longbibliography,
showpacs,preprintnumbers,
bibnotes,
 amsmath,amssymb,
prb,
]{revtex4-2}

\usepackage[normalem]{ulem}
\usepackage{graphicx}
\usepackage[english]{babel}
\usepackage{microtype}       
\usepackage[utf8]{inputenc}
\usepackage[breaklinks=true,colorlinks=true,linkcolor=blue,urlcolor=blue,citecolor=blue]{hyperref}

\usepackage[dvipsnames]{xcolor}      
\usepackage{mathtools}
\usepackage{braket}
  \usepackage{booktabs}
\usepackage{tikz}
\usepackage{hhline}
\usepackage{bbm}
\usetikzlibrary{shapes.geometric, arrows}
\usetikzlibrary{positioning, arrows.meta, calc, fit}
\tikzstyle{box} = [rectangle, rounded corners, 
minimum width=3cm, 
minimum height=1cm, 
text centered, 
text width=7cm, 
draw=black, 
fill=white]
\tikzstyle{arrow} = [thick,->,>=stealth]
\tikzstyle{darrow} = [thick,<->,>=stealth]

 \usepackage{longtable}

 \usepackage{siunitx}
 \usepackage{csquotes}
 
 \usepackage{multirow}
\usepackage{amssymb}

\newcommand{\eqrefr}[2]{Eqs.~\eqref{#1} -- \eqref{#2}}

\newcommand{\CC}{\mathbb{C}}

\usepackage{dcolumn}
\usepackage{bm}

\makeatletter
\g@addto@macro\bfseries{\boldmath}
\newcommand*{\balancecolsandclearpage}{%
   \close@column@grid
   \clearpage
   \twocolumngrid
 }
\makeatother

\begin{document}

\title{Nonequilibrium Green Functions Simulations for Large Correlated Systems}

\author{Erik Schroedter, Michael Bonitz, and Jan-Philip Joost
 \email{joost@theo-physik.uni-kiel.de}}
\affiliation{
Institute for Theoretical Physics and Astrophysics, 
Kiel University, 24098 Kiel, Germany \\ and Kiel Nano, Surface and Interface Science KiNSIS, Kiel University, Kiel, Germany
}

\date{\today}

\begin{abstract}
Correlated real-time dynamics in large, spatially inhomogeneous quantum systems remain difficult to access with nonequilibrium many-body methods. Existing large-scale approaches often rely on effective single-particle, static, or adiabatic descriptions, whereas two-time nonequilibrium Green functions retain dynamical correlations and memory but their computational runtime grows cubically with the number of time steps $N_\mathrm{t}$.
This scaling bottleneck could recently be overcome by introducing the G1--G2 scheme [Schlünzen et al., Phys. Rev. Lett. \textbf{124}, 076601 (2020)]---a time-local reformulation of the Generalized Kadanoff-Baym ansatz with Hartree-Fock propagators (HF-GKBA)---that is linear in $N_\mathrm{t}$, but still requires the propagation of a two-particle correlation function and may suffer from numerical instabilities.
This has restricted time-dependent dynamical NEGF simulations to small systems with $N_\mathrm{b} \sim 10^2$ basis states.
Here we introduce a quantum-fluctuation formulation of nonequilibrium Green functions, denoted $\delta$NEGF [the concept was originally formulated in Cond. Matt. Phys. \textbf{25}, 23401 (2022)], that represents dynamical two-particle correlations through fluctuations of field-operator products, $\delta \hat G$.
This factorized representation guarantees stable dynamics by preserving the positivity of the reduced density matrices, avoids the explicit storage of the two-particle Green function, and reduces the propagation to a finite ensemble of Hartree-Fock-like fluctuation trajectories that can be evolved independently and in parallel.
Combined with a stochastic low-rank decomposition of the correlation functions, the method retains time-linear scaling while extending dynamical $GW$ and particle-particle and particle-hole $T$-matrix simulations to basis sizes of order $N_\mathrm{b}\sim 10^4$, roughly two orders of magnitude beyond previous time-dependent dynamical NEGF calculations.
We benchmark $\delta$NEGF against exact and HF-GKBA results for lattice systems, finding stable correlated dynamics also at strong coupling. We further demonstrate large-scale simulations of diffusion in two-dimensional Hubbard lattices and, as an outlook beyond local Hubbard interactions, ultrafast relaxation in graphene nanoribbon heterostructures with long-range Coulomb interactions.
These results establish $\delta$NEGF as a scalable route to dynamical self-energy simulations of large, spatially inhomogeneous correlated quantum systems beyond the reach of existing NEGF implementations.
\end{abstract}

\maketitle

\section{Introduction}\label{s:intro}
Correlated quantum many-body systems are of high current interest in many fields. This includes quantum materials \cite{Keimer2017, Dzsaber2022}, ultracold atoms \cite{Hart2015, Onofrio2024}, warm dense matter \cite{vorberger_wdm_26, dornheim_physrep_18} or the quark-gluon plasma \cite{Gross2023}.
Many recent experimental and theoretical works have investigated the properties of 
quantum many-particle systems 
 following a rapid external excitation, such as the impact of highly charged ions \cite{niggas_prl_22} or rapid changes of interaction or confinement potentials \cite{schneider_fermionic_2012,schluenzen_prb16,schluenzen_cpp16}. Of particular current interest is the excitation of quantum materials by short laser pulses \cite{lisowski_05, Cavalieri2007, schuette_prl_12,Torre_2021, joost_prr_25}, which may initiate metal-insulator transitions \cite{li_prl_26}, high-temperature superconductivity \cite{Fausti_2011,buzzi_PRX_20}, anomalous Hall effect \cite{mciver_light-induced_2020} or high-temperature ferromagnetism \cite{disa_nat_23};  
 for a recent overview on ultrafast dynamics in materials, see \cite{caruso_jpmat_25}.

Beyond spatially homogeneous settings, an especially important class of nonequilibrium problems concerns systems in which spatial inhomogeneity is not a perturbation, but an essential part of the physics. Quantum materials, for example, frequently contain defects, impurities, edges, surfaces, interfaces, and disorder, which can strongly influence their electronic, magnetic, and topological properties~\cite{joost_19_nanolett,Liu_2019,Wolfowicz_2021}. Related examples include photoionized molecules, where attosecond spectroscopy has enabled the real-time investigation of ultrafast charge migration and charge-transfer dynamics~\cite{Calegari_2014,Woerner_2017,Nisoli_2017,Palacios_2020,Calegari_2023}, molecule--metal interfaces and adsorbate-covered surfaces, where photoinduced charge transfer, dynamical screening, and relaxation into substrate degrees of freedom can occur on femtosecond or even sub-femtosecond time scales~\cite{Foehlisch_2005,Lindstrom_2006,Diez-Muino_2011,Aguilar-Galindo_2021,Inzani_2025}, and van der Waals heterostructures, in which optical excitation can trigger interlayer charge transfer, exciton formation, and energy redistribution across structurally distinct layers~\cite{Hong_2014,Jin_2018,Rivera_2018}.
In such systems, the nonequilibrium dynamics may involve memory effects and interaction-driven processes such as multiple excitations, charge-transfer resonances, dynamical screening, and transient correlated states on multiple competing time scales~\cite{CEDERBAUM_1999,Calegari_2016,Ovesen_2019}.
This poses a severe theoretical challenge. Accurate simulations of heterogeneous systems generally require large single-particle basis sets, while the absence of exploitable spatial symmetries prevents the dimensional reduction that is often crucial for correlated nonequilibrium calculations in homogeneous solids or surfaces~\cite{balzer-book,schluenzen_cpp16}. As a result, large-scale real-time simulations frequently rely on time-dependent density-functional theory (TDDFT) in adiabatic, memory-less approximations and related effective single-particle approaches, which are computationally efficient and applicable to complex systems~\cite{Li_2020,Tancogne-Dejean_2020,Wang_2016,Shubhadeep_2025,GuiotduDoignon_2025}. However, their practical accuracy is limited by the available exchange-correlation approximations, which commonly  do not explicitly retain memory dependence and may be problematic for charge-transfer dynamics, double or multiple excitations, resonantly driven dynamics, and interaction-induced relaxation~\cite{Maitra_2022,Lacombe_2023}. These limitations motivate the development of nonequilibrium many-body methods that retain dynamical correlations and memory while remaining scalable to large, spatially inhomogeneous systems.

Nonequilibrium Green functions (NEGF) provide a powerful framework for treating such correlated nonequilibrium dynamics, e.g.,~\cite{keldysh_diagram_1964, kadanoff_quantum_1962, stefanucci_nonequilibrium_2013, balzer-book}. Their attractive features—the field-theoretical foundation, systematic approximation schemes based on Feynman diagrams, direct access to both statistical and spectral observables, and applicability to many-particle dynamics over a wide range of time scales—have led to the implementation of NEGF techniques in many areas of physics over the past three decades, e.g.,~\cite{pngf1,pngf2, pngf7, pngf8_preface}, as well as to public codes~\cite{koehler_cpc_99, schuler_cpc_20, marini_yambo_2009}.

At the same time, the systematic treatment of correlations in NEGF theory comes with a substantial computational cost.
In standard real-time implementations, the two-time structure of the Green function leads to rapidly increasing memory and runtime requirements, which become particularly restrictive for large basis sets and in the absence of symmetries. Consequently, correlated NEGF calculations have traditionally been limited to very short times and to small model systems, compact basis representations, or systems where translational and other symmetries can be exploited~\cite{dahlen_solving_2007,Stan_2009,von_friesen_successes_2009,balzer_doublon_2018,joost_femtosecond_2019,joost_cpp_21}.

One major computational disadvantage of conventional two-time NEGF---the cubic scaling of the runtime with the number of time steps $N_\mathrm{t}$---could recently be overcome by introducing the G1--G2 scheme \cite{schluenzen_prl_20}---a time-local exact reformulation of the Generalized Kadanoff Baym ansatz \cite{lipavsky_generalized_1986} with Hartree-Fock propagators (HF-GKBA) that is linear in $N_\mathrm{t}$. It could be shown that this property is achieved for weak and strong interactions with a broad variety of self-energies, including the $GW$ and $T$-matrix approximations \cite{joost_prb_20} and even combinations of both---the dynamically screened ladder approximation \cite{joost_prb_22,donsa_prr_23}.
Due to its exceptional performance, the G1--G2 scheme has been picked up by many groups, included in software packages~\cite{pavlyukh_pss_23} and applied to a variety of physical problems, including ion stopping in correlated materials~\cite{borkowski_pss_22, lovato_pssb_25}, quantum transport and open systems~\cite{tuovinen_prl_23,balzer_prb_23,cosco_pss_24,pavlyukh_prb_25,pavlyukh_epjst_25}, 
 coupled electron-boson dynamics~\cite{karlsson_fast_2021, pavlyukh_interacting_2022,pavlyukh_time-linear_2022-1, pavlyukh_time-linear_2022}, plasmas \cite{makait_cpp_23}, benchmark model systems~\cite{pavlyukh_pss_24,verdozzi_pss_24,pavlyukh_epjst_25}, molecular junctions~\cite{tuovinen_prx_25}, and ultrafast carrier, exciton, and laser-driven dynamics in low-dimensional materials ~\cite{perfetto_prl_22,joost_prr_25}.

Despite these advances, important limitations remain. First, simulations based on the HF-GKBA may become unstable, in particular at increased interaction strength or for long propagation times. This instability is not a consequence of the G1--G2 reformulation itself, but is inherited from the use of Hartree--Fock propagators in the HF-GKBA. It can be mitigated by various regularization procedures (``purification'') \cite{lackner_propagating_2015,lackner_high-harmonic_2017, joost_prb_22, donsa_prr_23} which, however, may introduce additional numerical overhead and can substantially increase the memory requirements, thereby reducing part of the computational advantage gained by the time-local formulation.

Second, an independent limitation arises from the scaling with the size of the single-particle basis. In the G1--G2 scheme, the time-nonlocal collision integral is replaced by coupled equations of motion for the single-particle Green function and the equal-time two-particle correlation function. While this removes the unfavorable cubic scaling with the number of time steps $N_\mathrm{t}$, it requires the storage of a two-particle object that scales as $N_\mathrm{b}^4$, where $N_\mathrm{b}$ denotes the number of single-particle basis states~\cite{joost_prb_20,joost_phd_2022,bonitz_pssb23}. Thus, the G1--G2 scheme has enabled correlated NEGF simulations on unprecedented time scales, but, due to its high memory requirements, it has not removed the basis-size bottleneck that obstructs applications to large heterogeneous systems where translational or other spatial symmetries cannot be exploited. In practice, time-dependent G1--G2 simulations have remained restricted to basis sizes of order $N_\mathrm{b}\sim 10^2$, comparable to conventional two-time NEGF calculations~\cite{borkowski_pss_22,pavlyukh_time-linear_2022,Brezinova2024,joost_prr_25}.

A different route is taken by real-time approaches based on static or adiabatic $GW$-type self-energies and kernels, including time-dependent COHSEX~\cite{Sangalli_2019}, real-time Bethe--Salpeter equation~\cite{Attaccalite_2011,Marek2025}, and time-dependent adiabatic $GW$~\cite{Chan_2021,Hu_2023,ChangLee_2024}.
Like the G1--G2 scheme, these formulations lead to time-local equations of motion and hence avoid the cubic time-step scaling of conventional two-time NEGF. In contrast to G1--G2, however, they do not require the propagation of an explicit two-particle correlation function and are therefore not subject to the same $N_\mathrm{b}^4$ memory bottleneck. This makes them useful for first-principles simulations of optical and ultrafast response in regimes where static or adiabatic descriptions of screened exchange, Coulomb-hole contributions, and electron--hole interactions are sufficient. However, their efficiency relies precisely on these static or adiabatic approximations: the full time-nonlocal structure of the dynamical nonequilibrium $GW$ self-energy is replaced by static or adiabatic effective interactions. They therefore enable simulations of large systems, but do so by giving up the full memory structure of dynamical NEGF.

Here, we pursue a different strategy to overcome the scaling bottleneck of NEGF simulations and to make correlated calculations with dynamical self-energies accessible both for long propagation times and for large, spatially inhomogeneous systems. To this end, we introduce an NEGF-based quantum-fluctuations approach, denoted $\delta$NEGF, which was first formulated in Ref.~\cite{schroedter_cmp_22} and is substantially extended in the present work. We develop $\delta$NEGF versions of three standard self-energy approximations: $GW$, the particle--particle $T$-matrix, and the particle--hole $T$-matrix. This approach solves the open scaling and stability problems of the G1--G2 scheme by avoiding the explicit propagation of the full two-particle correlation function while preserving time-linear propagation.

The method is based on the fact that two-particle correlation functions can be represented by tensor decompositions in terms of single-particle quantum fluctuations, a concept introduced in equilibrium by Garrod and Percus in 1964~\cite{garrod_reduction_1964}. In nonequilibrium, this representation allows the correlated dynamics to be formulated in terms of fluctuation trajectories rather than by explicitly propagating the full two-particle correlation function. This has several important consequences. First, the fluctuation representation preserves the positivity structure of the two-particle correlations, thereby avoiding the instability problem of HF-GKBA simulations without purification or other regularization procedures. This opens the way to stable simulations also at stronger coupling. Second, the decisive reduction of the basis-size scaling compared to G1--G2 is achieved by combining the fluctuation formulation with a low-rank representation of the initial state. In practice, such rank-reduced initial states can be generated efficiently by random sampling, so that only a finite number $N_\mathrm{s}$ of single-particle fluctuation trajectories has to be propagated. The resulting equations remain time-linear and reduce the numerical task to a set of Hartree--Fock-like propagations that can be performed almost independently and in parallel. Finally, expressing two-particle correlation functions in terms of fluctuations provides direct access to two-time response and correlation functions from a single-time propagation, enabling the calculation of two-particle spectra in large systems.

We benchmark the resulting $\delta$NEGF self-energy approximations against HF-GKBA and exact results and systematically analyze the errors introduced by the low-rank representation. We demonstrate that $\delta$NEGF enables correlated calculations with dynamical self-energies simultaneously for long propagation times and for large heterogeneous systems with basis sizes exceeding $N_\mathrm{b}=10^4$. This constitutes an improvement of more than two orders of magnitude over the basis sizes accessible to dynamical NEGF simulations to date. As representative large-scale applications, we consider two inhomogeneous lattice systems. First, we simulate the nonequilibrium dynamics after a sudden confinement quench in a two-dimensional Hubbard model with more than $10^4$ lattice sites, a setting motivated by ultracold-atom experiments. Second, we study the ultrafast relaxation dynamics of a laser-excited graphene-nanoribbon heterostructure described by an extended Hubbard model with long-range Coulomb interactions and more than 1700 basis functions.

This paper is structured as follows: In Sec.~\ref{s:negf} we recall basic ideas of NEGF theory and also introduce the HF-GKBA and the G1--G2 scheme.
In Sec.~\ref{s:theory} we introduce the NEGF-based quantum fluctuations ($\delta$NEGF) approach, including main concepts and approximations. In Sec.~\ref{s:results} we illustrate the theoretical concepts by numerical results for the standard Hubbard model and an extended Hubbard model with long-range Coulomb interactions.
Sec.~\ref{s:discussion} concludes this paper with a summary and an outlook.

\section{Nonequilibrium Green Functions}\label{s:negf}
We consider a nonequilibrium quantum many-body system of fermions \footnote{The following considerations can all be extended to bosonic systems following some adjustments.} described by the following Hamiltonian,
\begin{equation}
    \hat{H}(t) = \sum_{ij}h_{ij}(t)\hat{a}^\dagger_i \hat{a}_j +\frac{1}{2}\sum_{ijkl}w_{ijkl}(t)\hat{a}^\dagger_i\hat{a}^\dagger_j \hat{a}_l \hat{a}_k,
\end{equation}
where $h(t)$ includes all single-particle contributions, i.e., the kinetic part and a (time-dependent) external potential, and $w(t)$ describes the pair interaction that is allowed to be time-dependent. This is, for example, the case for interaction quenches or the generation of a correlated initial state using the ``adiabatic switching method'' \cite{hermanns_hubbard_2014, schluenzen_jpcm_19}.  Further, $\hat{a}_i $ $(\hat{a}^\dagger_i)$ denotes the annihilation (creation) operator with respect to an element $\phi_i$ of some complete orthonormal basis $(\phi_i)_{i\in B}$ of the underlying single-particle Hilbert space $\mathcal{H}$. Here, we assume that the single-particle Hilbert space is finite, i.e., $N_\mathrm{b}\coloneqq  |B| = \mathrm{dim}(\mathcal{H})  < \infty$. Moreover, in the following, all sums run over the elements of the set $B$ unless stated otherwise.\\
We begin the rigorous derivation of the $\delta$NEGF approach by introducing the single-particle nonequilibrium Green function (NEGF). It is defined as
\begin{equation}
    G_{ij}(z,z') \coloneqq \frac{1}{\mathrm{i}\hbar} \big\langle \mathcal{T}_\mathcal{C}\big\{ \hat{a}_i(z)\hat{a}^\dagger_j(z')\big\}\big\rangle,
\end{equation}
where the expectation value is taken with respect to the density operator $\hat{\rho}$ describing the initial state of the system with a fixed particle number. Further, $\mathcal{T}_\mathcal{C}$ denotes the time-ordering operator on the Keldysh contour $\mathcal{C}$, which we assume to be of the form $\mathcal{C}= [t_0,\infty) \oplus (\infty, t_0]$ for some initial time $t_0$. The dynamics of the single-particle NEGF are described by the (Keldysh-)Kadanoff-Baym equations (KBE)
\begin{align}
    \mathrm{i}\hbar \partial_z G(z,z') &= \delta_\mathcal{C}(z,z')\mathbbm{1} +h(z)G(z,z') +I(z,z'),\\
    I_{ij}(z,z') &\coloneqq - \mathrm{i}\hbar \sum_{klm}w_{iklm}(z) G^{(2)}_{lmjk}(z,z,z',z^+),
\end{align}
where $\mathbbm{1}\in  \mathbb{C}^{N_\mathrm{b}\times N_\mathrm{b}}$ denotes the identity matrix and we set $z^+\coloneqq z + \epsilon$ for an infinitesimally small positive constant $\epsilon$ in order to avoid ambiguities in the time ordering. Further, $I$ denotes the collision term that describes the coupling to the two-particle NEGF defined as 
\begin{align}
   & G^{(2)}_{ijkl}(z_1, z_2, z'_1,z'_2)\coloneqq \nonumber \\&\qquad -\frac{1}{\hbar^2} \big\langle \mathcal{T}_\mathcal{C}\big\{ \hat{a}_i(z_1) \hat{a}_{j}(z_2) \hat{a}^\dagger_{l}(z'_2) \hat{a}^\dagger_{k}(z'_1)\big\}\big\rangle. \label{eq:2p-NEGF}
\end{align}
As the equation of motion for the two-particle NEGF, in turn, couples to the three-particle NEGF, the full solution of the many-body problem requires the solution of a hierarchy of equations, the Martin-Schwinger hierarchy. In practice, one requires some cutoff to the hierarchy by applying sensible approximations. A common starting point is the decomposition of the two-particle NEGF into a mean-field (Hartree), $G^{(2),\mathrm{H}}$, an exchange (Fock), $G^{(2),\mathrm{F}}$, and a correlation contribution, $\mathcal{G}$,
\begin{align}
    G^{(2)} &= G^{(2),\mathrm{H}}+G^{(2),\mathrm{F}}+\mathcal{G}\,.\label{eq:g2-structure}
\end{align}
The first two are defined in terms of the single-particle NEGF,
\begin{subequations}
\begin{align}
    G^{(2),\mathrm{H}}_{ijkl}(z_1,z_2,z'_1,z'_2) &\coloneqq G_{ik}(z_1,z'_1) G_{jl}(z_2,z'_2),\\
    G^{(2),\mathrm{F}}_{ijkl}(z_1,z_2,z'_1,z'_2) &\coloneqq - G_{il}(z_1,z'_2) G_{jk}(z_2,z'_1)\,.
\end{align}
\end{subequations}
In this paper we focus on correlation effects, such as scattering, dynamical screening or bound states that require an accurate and efficient treatment of $\mathcal{G}$.

\subsection{Self-energy approximations and Bethe-Salpeter equation}\label{ss:sigma_BSE}
An alternative expression for the collision integral $I$ is 
obtained by eliminating  the two-particle NEGF in favor of the  the self-energy $\Sigma = \Sigma[G]$, 
\begin{equation}
    I(z,z') = \int_\mathcal{C}\Sigma(z,\bar{z})G(\bar{z},z') \,\mathrm{d}\bar{z}\, ,\label{eq:I_Sigma}
\end{equation}
where the structure of $G^{(2)}$ [cf. Eq.~\eqref{eq:g2-structure}]  
implies
\begin{align}\label{eq:sigma-allgemein}
    \Sigma =\Sigma^\mathrm{HF}+\Sigma^\mathrm{cor}\coloneqq  \Sigma^\mathrm{H}+\Sigma^\mathrm{F}+\Sigma^\mathrm{cor},
\end{align}
with the mean-field (Hartree), $\Sigma^\mathrm{H}$, and exchange (Fock) self-energies, $\Sigma^\mathrm{F}$, given by
\begin{subequations}
\begin{align}
    \Sigma^\mathrm{H}_{ij}(z,z')&\coloneqq -\mathrm{i}\hbar \sum_{mn}w_{imjn}(z,z') G_{nm}(z,z^+), \label{eq:Sigma_H}\\
    \Sigma^\mathrm{F}_{ij}(z,z')&\coloneqq  \mathrm{i}\hbar \sum_{mn}w_{imnj}(z,z')G_{nm}(z,z^+) \label{eq:Sigma_F},
\end{align}
\end{subequations}
and we denote $w(z,z') \coloneqq w(z) \delta_\mathcal{C}(z,z')$. While the Hartree--Fock self-energy is singular, i.e., $\Sigma^\mathrm{HF}(z,z') \propto \delta_\mathcal{C}(z,z')$, the correlation self-energy $\Sigma^\mathrm{cor}$ is non-singular and thus leads to the emergence of memory effects. Using, for example, diagram techniques, it is possible to systematically derive approximations for the self-energy, such as the well-known second-order Born (2B), the $GW$, or the particle-particle (TPP) and particle-hole (TPH) $T$-matrix approximations. The latter three arise from diagram resummation techniques where the interaction is included up to infinite order, for special classes of scattering processes. More specifically, these processes can be divided into three channels: the particle-particle channel ``$\mathrm{pp}$,'' the longitudinal particle-hole channel ``$\mathrm{ph}$,'' and the transversal particle-hole channel ``$\overline{\mathrm{ph}}$,'' corresponding to the TPP, TPH and $GW$ approximations, respectively. 

For the different channels ``$c$'', it is advantageous to decompose the self-energy into mean-field and correlation parts in different ways other than the standard form \eqref{eq:sigma-allgemein}:
\begin{align}
    \Sigma = \Sigma^{0,c}+\Sigma^c,
\end{align}
with the channel-specific mean-field part
\begin{align}\label{eq:def-sigma0}
    \Sigma^{0,c} \coloneqq \begin{cases}
        0, & c = \mathrm{pp},\\
        \Sigma^\mathrm{F}, & c = \mathrm{ph}, \\
        \Sigma^\mathrm{H}, & c = \overline{\mathrm{ph}},
    \end{cases}
\end{align}
and the channel-specific correlation part $\Sigma^c$ given by
\begin{subequations}
\begin{align}
    \Sigma^\mathrm{pp}_{ij}(z,z') &\coloneqq -\mathrm{i}\hbar\sum_{mn}T^\mathrm{pp}_{imjn}(z,z') G_{nm}(z',z),\\
    \Sigma^\mathrm{ph}_{ij}(z,z')& \coloneqq -\mathrm{i}\hbar\sum_{mn}T^\mathrm{ph}_{imjn}(z,z') G_{nm}(z,z'),\\
    \Sigma^{\overline{\mathrm{ph}}}_{ij}(z,z') &\coloneqq \mathrm{i}\hbar \sum_{mn}W_{imnj}(z,z') G_{nm}(z,z'),
\end{align}
\end{subequations}
where, $T^\mathrm{pp}$,  $T^\mathrm{ph}$, and  $W$ denote the particle-particle $T$-matrix, the particle-hole $T$-matrix and the screened interaction, respectively. 

To achieve a compact and unified description of all channels, we introduce the channel vertex $K^c$
\begin{align}
    K^c_{ijkl}(z,z') \coloneqq \begin{cases}
        T^\mathrm{pp}_{ijkl}(z,z'), & c= \mathrm{pp},\\[1ex]
        T^\mathrm{ph}_{ijkl}(z,z') , &c= \mathrm{ph}, \\[1ex]
        W_{ijkl}(z,z'), & c = \overline{\mathrm{ph}}.
    \end{cases}
\end{align}
Furthermore, we identify rank-4 tensors in the single-particle space, $A \in \CC^{N_\mathrm{b}\times N_\mathrm{b}\times N_\mathrm{b}\times N_\mathrm{b}}$, with matrices in the channel-specific two-particle spaces, $\bm{A}^c \in \CC^{N_\mathrm{b}^2\times N_\mathrm{b}^2}$, as outlined in appendix~\ref{a:tensor_algebra} [indicated by bold letters and the channel-specific superscript ``$c$'']. Using the tensor notation, the channel vertex $K^c$ is given by
\begin{align}
   \bm{K}^c(z,z') = \tilde{\bm{w}}^c(z,z')+\bm{w}^c(z)\bm{\chi}^c(z,z')\bm{w}^c(z'), \label{eq:K}
\end{align}
and involves a generalized susceptibility, 
\begin{align}
    &\chi^c_{ijkl}(z,z')\coloneqq \nonumber
    \label{eq:def-chi-2times}
    \\&\,\mathrm{i}\hbar \begin{cases}
        G_{ijkl}^{(2)}(z,z,z',z'), &  c= \mathrm{pp},\\[1ex]
        G^{(2)}_{ijkl}(z,z',z'^+,z^+) -  G^{(2),\mathrm{F}}_{ijkl}(z,z',z'^+,z^+), & c = \mathrm{ph},\\[1ex]
        G^{(2)}_{ijkl}(z,z',z^+,z'^+) -  G^{(2),\mathrm{H}}_{ijkl}(z,z',z^+,z'^+), & c = \overline{\mathrm{ph}} ,
    \end{cases}
\end{align}
and modified pair interaction, 
\begin{align}
    \tilde{w}^c_{ijkl} \coloneqq \begin{cases}
        w^-_{ijkl}, & c =\mathrm{pp}, \\[1ex]
        w_{ijkl}, & c= \mathrm{ph},\overline{\mathrm{ph}}\,,
    \end{cases} \label{eq:def_w_tilde}
\end{align}
with $w^-_{ijkl}\coloneqq w_{ijkl}-w_{ijlk}$. It is important to emphasize at this point that certain quantities, such as the channel vertex $K^c$ or the generalized susceptibilities $\chi^c$, are defined for specific channels. For these quantities, we omit the additional superscript ``$c$'' for simplicity when identifying them with matrices in the respective two-particle spaces. Moreover, products of matrices in the channel-specific two-particle spaces correspond to different tensor contractions depending on the specific channel under consideration, see again appendix~\ref{a:tensor_algebra}.

The susceptibility, $\chi^c$, satisfies the following Bethe--Salpeter equation (BSE) \footnote{The generalized susceptibility $\chi^c$ can be eliminated in favor of the vertex $K^c$ as its BSE can be used to derive a BSE for $K^c$ of the form $K^c = \tilde{w}^c+\int_\mathcal{C} w^c \chi^{0,c} K^c$. This turns out to be advantageous, for example, when considering diagonal interactions, $w_{ijkl}= v_{ij}\delta_{ik}\delta_{jl}$, as the screened interaction $W$ in the $GW$ approximation also becomes diagonal.  This is, however, not the case for this example when considering $GW$+X or the $T$ matrices.}
\begin{align}
    \bm{\chi}^c(z,z') =&\, \tilde{\bm{\chi}}^{0,c}(z,z')\nonumber\\&+\int_\mathcal{C} \bm{\chi}^{0,c}(z,\bar z) \bm{w}^c(\bar z) \bm{\chi}^c(\bar z,z')\mathrm{d}\bar z,\label{eq:BSE_chi}
\end{align}
with the ideal generalized susceptibilities,
\begin{align}
    \chi^{0,c}_{ijkl}(z,z') \coloneqq \mathrm{i}\hbar \begin{cases}
        G^{(2),\mathrm{H}}_{ijkl}(z,z,z',z'), & c = \mathrm{pp},\\[1ex]
        G^{(2),\mathrm{H}}_{ijkl}(z,z',z'^+,z^+), & c = \mathrm{ph} ,\\[1ex]
        G^{(2),\mathrm{F}}_{ijkl}(z,z',z^+,z'^+), & c = \overline{\mathrm{ph}},
    \end{cases}
    \label{eq:chi0_cases}
\end{align}
and $\tilde{\chi}^{0,c}$ being defined analogously to $\tilde{w}^c$ [cf. Eq.~\eqref{eq:def_w_tilde}],
\begin{align}
    \tilde{\chi}^{0,c}_{ijkl}(z,z') \coloneqq \begin{cases}
        \chi^{0,\mathrm{pp}}_{ijkl}(z,z') - \chi^{0,\mathrm{pp}}_{ijlk}(z,z'), & c = \mathrm{pp},\\[1ex]
        \chi^{0,c}_{ijkl}(z,z'), & c = \mathrm{ph},\overline{\mathrm{ph}},
    \end{cases}
\end{align}
This scheme gives access to three additional important self-energy approximations. The first two are obtained by replacing $w\rightarrow w^-$ in Eq.~\eqref{eq:BSE_chi} in the particle-hole channels. This gives rise to additional exchange contributions, and  the corresponding approximations will be denoted as TPH+X and $GW$+X. Finally, the limiting case of the 2B approximation is recovered by using, in Eq.~\eqref{eq:K}, $\chi^{0,c}$ instead of $\chi^c$, yielding the self-energy $\Sigma^\mathrm{2B}$ that is  channel-independent. In the following, we will focus on the more advanced TPP, TPH and $GW$ approximations, unless stated otherwise.

With these self-energies it is possible to capture important many-body effects that are neglected by a mean-field (Vlasov/Hartree or Hartree-Fock) description,  such as collisions, bound states, and dynamical screening. 

At the same time, these advanced self-energies impose a drastic  computational overhead for the description of nonequilibrium systems, leading to a cubic scaling of the runtime with the number of time steps, $N_\mathrm{t}$, and, therefore, correlated NEGF simulations are typically limited to very short times and small systems. In the remainder of this article, we present strategies to overcome the runtime and memory bottlenecks.
\subsection{Real-time components and time-diagonal KBE}\label{ss:def-ghat}
For practical applications, we now make the transition from  the contour-time-ordered NEGFs to their real-time counterparts. For the single-particle NEGF, these are given by the greater and lesser components respectively defined as the expectation value of products of two field operators:
\begin{subequations}
\begin{align}
    G^\gtrless_{ij}(t,t') & \coloneqq \big\langle\hat{G}^\gtrless_{ij}(t,t')\big\rangle,
    \label{eq:ggl-def}\\
    \hat{G}^>_{ij}(t,t')& \coloneqq \frac{1}{\mathrm{i}\hbar}\hat{a}_i(t)\hat{a}^\dagger_j(t'),\label{eq:gg-op-def}\\
    \hat{G}^<_{ij}(t,t') & \coloneqq - \frac{1}{\mathrm{i}\hbar} \hat{a}^\dagger_j(t') \hat{a}_i(t)\,.\label{eq:gl-op-def}    
\end{align}
\end{subequations}
Further, we define additional components as linear combinations of the greater and lesser components: the spectral ($\Delta$) and the Keldysh ($\mathrm{K}$) components:
\begin{subequations}
\begin{align}
    G^\Delta(t,t') &\coloneqq G^>(t,t') - G^<(t,t')\,, \label{eq:def-g-Delta_component} \\
    G^\mathrm{K}(t,t') &\coloneqq G^>(t,t') +G^<(t,t')\,, \label{eq:def-K_component}
\end{align}
\end{subequations}
as well as the retarded and advanced components:
\begin{subequations}
\begin{align}
    G^\mathrm{R}(t,t') &\coloneqq \Theta(t,t') G^\Delta(t,t')\,, \label{eq:def-R_component}\\
    G^\mathrm{A}(t,t') &\coloneqq -\Theta(t',t)G^\Delta(t,t')\,,\label{eq:def_A-component}
\end{align}
\end{subequations}
where $\Theta$ denotes the Heaviside function. Analogously, the greater or lesser components of the susceptibility, $\chi^{c,\gtrless}$, follow by taking the greater and lesser components of Eq.~\eqref{eq:def-chi-2times}, and the spectral and Keldysh components of $\chi^c$ are defined as
\begin{subequations}
\begin{align}
    \chi^{c,\Delta}(t,t') &\coloneqq \chi^{c,>}(t,t') - \chi^{c,<}(t,t')\,, \label{eq:def-Delta_component} \\
    \chi^{c,\mathrm{K}}(t,t') &\coloneqq \chi^{c,>}(t,t') +\chi^{c,<}(t,t')\,, \label{eq:def-K_component-chi}
\end{align}
\end{subequations}
and analogously for $\chi^{c,0}$. 

On the time diagonal, $t=t'$, we use the compact notation $G^\gtrless(t)\coloneqq G^\gtrless(t,t)$ and do the same for all other quantities.
In the following, we will need the 
the time-diagonal KBE, which is given by \cite{haug_quantum_2007,bonitz_qkt}
\begin{align}
     \mathrm{i}\hbar \partial_t G^<(t) &= \big[h,G^<\big](t)+ I^<(t)+I^{<,\dagger}(t), \label{eq:i-sigma}
\end{align}
and the collision integral
\begin{align}
   I^<(t) = \int_{t_0}^\infty& \big\{\Sigma^{<}(t,\bar{t})G^\mathrm{A}(\bar{t},t)+ \Sigma^\mathrm{R}(t,\bar{t})G^<(\bar{t},t)\big\}\mathrm{d}\bar{t}\,,
\end{align}
where the retarded component of the self-energy, $\Sigma^\mathrm{R},$ includes a singular part in the form of the Hartree-Fock self-energy $\Sigma^\mathrm{HF}$. The appearing time integral is the main cause of the large computational overhead that was mentioned above. It can be eliminated by introducing the G1--G2 scheme \cite{schluenzen_prl_20}, cf. Sec.~\ref{ss:f1-g2}, where, in addition to the equation for $G^<$, a time-local equation for the correlated part $\mathcal{G}$ of $G^{(2)}$ is solved, cf. Eq.~\eqref{eq:g2-structure},
\begin{align}
     \mathrm{i}\hbar \partial_t G^<(t) &= \big[h^\mathrm{HF},G^<\big](t)+\mathcal{I}^\mathcal{G}(t), \label{eq:G_G2}
\end{align}
where we introduced the effective single-particle Hartree--Fock Hamiltonian,
\begin{align}
    h^\mathrm{HF}(t) \coloneqq h(t)+ \Sigma^\mathrm{HF}(t),
\end{align}
and combined the collision term and its adjoint without the singular contributions and expressed it in terms of the correlated part of the two-particle NEGF,
\begin{align}
    \mathcal{I}^\mathcal{G}_{ij}(t)\coloneqq -\mathrm{i}\hbar \sum_{klm}\big\{w_{iklm}(t)\mathcal{G}^<_{lmjk}(t)- \mathcal{G}^<_{iklm}w_{lmjk}(t)(t)\big\}. \label{eq:def_I^G}
\end{align}
Despite the success of the G1--G2 scheme in terms of computational speedup, the storage of the two-particle correlation function $\mathcal{G}$ gives rise to a new bottleneck severely limiting the size of the systems that can be treated. As we will show below, this bottleneck can be removed by replacing $\mathcal{G}$ with the generalized susceptibilities $\chi^c$, Eq.~\eqref{eq:def-chi-2times}.

The first step is to relate the greater and lesser components of $\mathcal{G}$ and $\chi^c$ to each other. Using Eqs.~\eqref{eq:g2-structure} and \eqref{eq:def-chi-2times} we obtain
\begin{align}
   \mathrm{i}\hbar\, \mathcal{G}^\gtrless(t) = \chi^{c,\gtrless}(t)-\tilde{\chi}^{0,c,\gtrless}(t), \label{eq:G2-chi_relation}
\end{align}
where the susceptibilities are channel-dependent. Moreover, on the time-diagonal, $\mathcal{G}^>(t)\equiv \mathcal{G}^<(t)\equiv \mathcal{G}(t)$, as a consequence of particle-hole symmetry. We, therefore, replace $\mathcal{G}^<(t)$, in the collision integral of Eq.~\eqref{eq:G_G2}, identically by the symmetrized expression
\begin{align}
  \mathcal{G}^<(t) = \frac{\mathcal{G}^>(t) + \mathcal{G}^<(t)}{2} = \frac{\chi^{c,\mathrm{K}}(t) - \tilde \chi^{0,c,\mathrm{K}}(t)}{2\mathrm{i}\hbar} \,,\quad
 \label{eq:g2-sym-chi}
\end{align}
where, in the last step, we used the result \eqref{eq:G2-chi_relation} and definition \eqref{eq:def-K_component-chi}.

Then the kinetic equation~\eqref{eq:G_G2} becomes
\begin{align}
    \mathrm{i}\hbar\partial_t G^<(t) &= \big[h^c+w^c_\mathrm{loc},G^<\big](t)  +\mathcal{I}^{c}(t), \label{eq:EOM_G_GWX-QF}
\end{align}
where we introduced the channel-dependent effective single-particle Hamiltonian, 
\begin{align}\label{eq:def_eff_h_c}
    h^c(t) &\coloneqq h(t)+\Sigma^{0,c}(t),
\end{align}
the local pair potential,
\begin{align}
        w^c_\mathrm{loc}(t) &\coloneqq \begin{cases}
           \frac{1}{2} \sum_k w^-_{ikjk}(t), & c = \mathrm{pp},\\[1ex]
           \frac{1}{2}\sum_k w_{ikjk}(t), & c = \mathrm{ph},\\[1ex]
           -\frac{1}{2}\sum_{k }w_{ikkj}(t) , & c = \overline{\mathrm{ph}}\,,
         \end{cases} 
    \label{eq:def_w_loc}
\end{align}
and the collision integral, 
\begin{align}
    \mathcal{I}^{c}(t) &\coloneqq \frac{1}{2}
   \sum_{klm}\big\{ w_{iklm}(t) \chi^{c,\mathrm{K}}_{lmjk}(t) - \chi^{c,\mathrm{K}}_{iklm}(t) w_{lmjk}(t) \big\}\nonumber\\
   &= \frac{1}{2}\mathrm{Tr}_2\big[\bm{\chi}^{c,\mathrm{K}},\bm{w}^c\big](t) \label{eq:i-def-chi-k}.
\end{align}
As we will see in Sec.~\ref{sss:conservation_laws}, the definition \eqref{eq:g2-sym-chi} is the basis for the derivation of conserving approximations in our quantum fluctuations approach.

\subsection{Hartree--Fock GKBA and G1--G2 scheme}\label{ss:f1-g2}
As is well-known, the cubic scaling with $N_\mathrm{t}$ can be reduced by first invoking the  generalized Kadanoff-Baym ansatz (GKBA) \cite{lipavsky_generalized_1986}, where the time-off-diagonal NEGFs are reconstructed using the retarded and advanced components, and second, the off-diagonal dynamics are treated on the Hartree-Fock level (HF-GKBA):
\begin{subequations}
\begin{align}
   \mathrm{i}\hbar \partial_t G^\gtrless(t,t') &=h^\mathrm{HF}(t) G^\gtrless(t,t'), &&t >t',\label{eq:HFGKBA_1} \\
  - \mathrm{i}\hbar \partial_{t'} G^\gtrless(t,t') &= G^\gtrless(t,t') h^\mathrm{HF}(t'), &&t<t'\,.\label{eq:HFGKBA_2}
\end{align}
\end{subequations}
The HF-GKBA reduces the  runtime to $\mathcal{O}(N^2_\mathrm{t})$, however, only for the 2B self-energy, whereas for the $GW$ approximation and the $T$-matrix approximations, the $\mathcal{O}(N^3_\mathrm{t})$ scaling persists due to the necessity of solving the BSE for the generalized susceptibility, cf. Eq.~\eqref{eq:BSE_chi}. 
Recently it was observed 
 that the HF-GKBA scaling can be brought even to $\mathcal{O}(N^1_\mathrm{t})$, and this not just for 2B, but also for $GW$ and $T$-matrix and combinations thereof, such as the dynamically screened ladder approximation \cite{joost_prb_22}. This is achieved with the G1--G2 scheme \cite{schluenzen_prl_20,joost_prb_20}, by reformulating the set of integro-differential equations into a set of time-local differential equations.

 This time-local structure becomes particularly transparent when the present channel notation is applied. Then, with the HF-GKBA the BSE~\eqref{eq:BSE_chi} simplifies to a mean-field dynamics of $\chi^{c}(t,t')$,
\begin{subequations}
\begin{align}
    \mathrm{i}\hbar\partial_t \bm{\chi}^{c,\gtrless}(t,t') &= \bm{\mathfrak{h}}^{(-),c}(t)\bm{\chi}^{c,\gtrless}(t,t'), &  &t>t', \label{eq:2t-chi_1}\\
    -\mathrm{i}\hbar\partial_{t'} \bm{\chi}^{c,\gtrless}(t,t') &= \bm{\chi}^{c,\gtrless}(t,t')\bm{\mathfrak{h}}^{(-),c\dagger}(t'), & &t<t', \;\label{eq:2t-chi_2}
\end{align}
\end{subequations}
with the effective two-particle Hamiltonian,
\begin{equation}
   \bm{\mathfrak{h}}^{(-),c}(t)\coloneqq  \bm{\mathfrak{h}}^{\mathrm{HF},c}(t)+\bm{\mathfrak{h}}^{\mathrm{cor},(-),c}(t), \label{eq:hfrak-def}
\end{equation}
where the first term is an effective two-particle Hartree-Fock Hamiltonian,
\begin{align}
    \bm{\mathfrak{h}}^{\mathrm{HF},c}(t) &\coloneqq\begin{cases}
        h^\mathrm{HF}(t)\otimes_c\mathbbm{1}+\mathbbm{1}\otimes_c h^\mathrm{HF}(t), & c = \mathrm{pp},\\[2ex]
        h^\mathrm{HF}(t)\otimes_c \mathbbm{1}-\mathbbm{1}\otimes_c h^\mathrm{HF,T}(t) , & c = \mathrm{ph},\overline{\mathrm{ph}}\,,
    \end{cases}\quad \label{eq:2p-HF-Hamiltonian}
\end{align}
and ``$\mathrm{T}$'' denotes the matrix transpose.
The second term in Eq.~\eqref{eq:hfrak-def} is a correlation part of the effective Hamiltonian:
\begin{align}
    \bm{\mathfrak{h}}^{\mathrm{cor},(-),c}(t) \coloneqq \mathrm{i}\hbar\bm{\chi}^{0,c,\Delta}(t)\bm{w}^{(-),c}(t). \label{eq:corr_h}
\end{align}
Finally, the dynamics of the generalized susceptibilities on the time diagonal, which complements equations \eqref{eq:2t-chi_1} and \eqref{eq:2t-chi_2}, are
\begin{align}
    \mathrm{i}\hbar\partial_t \bm{\chi}^{c,\gtrless}(t) =&\,\bm{\mathfrak{h}}^{(-),c}(t)\bm{\chi}^{c,\gtrless}(t) - \bm{\chi}^{c,\gtrless}(t)\bm{\mathfrak{h}}^{(-),c\dagger}(t)\nonumber\\&+ \tilde{\bm{\mathcal{R}}}^{c,\gtrless}(t), \label{eq:eom_chi_HFGKBA}
\end{align}
where $\tilde{\mathcal{R}}^c$ is defined by
\begin{align}
    \tilde{\mathcal{R}}^{c,\gtrless}_{ijkl}(t) \coloneqq \begin{cases}
        \mathcal{R}^{\mathrm{pp},\gtrless}_{ijkl}(t)-\mathcal{R}^{\mathrm{pp},\gtrless}_{ijlk}(t), & c =\mathrm{pp}, \\[1ex]
        \mathcal{R}^{c,\gtrless}_{ijkl}(t), & c= \mathrm{ph},\overline{\mathrm{ph}}\,,
    \end{cases} \label{eq:def_r_tilde}
\end{align}
and $\mathcal{R}$, in the three channels, is given by
\begin{align}
 \mathcal{R}_{ijkl}^{c,\gtrless}(t)\coloneqq  \mathrm{i}\hbar\begin{cases}
        G^\gtrless_{ik}(t) \mathcal{I}_{jl}^\mathcal{G}(t)+ G^\gtrless_{jl}(t)\mathcal{I}_{ik}^\mathcal{G}(t), & c\! = \mathrm{pp},\\[1.5ex]
        G^\gtrless_{ik}(t) \mathcal{I}_{jl}^{\mathcal{G}}(t) + G^\lessgtr_{jl}(t) \mathcal{I}_{ik}^{\mathcal{G}}(t) , &\! c = \mathrm{ph}, \\[1.5ex]
        -G^\gtrless_{il}(t) \mathcal{I}_{jk}^{\mathcal{G}}(t)- G^\lessgtr_{jk}(t)\mathcal{I}_{il}^{\mathcal{G}}(t), &\! c=\overline{\mathrm{ph}}\,.
    \end{cases}
\end{align}
This residual term can be understood as the correlation contribution to the dynamics of the time-diagonal ideal generalized susceptibility, $\chi^0(t)$, cf.~Eq.~\eqref{eq:BSE_chi}. \\
To compare with the G1--G2 scheme, we present the channel-specific equation of motion for the correlated part of the two-particle NEGF: 
\begin{align}
    \mathrm{i}\hbar \partial_t \bm{\mathcal{G}}^c(t) =&\, \bm{\mathfrak{h}}^{(-),c}(t)\bm{\mathcal{G}}^c(t)- \bm{\mathcal{G}}^c(t)\bm{\mathfrak{h}}^{(-),c\dagger}(t)\nonumber\\&+\bm{\Psi}^{(-),c}(t),
\label{eq:eom-cal-g}
\end{align}
with the second-order scattering contribution, 
\begin{align}
    \bm{\Psi}^{(-),c}(t) = \big\{ \bm{\chi}^{0,c,>}(t)\bm{w}^{(-),c}(t)\bm{\chi}^{0,c,<}(t)-(>\leftrightarrow <)\big\}.
\end{align}
Notice that the scattering term $\Psi$ (or $\Psi^-$) is the same for all approximations (except for exchanging $w \leftrightarrow w^-$), and the same holds for $\bm{\mathfrak{h}}^{\rm HF,c}$. Thus, different channels differ on the time diagonal only in $\bm{\mathfrak{h}}^{\mathrm{cor},(-),c}$.

\begin{table}[t]
\caption{\label{t:approximations_G1-G2}
List of  self-energy approximations ($\Sigma$) within the G1--G2 scheme (equivalent to NEGF within the HF-GKBA) and their respective channels for the two-particle correlated NEGF $\mathcal{G}$. Included are the $T$-matrix approximation in the particle-particle channel (TPP) and the particle-hole channel (TPH) as well as the $GW$ approximation. X denotes exchange contributions.} 
\begin{ruledtabular}
\renewcommand{\arraystretch}{1.5}
\begin{tabular}{ c c c c }
Channel & $\Sigma$ & \begin{tabular}{c} Effective \\ Hamiltonian \end{tabular} & \begin{tabular}{c} Scattering \\ Term \end{tabular} \\ 
\colrule

$\mathrm{pp}$ & TPP & $\bm{\mathfrak{h}}$ & $\bm{\Psi}\mathrlap{^-}$ \\ 
\colrule

$\mathrm{ph}$ & 
\begin{tabular}{@{}c@{}} 
    TPH \\ 
    TPH+X 
\end{tabular} & 
\begin{tabular}{@{}c@{}} 
    $\bm{\mathfrak{h}}$ \\ 
    $\bm{\mathfrak{h}}\mathrlap{^-}$ 
\end{tabular} & 
\begin{tabular}{@{}c@{}} 
    $\bm{\Psi}$ \\ 
    $\bm{\Psi}\mathrlap{^-}$ 
\end{tabular} \\ 
\colrule

$\overline{\mathrm{ph}}$ & 
\begin{tabular}{@{}c@{}} 
    $GW$ \\ 
    $GW$+X 
\end{tabular} & 
\begin{tabular}{@{}c@{}} 
    $\bm{\mathfrak{h}}$ \\ 
    $\bm{\mathfrak{h}}\mathrlap{^-}$ 
\end{tabular} & 
\begin{tabular}{@{}c@{}} 
    $\bm{\Psi}$ \\ 
    $\bm{\Psi}\mathrlap{^-}$ 
\end{tabular} \\ 

\end{tabular}
\end{ruledtabular}
\end{table}

Table~\ref{t:approximations_G1-G2} lists a number of relevant approximations within the G1--G2 scheme and the associated BSE. 
Note that, in contrast to the TPP approximation, the $GW$(+X) and TPH(+X) self-energies separately do not obey the correct exchange symmetry of the two-particle NEGF, i.e.,
\begin{equation}
    \mathcal{G}_{ijkl} \neq - \mathcal{G}_{ijlk}. \label{eq:broken_exchange_symmetry}
\end{equation}
This symmetry is restored by the sum of TPH+X  and $GW$+X \cite{joost_prb_22}.\\
Even though the reformulation of the HF-GKBA, with the help of the G1--G2 scheme, has allowed to drastically reduce the runtime scaling to $\mathcal{O}(N_\mathrm{t})$ \cite{schluenzen_prl_20}, the appearance of the two-particle NEGF $\mathcal{G}$ (a rank-4 tensor of size $N_\mathrm{b}\times N_\mathrm{b}\times N_\mathrm{b} \times N_\mathrm{b}$) comes with considerable disadvantages. While the dimensionality of single-particle quantities (single-particle NEGF, self-energies, which are rank-2 tensors of size $N_\mathrm{b}\times N_\mathrm{b}$)  
can often be strongly reduced by taking advantage of the specific structure of the pair interaction or symmetries of the Hamiltonian, this is not necessarily the case for the two-particle NEGF. Consequently, the unfavorable scaling of $\mathcal{G}$ with the basis size $N_\mathrm{b}$ turns out to be a severe bottleneck for this approach, even for systems in one and two dimensions. Further, introducing $\mathcal{G}$ does not fix the instabilities already present within the HF-GKBA. Instead, the ability to treat advanced self-energies, such as the dynamically screened ladder approximation, even amplifies the problem~\cite{joost_prb_22}. 
These instability issues can often be circumvented by enforcing some of the so-called $N$-representability conditions (see, e.g., Refs.~\cite{coleman_reduced_2000,mazziotti_reduced-density-matrix_2007,lackner_time-dependent_2017} for an overview), namely the trace consistency and positive semi-definiteness of the two-particle and two-hole reduced density matrix. Algorithms used for this (``purification'' \cite{joost_prb_22,donsa_prr_23}), however, introduce a considerable computational overhead and significantly worsen the runtime scaling again. As a consequence, the G1--G2 scheme with advanced self-energies is currently only applicable to nonequilibrium simulations for systems with $N_\mathrm{b} \lesssim  150$, see, e.g., Refs.~\cite{borkowski_pss_22,pavlyukh_time-linear_2022,Brezinova2024,joost_prr_25}. \\

We will now show that these problems can be overcome with an alternative approach to nonequilibrium dynamics that uses NEGF-based quantum fluctuations.

\section{NEGF-based quantum fluctuations  ($\delta$NEGF) approach}
\label{s:theory}
\subsection{Definitions. Equations of motion}\label{ss.eom-dg}
The idea of the NEGF-based quantum fluctuations ($\delta$NEGF) approach is to consider fluctuations of operators that correspond to specific real-time components of single-particle NEGFs. We define these operators for the three channels in the following way:
\begin{align}
   \mathrm{i}\hbar \hat{G}^c_{ij}(t) \coloneqq \begin{cases}
        \hat{a}_i(t) \hat{a}_j(t), & c = \mathrm{pp},\\[1ex]
        \hat{a}_i(t)\hat{a}^\dagger_j(t), & c = \mathrm{ph},\overline{\mathrm{ph}}.
    \end{cases}
    \label{eq:def-gf-operator}
\end{align}
In the particle-particle channel, the expectation value of $\hat{G}^{c}$ gives rise to anomalous NEGFs on the time diagonal, whereas, in the particle-hole channels, $\hat{G}^c$ corresponds to the operator of the time-diagonal greater component of the regular single-particle NEGF. We define the channel-specific fluctuations as deviations of the operators \eqref{eq:def-gf-operator} from the standard NEGF,
\begin{align}
    \delta\hat{G}^c(t)\coloneqq \hat{G}^c(t) - \big\langle \hat{G}^c(t)\big\rangle.
    \label{eq:def-fluctuations}
\end{align}
As we are considering a system with conserved particle number, anomalous NEGFs vanish so that $\delta\hat{G}^\mathrm{pp}= \hat{G}^\mathrm{pp}$. Notice that, due to the anticommutation relations of the field operators, the fluctuations of the greater and  lesser components coincide on the time diagonal [cf. Eqs.~ \eqref{eq:ggl-def}, \eqref{eq:gg-op-def}, and  \eqref{eq:gl-op-def} and, analogously, for the anomalous NEGFs]. 

The fluctuations $\delta\hat{G}^c$ are the key quantities of the $\delta$NEGF approach, as their correlation functions correspond to the different channel-specific generalized susceptibilities,
\begin{align}
    \bm{\chi}^c_{(ij)(kl)}(z,z') = \alpha_c \big\langle \mathcal{T}_\mathcal{C}\big\{\delta\hat{G}^c_{ij}(z) \delta\hat{G}_{kl}^{c,\dagger}(z')\big\}\big\rangle,
\end{align}
where 
\begin{align}
    \alpha_c \coloneqq \mathrm{i}\hbar\times \begin{cases}
        -1, & c = \mathrm{pp},\overline{\mathrm{ph}},\\[1ex]
       1, & c = \mathrm{ph}.
    \end{cases}
\end{align}
This yields the following greater and lesser components
\begin{subequations}
\begin{align}
    \bm{\chi}^{c,>}_{(ij)(kl)}(t,t') &= \alpha_c  \big\langle \delta \hat{G}_{ij}^c(t) \delta\hat{G}_{kl}^{c\dagger}(t') \big\rangle,\label{eq:chi>_dG}\\[1ex]
    \bm{\chi}^{c,<}_{(ij)(kl)}(t,t') &= \alpha_c \big\langle   \delta\hat{G}_{kl}^{c\dagger}(t')\delta\hat{G}_{ij}^c(t) \big\rangle. \label{eq:chi<_dG}
\end{align}
\end{subequations}
The key idea of the $\delta$NEGF theory is to solve Eq.~\eqref{eq:EOM_G_GWX-QF} without solving an equation for the generalized susceptibilities, which are two-particle quantities as well. Instead, we evaluate these functions by solving an equation  of motion for the single-particle fluctuations \eqref{eq:def-fluctuations}. Their equation of motion follows by subtracting from the equation of motion of the two-operator product $\hat G^c$ the equation for the single-particle density matrix \cite{schroedter_cmp_22}. 
In the particle-hole channels, we obtain 
\begin{align}
    \mathrm{i}\hbar \partial_t \delta\hat{G}^c(t) =& \big[h^c, \delta\hat{G}^c\big](t)+\big[\delta\hat{\Sigma}^{0,c},G^<\big](t)
    \nonumber\\&+ \delta\big\{ \big[\delta\hat{\Sigma}^{0,c},\delta\hat{G}^c\big]\big\}(t), \quad
    c = \mathrm{ph},\overline{\mathrm{ph}},\label{eq:eom-dg}
\end{align}
where $\delta\hat{\Sigma}^{0,c}\coloneqq \Sigma^{0,c}[\delta\hat{G}^c]$. The first term in Eq.~\eqref{eq:eom-dg} describes the interaction of the particle-hole fluctuations with the mean-field of the particles, while the second term describes the interaction of the fluctuations of the mean-field with the particles. The final term includes second-order fluctuations, i.e., fluctuations of fluctuations.
It is evident that the latter term gives rise to a hierarchy of equations: the equation of motion for the correlation function of a product of two $\delta \hat G^c$ couples to the equation of motion for the correlation function of a product of three $\delta \hat G^c$, and so on \cite{schroedter_cmp_22}. Consequently, we have to find suitable approximations to decouple the hierarchy of fluctuations, which will be done in Sec.~\ref{s:approx}. 

But before doing this, we present the equation of the fluctuations in the particle-particle channel:
\begin{align}
    \mathrm{i}\hbar\partial_t \delta\hat{G}^\mathrm{pp}(t) =& h(t) \delta\hat{G}^\mathrm{pp}(t)+\hat{G}^<(t) \hat{\Delta}(t) -\big[\dots\big]^\mathrm{T}\nonumber\\&+ \frac{1}{\mathrm{i}\hbar}\hat{\Delta}(t), \label{eq:eom-F}
\end{align}
where ``$[\dots]^\mathrm{T}$'' denotes the matrix transpose of the previous terms,  and we introduced the pairing self-energy operator: 
\begin{equation}
    \hat{\Delta}_{ij}(t) \coloneqq \Delta_{ij}[\delta\hat{G}^\mathrm{pp}](t) \coloneqq \mathrm{i}\hbar \sum_{mn} w_{ijmn}(t)\delta\hat{G}^\mathrm{pp}_{mn}(t). \label{eq:pairing_self-energy}
\end{equation}
In Eq.~\eqref{eq:eom-F}, the first term and its transpose describe the interaction of the particle-particle fluctuations  with the single-particle Hamiltonian, whereas the second term and its transpose, together with the last term, describe the interaction of the particles and holes with the pairing field. Again, the operator product gives rise to a hierarchy of equations: the BBGKY hierarchy of reduced density matrices \cite{bonitz_qkt}.

\subsection{Approximations}
\label{s:approx}
In the following, we introduce two sets of approximations, for the particle-hole channels and the particle-particle channel, respectively, that are fundamental in terms of their physical content. Moreover, by construction, they guarantee the positive semi-definiteness of the (modified) particle-hole reduced density matrix or the two-particle/two-hole reduced density matrix for the corresponding channels.
These approximations are essentially based on an appropriate linearization of the equations of motion of the fluctuations, which enables a solution of the operator equations. At this point, we will first consider the different channels separately, due to the slight asymmetry of the equations of motion of the respective fluctuations [cf. Eqs.~\eqref{eq:eom-dg} and \eqref{eq:eom-F}].

\subsubsection{Particle-hole approximations}\label{ss:spa}
\textbf{Approximation of second moments}: The simplest approximation is to neglect the second-order fluctuations (the last term) in Eq.~\eqref{eq:eom-dg},
\begin{align}
    \delta\big\{\delta \hat{G}^c_{ik}(t)\delta\hat{G}^c_{jl}(t)\big\} \to 0\,,\label{eq:2M}
\end{align}
giving rise to the equation of motion 
\begin{align}
    \mathrm{i}\hbar \partial_t \delta\hat{G}^c(t) =& \big[h^c, \delta\hat{G}^c\big](t)+\big[\delta\hat{\Sigma}^{0,c},G^<\big](t).
\label{eq:eom-dg_2M}
\end{align}
The neglect of the second-order fluctuations leads to different equations in the two channels. In the $\overline{\mathrm{ph}}$ channel, only the interactions of the particle-hole fluctuations with the Hartree mean-field and of the  Hartree field induced by the fluctuations with the single-particle NEGF appear. On the other hand, in the $\mathrm{ph}$ channel, these processes are considered with their corresponding exchange parts instead. Consequently, this linear-order approximation describes the essential processes of the respective particle-hole channels only. All contributions from the other respective sub-channel are neglected, which represents a relatively drastic approximation. This is further illustrated by the fact that fluctuations do not vanish, even for uncorrelated quantum systems. Nevertheless, if exchange effects are weak, this approximation offers a significant numerical advantage, as the calculation of exchange terms is often very costly. 

In the following, we will, therefore, consider the approximation of second moments only within the $\overline{\mathrm{ph}}$ channel. This approximation is the basis for the $\delta$RPA approximation, in the $\overline{\mathrm{ph}}$ channel, cf. Sec.~\ref{sec:AoA}. More details of this approximation are summarized in Tab.~\ref{t:NEGF-QF_approximations}.

\textbf{Polarization approximation (PA)}:
  The PA does not neglect the second-order fluctuations, but treats them in a properly linearized form \cite{schroedter_cmp_22}:
\begin{align}
    &\delta\big\{\delta\hat{G}^c_{ik}(t) \delta\hat{G}^c_{jl}(t)\big\}\approx - \big\{ \delta\hat{G}^c_{il}(t) G^<_{jk}(t) + G^>_{il}(t) \delta \hat{G}^c_{jk}(t)\big\}\,, \label{eq:pa}
\end{align}
and can be understood as the analog of the Hartree-Fock approximation, at the level of single-particle fluctuations. The PA transforms Eq.~\eqref{eq:eom-dg} into
\begin{align}
    \mathrm{i}\hbar \partial_t \delta \hat{G}^c(t) = \big[h^\mathrm{HF},\delta\hat{G}^c\big](t) +\big[\delta\hat{\Sigma}^{\mathrm{HF}}, G^<\big](t), \label{eq:dg-pa}
\end{align}
where $\delta\hat{\Sigma}^{\mathrm{HF}}\coloneqq \Sigma^\mathrm{HF}[\delta\hat{G}^c]$.

The PA \eqref{eq:pa} essentially adds self-energy contributions originating from the respective opposite particle-hole channel, as we have $\Sigma^\mathrm{HF} = \Sigma^{0,\mathrm{ph}} + \Sigma^{0,\overline{\mathrm{ph}}}$, cf. definition \eqref{eq:def-sigma0}.\\
At this point, it is important to emphasize that, even though Eq.~\eqref{eq:dg-pa} is symmetric with respect to the two different particle-hole channels, it leads to two distinct approximations. This is due to the fact that, the PA violates exchange symmetries of the susceptibilities [see also Eq.~\eqref{eq:broken_exchange_symmetry} for $GW$ and TPH]. Thus, it is not possible to interchange $\chi^\mathrm{ph}\leftrightarrow \chi^{\overline{\mathrm{ph}}}$ in the equation of motion for the single-particle NEGF \eqref{eq:EOM_G_GWX-QF}, giving rise to different dynamics depending on the chosen particle-hole channel. This approximation is the basis for the $\delta GW$+X approximation, in the $\overline{\textnormal{ph}}$ channel and, for the $\delta$TPH+X approximation, in the ph channel, respectively, cf. Sec.~\ref{sec:AoA} and Tab.~\ref{t:NEGF-QF_approximations}. 

\textbf{Reduced polarization approximation}: This approximation is intermediate between the PA and the approximation of second moments and given by
\begin{align}
    \delta\big\{\delta \hat{G}^c_{ik}(t)\delta\hat{G}^c_{jl}(t)\big\} \approx - G^>_{il}(t) \delta \hat{G}^c_{jk}(t), \label{eq:approx_GW}
\end{align}
which corresponds to neglecting the first term on the r.h.s. of Eq.~\eqref{eq:pa}. This, in turn, breaks the symmetry of Eq.~\eqref{eq:eom-dg} for the different particle-hole channels, leading to two different equations of motion for $\delta\hat{G}^c$. Applying this approximation gives rise to
\begin{equation}
    \mathrm{i}\hbar\partial_t \delta\hat{G}^c(t) = \big[h^\mathrm{HF}, \delta\hat{G}^c\big](t) + \big[\delta\hat{\Sigma}^\mathrm{0,c}, G^<\big](t). \label{eq:EOM_dG_GW}
\end{equation}

This approximation is the basis for the $\delta GW$ approximation, in the $\overline{\textnormal{ph}}$ channel and for the $\delta$TPH approximation, in the $\mathrm{ph}$ channel, respectively, cf. Sec.~\ref{sec:AoA} and Tab.~\ref{t:NEGF-QF_approximations}. 

It is possible to construct numerous further approximations via proper linearization of the equation of motion for the fluctuations \eqref{eq:eom-dg}. 

\subsubsection{Particle-particle approximations}\label{ss:stma}
Analogously to the PA \eqref{eq:pa} for the particle-hole channels, we now introduce the 

\textbf{Ladder approximation (LA)}: It is obtained by  
applying the following mean-field approximation in the particle-particle channel:
\begin{align}
    \hat{G}^<_{ij}(t) \delta\hat{G}^\mathrm{pp}_{kl} (t)\approx \,&G^<_{ij}(t) \delta\hat{G}^\mathrm{pp}_{kl}(t) - G^<_{kj}(t)\delta\hat{G}^\mathrm{pp}_{il}(t)\nonumber\\&+G^<_{lj}(t) \delta\hat{G}^\mathrm{pp}_{ik}(t). \label{eq:LA}
\end{align}
Applying the LA \eqref{eq:LA} to Eq.~\eqref{eq:eom-F} gives rise to
\begin{align}
    \mathrm{i}\hbar \partial_t  \delta\hat{G}^\mathrm{pp}(t) =\,& h^\mathrm{HF}(t) \delta\hat{G}^\mathrm{pp}(t) + G^< (t)\hat{\Delta}(t)-  \big[\dots\big]^\mathrm{T}\nonumber \\&+\frac{1}{\mathrm{i}\hbar} \hat{\Delta}(t), \label{eq:eom_F_TPP}
\end{align}
where the ideal single-particle Hamiltonian $h$ is replaced by the effective single-particle Hartree-Fock Hamiltonian, whereas the single-particle NEGF operator $\hat{G}^<$ is replaced by $G^<$. The LA is the basis for the $\delta$TPP approximation, 
cf. Sec.~\ref{sec:AoA} and Tab.~\ref{t:NEGF-QF_approximations}.

\subsubsection{Assessment of approximations}\label{sec:AoA}
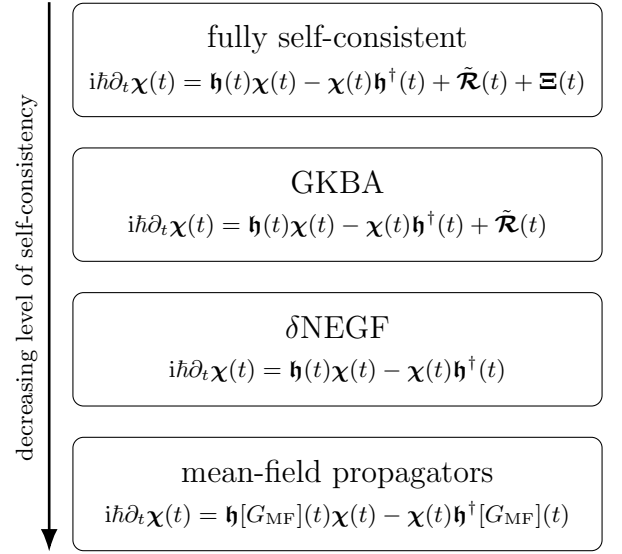
\begin{figure}[t]
    \centering
    \begin{tikzpicture}[
  box/.style={draw, minimum width=\boxwidth, minimum height=\boxheight, align=center, inner sep=2mm, rounded corners},
  arrowtext/.style={font=\small, rotate=90, anchor=center}, 
  every node/.style={font=\small}
]
\def\boxwidth{7cm}     
\def\boxheight{1.5cm}     
\def\boxsep{4mm}         
\def\arrowoffset{3mm}    
\def\textoffset{3mm}     

\node[box] (b1) {{\large fully self-consistent}\\[1.5mm] {$\mathrm{i}\hbar \partial_t \bm\chi(t) =\bm{\mathfrak{h}}(t)\bm{\chi}(t)-\bm{\chi}(t)\bm{\mathfrak{h}}^{\dagger}(t)+\tilde{\bm{\mathcal{R}}}(t) + \bm{\Xi}(t) $}};
\node[box, below=\boxsep of b1] (b2) {\large{GKBA}\\[1.5mm] $\mathrm{i}\hbar \partial_t \bm\chi(t) =\bm{\mathfrak{h}}(t)\bm{\chi}(t)-\bm{\chi}(t)\bm{\mathfrak{h}}^{\dagger}(t)+\tilde{\bm{\mathcal{R}}}(t)$};
\node[box, below=\boxsep of b2] (b3) {\large{$\delta$NEGF}\\[1.5mm] $\mathrm{i}\hbar \partial_t \bm\chi(t) =\bm{\mathfrak{h}}(t)\bm{\chi}(t)-\bm{\chi}(t)\bm{\mathfrak{h}}^{\dagger}(t)$};
\node[box, below=\boxsep of b3] (b4) {\large{mean-field propagators}\\[1.5mm] $\mathrm{i}\hbar \partial_t \bm\chi(t) =\bm{\mathfrak{h}}[G_\mathrm{MF}](t)\bm{\chi}(t)-\bm{\chi}(t)\bm{\mathfrak{h}}^{\dagger}[G_\mathrm{MF}](t)$};

\node[inner sep=0pt, fit=(b1) (b2) (b3) (b4)] (group) {};

\coordinate (arrowtop)    at ($(group.north west)+(-\arrowoffset,0)$);
\coordinate (arrowbottom) at ($(group.south west)+(-\arrowoffset,0)$);

\draw[line width=1pt, -{Latex[length=3mm]}] (arrowtop) -- (arrowbottom);

\node[arrowtext] at ($(arrowtop)!0.5!(arrowbottom)+(-\textoffset,0)$) {decreasing level of self-consistency};
\end{tikzpicture}

     \caption{Levels of self-consistency for the different approaches. Here, ``fully self-consistent'' refers to the self-consistent solution of the KBEs, for any of the discussed self-energy approximations.  $\Xi$ denotes the contribution arising from the collision term inside the integral term of the BSE \eqref{eq:BSE_chi} and is of the form $\Xi \sim \int  I Gw\chi$. Application of the GKBA is equivalent to neglecting $\Xi$. Within the $\delta$NEGF approach, the $\mathcal{R}$ term is neglected what constitutes an intermediate level of self-consistency. Finally,  ``mean-field propagators'' refers to inserting single-particle NEGFs at the mean-field level into the BSE (``one-shot'' scheme).}
    \label{fig:sc-hierarchy}
\end{figure}

In the following, we put the approximations of our $\delta$NEGF theory into the context of the general NEGF framework. To this end, we compare the equations of motion of the generalized susceptibilities $\chi^c$ for the $\delta$NEGF approach to the ones derived for the HF-GKBA in Sec.~\ref{ss:f1-g2}. Here, for simplicity, we will primarily focus on the reduced PA and LA, Eqs.~\eqref{eq:approx_GW} and  \eqref{eq:LA}. Then, using $\mathrm{i}\hbar\, G^\Delta(t) \equiv \mathbbm{1}$, the generalized susceptibilities obey the time-off-diagonal equations of motion
\begin{subequations}
\begin{align}
    \mathrm{i}\hbar\partial_t \bm{\chi}^{c,\gtrless}(t,t') &= \bm{\mathfrak{h}}^{c}(t)\bm{\chi}^{c,\gtrless}(t,t'), \label{eq:2t-chi_1'} \\
    -\mathrm{i}\hbar\partial_{t'} \bm{\chi}^{c,\gtrless}(t,t') &= \bm{\chi}^{c,\gtrless}(t,t')\bm{\mathfrak{h}}^{c\dagger}(t'),\label{eq:2t-chi_2'} 
\end{align}
\end{subequations}
hence, reproducing the expressions for $\chi^{c,\gtrless}$, within the HF-GKBA [cf. Eqs~\eqref{eq:2t-chi_1} and \eqref{eq:2t-chi_2}]. Further, considering the full PA, Eq.~\eqref{eq:pa}, recovers the exchange contributions corresponding to the replacement $\bm{\mathfrak{h}}\rightarrow \bm{\mathfrak{h}}^-$. If we consider the approximation of second-moments \eqref{eq:2M}, this corresponds to the replacement $h^\mathrm{HF}\rightarrow h^c$ in the effective two-particle Hamiltonian $\mathfrak{h}^{\mathrm{HF},c}$ [cf. Eq.~\eqref{eq:2p-HF-Hamiltonian}]. Within the framework of the GKBA, this would then be equivalent to considering a  lower-level GKBA in the form of a Hartree- or Fock-GKBA.

Next, we consider the time-diagonal dynamics:
\begin{align}
    \partial_t \chi^{c,\gtrless}(t) = \big\{\partial_t\chi^{c,\gtrless}(t,t') + \partial_{t'}\chi^{c,\gtrless}(t,t')\big\}\Big|_{t=t'}, \label{eq:continuity}
\end{align}
leading to the following equation of motion
\begin{align}
    \mathrm{i}\hbar \partial_t \bm{\chi}^{c,\gtrless}(t) = &\bm{\mathfrak{h}}^{c}(t)\bm{\chi}^{c,\gtrless}(t)-\bm{\chi}^{c,\gtrless}(t)\bm{\mathfrak{h}}^{c\dagger}(t)\,.\label{eq:EOM_L_PA}
\end{align}
This differs from the equation of motion for ${\chi}^{c,\gtrless}$ within the HF-GKBA, only by the $\mathcal{R}$ term [cf. Eq.~\eqref{eq:eom_chi_HFGKBA}], which emerges from the time derivative of the BSE [cf. Eq.~\eqref{eq:BSE_chi}] when treating the time diagonal of the leading ${\tilde\chi}^0$ as correlated. Hence, we conclude that the $\delta$NEGF approach differs from the HF-GKBA only in terms of the self-consistent treatment of $\tilde\chi^0$, whereas the self-energy approximations remain unchanged. 
Note that the same reduced self-consistency does not apply to the ${\chi}^0$ entering in the time integral of Eq.~\eqref{eq:BSE_chi}. Such a treatment would correspond to replacing all NEGFs by mean-field NEGFs in the equation of motion for ${\chi}$, which is not the case for $\delta$NEGF, cf. Fig.~\ref{fig:sc-hierarchy}.  

The notion that a change in the degree of self-consistency leads to the emergence of additional terms might seem unintuitive, as this is not a common theme when working with the Dyson equation or the KBEs. In the equation of motion of $\chi$, however, self-consistency is directly connected to individual terms. For instance, a fully self-consistent treatment of ${\chi}^0$ and ${\tilde\chi}^0$ also for off-diagonal times, which corresponds to the self-consistent solution of the KBEs, leads to the appearance of an additional contribution, $\Xi \sim \int I G w \chi$, in the equation of motion of ${\chi}$, cf. Fig.~\ref{fig:sc-hierarchy}. 

In summary, the reduced PA (full PA) of the particle-hole channels corresponds to a $\delta$NEGF version of the $GW$(+X) and TPH(+X) self-energies with an intermediate level of self-consistency between the HF-GKBA and a full mean-field treatment at the level of the BSE. In the following, they will be denoted as $\delta$$GW$(+X) and $\delta$TPH(+X), short for $\delta$NEGF+$GW$(+X) and $\delta$NEGF+TPH(+X). Likewise, the LA of the pp channel is associated with the TPP self-energy and will be denoted as $\delta$TPP. Additionally, we denote the approximation of second moments for the $\overline{\mathrm{ph}}$ channel as $\delta$RPA as it emerges from $GW$ when treating the single-particle Hamiltonian in the equation of motion for the fluctuations on a mean-field (Hartree) level, cf. Eq.~\eqref{eq:dg-pa}.  

\begin{table*}[t!]
\caption{\label{t:NEGF-QF_approximations}  Overview of the NEGF-based quantum fluctuations ($\delta$NEGF) approach. Shown are the generalized susceptibilities $\chi^c$ and $\chi^{0,c}$ on the Keldysh contour $\mathcal{C}$ giving rise to the associated real-time components $\chi^{c,\gtrless}$ and $\chi^{c,\mathrm{K}} = \chi^{c,>}+\chi^{c,<}$ (analogously for $\chi^{0,c}$).  Further shown are the self-energy approximations, $\Sigma$, with their corresponding equations of motion for the different channels. The equation of motion for the single-particle Green function $G^<$ is fixed for each channel and only the equation for the fluctuations $\delta\hat{G}^c$ varies, within each channel, according to the approximation. $\delta$TPP denotes the TPP approximation within the $\delta$NEGF approach and so. The equations contain the channel-specific single-particle Hamiltonian $h^c$ [Eq.~\eqref{eq:def_eff_h_c}] with $h^\mathrm{H} =h^{\overline{\mathrm{ph}}} = h+\Sigma^\mathrm{H}$ [Eq.~\eqref{eq:Sigma_H}] and $h^\mathrm{F}= h^\mathrm{ph} = h+\Sigma^\mathrm{F}$ [Eq.~\eqref{eq:Sigma_F}]. Moreover, $h^\mathrm{HF} = h+\Sigma^\mathrm{HF} = h+\Sigma^\mathrm{H}+\Sigma^\mathrm{F}$ denotes the effective single-particle Hartree-Fock Hamiltonian and $\hat{\Delta}$ the pairing self-energy~\eqref{eq:pairing_self-energy}. The local interaction, $w^c_\mathrm{loc}$, is defined in Eq.~\eqref{eq:def_w_loc} and the collision term $\mathcal{I}^{c}$, which involves the Keldysh component $\chi^{\textnormal{c,K}}$, in Eq.~\eqref{eq:i-def-chi-k}.}
\begin{ruledtabular}
\renewcommand{\arraystretch}{1.5} 
\begin{tabular}{ c c c c }
Channel & Susceptibilities & $\Sigma$ & Equations of Motion \\ 
\colrule

$\mathrm{pp}$& 
$\begin{aligned}
   \chi^\mathrm{pp}_{ijkl}(z,z') &= -\mathrm{i}\hbar\big\langle \mathcal{T}_\mathcal{C}\{\delta\hat{G}^\mathrm{pp}_{ij}(z)\delta\hat{G}^{\mathrm{pp},\dagger}_{kl}(z')\}\big\rangle \\ 
   \chi^{0,\mathrm{pp}}_{ijkl}(z,z') &= \mathrm{i}\hbar G_{ik}(z,z') G_{jl}(z,z')
\end{aligned}$ & 
\begin{tabular}{@{}c@{}}
     \\[0.75ex] 
    $\delta$TPP 
\end{tabular} & 
$\begin{aligned}
    \mathrm{i}\hbar\partial_t G^<(t) &= [h+w^\mathrm{pp}_\mathrm{loc}, G^<](t) + \mathcal{I}^{\mathrm{pp}}(t) \\[0.75ex] 
    \mathrm{i}\hbar \partial_t \delta\hat{G}^\mathrm{pp}(t) &= h^{\mathrm{HF}}(t) \delta\hat{G}^\mathrm{pp}(t) + G^<(t)\hat{\Delta}(t) - [\dots]^{\mathrm{T}}+ \tfrac{1}{\mathrm{i}\hbar} \hat{\Delta}(t)
\end{aligned}$ \\ 
\colrule

$\mathrm{ph}$ & 
$\begin{aligned}
   \chi^{\mathrm{ph}}_{ijkl}(z,z') &= \mathrm{i}\hbar\big\langle \mathcal{T}_\mathcal{C}\{\delta\hat{G}^\mathrm{ph}_{il}(z)\delta\hat{G}^{\mathrm{ph},\dagger}_{kj}(z')\}\big\rangle \\ 
   \chi^{0,\mathrm{ph}}_{ijkl}(z,z') &= \mathrm{i}\hbar G_{ik}(z,z') G_{jl}(z',z)
\end{aligned}$ & 
\begin{tabular}{@{}c@{}}
     \\[0.75ex] 
    $\delta$TPH \\[0.75ex]
    $\delta$TPH+X
\end{tabular} & 
$\begin{aligned}
    \mathrm{i}\hbar\partial_t G^<(t) &= \big[h^{\mathrm{F}}+w^\mathrm{ph}_\mathrm{loc}, G^<\big](t)  + \mathcal{I}^{\mathrm ph}(t) \\[0.75ex] 
    \mathrm{i}\hbar \partial_t \delta \hat{G}^\mathrm{ph}(t) &= \big[h^{\mathrm{HF}}, \delta\hat{G}^\mathrm{ph}\big](t) + \big[\delta\hat{\Sigma}^{\mathrm{F}}, G^<\big](t) \\[0.75ex]
    \mathrm{i}\hbar \partial_t \delta\hat{G}^\mathrm{ph}(t) &= \big[h^{\mathrm{HF}}, \delta\hat{G}^\mathrm{ph}\big](t) + \big[\delta\hat{\Sigma}^{\mathrm{HF}}, G^<\big](t)
\end{aligned}$ \\ 
\colrule

$\overline{\mathrm{ph}}$ & 
$\begin{aligned}
   \chi^{\overline{\mathrm{ph}}}_{ijkl}(z,z') &= -\mathrm{i}\hbar\big\langle \mathcal{T}_\mathcal{C}\{\delta\hat{G}^{\overline{\mathrm{ph}}}_{ik}(z)\delta\hat{G}^{\overline{\mathrm{ph}},\dagger}_{lj}(z')\}\big\rangle \\ 
   \chi^{0,\overline{\mathrm{ph}}}_{ijkl}(z,z') &= -\mathrm{i}\hbar G_{il}(z,z') G_{jk}(z',z)
\end{aligned}$ & 
\begin{tabular}{@{}c@{}}
     \\[0.75ex] 
    $\delta$RPA\\[0.75ex]
    $\delta GW$ \\[0.75ex]
    $\delta GW$+X 
\end{tabular} & 
$\begin{aligned}
    \mathrm{i}\hbar\partial_t G^<(t) &= \big[h^{\mathrm{H}} +w^{\overline{\mathrm{ph}}}_\mathrm{loc}, G^<\big](t)  + \mathcal{I}^{\overline{\mathrm{ph}}}(t) \\[0.75ex] 
    \mathrm{i}\hbar \partial_t \delta\hat{G}^{\overline{\mathrm{ph}}}(t) &= \big[h^{\mathrm{H}}, \delta\hat{G}^{\overline{\mathrm{ph}}}\big](t) + \big[\delta\hat{\Sigma}^{\mathrm{H}}, G^<\big](t) \\[0.75ex]
    \mathrm{i}\hbar \partial_t \delta\hat{G}^{\overline{\mathrm{ph}}}(t) &= \big[h^{\mathrm{HF}}, \delta\hat{G}^{\overline{\mathrm{ph}}}\big](t) + \big[\delta\hat{\Sigma}^{\mathrm{H}}, G^<\big](t) \\[0.75ex]
    \mathrm{i}\hbar \partial_t \delta\hat{G}^{\overline{\mathrm{ph}}}(t) &= \big[h^{\mathrm{HF}}, \delta\hat{G}^{\overline{\mathrm{ph}}}\big](t) + \big[\delta\hat{\Sigma}^{\mathrm{HF}}, G^<\big](t) 
\end{aligned}$ \\ 

\end{tabular}
\end{ruledtabular}
\end{table*}

\subsubsection{Conservation laws and symmetries} \label{sss:conservation_laws}

Conservation laws play a pivotal role for reliable many-particle simulations and  in the construction and assessment of self-energy approximations.  
In NEGF theory, conservation laws are linked to the exchange symmetry of the two-particle Green function \cite{baym_conservation_1961}. Since the Hartree-Fock part of $G^{(2)}$ is conserving, conserving approximations require the symmetry 
\begin{align}
    \mathcal{G}_{ijkl}(t) = \mathcal{G}_{jilk}(t)\,,
\end{align}
where we may restrict ourselves to the time diagonal. This criterion, however, is not sufficient by itself to guarantee, for example, energy conservation. An additional condition on the time diagonal to ensure energy conservation is given by considering the analogous exchange symmetries of the approximate three-particle NEGF \cite{bonitz_qkt}. We emphasize that, within the $\delta$NEGF approach, the necessary symmetry requirements at the two-particle level are fulfilled by construction, as it holds for the Keldysh component of the generalized susceptibilities for all channels:
\begin{align}
    \chi^{c,\mathrm{K}}_{ijkl}(t) = \chi^{c,\mathrm{K}}_{jilk}(t)\,.
\end{align}
As the result, all approximation we used for the Hubbard model ($\delta$TPP, $\delta$TPH and $\delta GW$) and extended Hubbard model ($\delta$RPA and $\delta GW$+X) are energy conserving, which we verified both numerically and analytically, as we demonstrate in detail in appendix \ref{a:symmetries-energy_conservation}.

\subsection{Factorization of correlation functions}\label{s:factorization}
The central idea of our $\delta$NEGF approach is to solve the equations of motion for the fluctuations, $\delta \hat G^c$, and to reconstruct the generalized susceptibilities from these solutions, rather than solving the equations for two-particle quantities, such as the two-particle Green function, its correlated part, $\mathcal{G}$, or the susceptibilities directly. In general, solving such operator equations requires representing the operators as matrices in the underlying Fock space, $\delta G^{c,\lambda \mu} \coloneqq \langle \Phi_\lambda | \delta\hat{G}^c| \Phi_\mu\rangle$ for an orthonormal basis $\{\Phi_\lambda\}_\lambda$,  which leads to exponential scaling due to the growth of the Fock space dimension. This difficulty can be avoided by linearizing the equations of motion for the fluctuations. Due to the form of $\alpha_c^{-1}\chi^{c,\gtrless}(t)$ as expectation values of the fluctuation operators and their adjoints [cf. Eqs.~\eqref{eq:chi>_dG} and \eqref{eq:chi<_dG}] and, consequently, also $\alpha_c^{-1}\chi^{c,\mathrm{K}}(t)$, as the sum of the former two components, these objects are positive semi-definite matrices. Consequently, for all times $t \geq t_0$, there exist matrices $\bm{\delta G}^{c,\mathrm{X}}(t) \in \mathbb{C}^{N_\mathrm{b}^2 \times N_\mathrm{b}^2}$ that can be expressed as linear combinations of the matrices associated with the fluctuation operators, $\delta G^{c,\lambda\mu}$, such that
\begin{align}
    \bm{\chi}^{c,\mathrm{X}}(t) &= \alpha_c \bm{\delta G}^{c,\mathrm{X}}(t) \bm{\delta G}^{c,\mathrm{X}\dagger}(t),
\end{align}
where $\mathrm{X}\in \{>,<,\mathrm{K}\}$ denotes the considered real-time component. These matrices are unique up to multiplication by unitary matrices. While this property is guaranteed for the exact solution of the many-body problem, approximations, such as the those within HF-GKBA [cf. Eq.~\eqref{eq:eom_chi_HFGKBA}], generally violate it, which may give rise to numerical instabilities. Preserving semi-definiteness is, however, essential, as it ensures the positivity of the two-particle, two-hole, and (modified) particle–hole reduced density matrices, which correspond to the real-time components of the susceptibilities in the respective channels. By contrast, within the $\delta$NEGF framework, the positive semi-definiteness of the corresponding reduced density matrices is preserved by construction. [cf. Eq.~\eqref{eq:EOM_L_PA}]. Within the $\delta$NEGF theory, the equation of motion for the factors is given by 
\begin{equation}
    \mathrm{i}\hbar \partial_t \bm{\delta G}^{c,\mathrm{X}}(t)  = \bm{\mathfrak{h}}^{(-),c}(t)\bm{\delta G}^{c,\mathrm{X}}(t) \label{eq:EOM_dG_PA_factorization}.
\end{equation}
Furthermore, the two-time susceptibilities can be reconstructed as 
\begin{align}
    \bm{\chi}^{c,\mathrm{X}}(t,t') &=\alpha_c \bm{\delta G}^{c,\mathrm{X}}(t) \bm{\delta G}^{c,\mathrm{X}\dagger}(t').
\end{align}
Such a reconstruction offers no immediate numerical advantage compared to a direct solution of the equation of motion for the generalized susceptibilities, both within the HF-GKBA and the $\delta$NEGF method, as the fluctuations are $N_\mathrm{b}^2\times N_\mathrm{b}^2$ matrices. However, within the $\delta$NEGF approach, only the initial state $\bm{\chi}^{c,\mathrm{X}}(t_0)$ has to be decomposed and, additionally, the factors can be chosen as $\bm{\delta G}^{c,\mathrm{X}}(t_0) \in \CC^{N_\mathrm{b}^2 \times N_\mathrm{r}^{c,\mathrm{X}}}$ with $N_\mathrm{r}^{c,\mathrm{X}} \coloneqq \mathrm{rank}\,  \bm{\chi}^{c,\mathrm{X}}(t_0)$. As $N_\mathrm{r}^{c,\mathrm{X}} \leq N_\mathrm{b}^2$, an exact decomposition of the initial state can offer a numerical advantage compared to the conventional approaches depending on the rank of the initial state. However, a significant advantage can be achieved for large systems by using low-rank approximations, i.e., fixing a rank $1\ll N_\mathrm{s} \ll N_\mathrm{r}^{c,\mathrm{X}} \leq N_\mathrm{b}^2$ and finding approximate solutions $\bm{\delta G}^{c,\mathrm{X}}(t_0) \in \CC^{N_\mathrm{b}^2 \times N_\mathrm{s}}$ such that 
\begin{align}
    \bm{\chi}^{c,\mathrm{X}}(t_0) &\approx  \alpha_c \bm{\delta G}^{c,\mathrm{X}}(t_0) \bm{\delta G}^{c,\mathrm{X}\dagger}(t_0).
\end{align}
More specifically, the low-rank approximation is constructed such that
\begin{align}
    \lVert \bm{\chi}^{c,\mathrm{X}}(t_0) - \alpha_c \bm{\delta G}^{c,\mathrm{X}}(t_0) \bm{\delta G}^{c,\mathrm{X}\dagger}(t_0) \rVert < \varepsilon_\mathrm{error}(N_\mathrm{s}),
\end{align}
where $\varepsilon_\mathrm{error}(N_\mathrm{s})$ is the rank-dependent error bound. This bound depends not only on the rank but also on the choice of the norm. Numerical tests indicate that commonly used norms, such as the Frobenius or spectral norm, are ill-suited for this purpose, whereas the maximum norm yields the best metric for the agreement with the exact decomposition. In practical applications, stochastic algorithms offer an efficient means to construct such low-rank approximations. By setting 
\begin{equation}
    \delta G^{c,\mathrm{X},(n)}_{ij}(t) \coloneqq \bm{\delta G}^{c,\mathrm{X}}_{(ij),n} (t)\,,
\end{equation}
for $n= 1,\dots, N_\mathrm{s}$, we can identify $\bm{\delta G}^{c,\mathrm{X}}$ with an ensemble of matrices in the single-particle space $\{\delta G^{c,\mathrm{X},(n)}\}_{n= 1,\dots,N_\mathrm{s}}$. Then, the ensemble of matrices can be constructed by generating random realizations according to some probability distribution, for example, a Gaussian distribution with mean zero and covariance matrix $\alpha^{-1}_c\bm{\chi}^{c,\mathrm{X}}(t_0)$. The reconstruction is then given by 
\begin{equation}
    \bm{\chi}^{c,\mathrm{X}}_{(ij)(kl)} (t,t') = \frac{\alpha_c}{N_\mathrm{s}} \sum_{n=1}^{N_\mathrm{s}} \delta G^{c,\mathrm{X},(n)}_{ij}(t) \delta G^{c,\mathrm{X},(n)*}_{kl}(t')\,.\label{eq:initial_sampling}
\end{equation}
By the central limit theorem, it holds $\varepsilon_\mathrm{error}(N_\mathrm{s}) \propto 1/\sqrt{N_\mathrm{s}}$.

Each stochastic realization of the ensemble, $\delta G^{c,\mathrm{X},(n)}$, obeys an equation of motion that is determined by the chosen approximation, as is summarized in Table~\ref{t:NEGF-QF_approximations} (replacing $\delta\hat{G}^c$ by $\delta G^{c,\mathrm{X},(n)}$). While the explicit form of these equations depends on the approximation, their numerical complexity is that of simple mean-field-type equations. Consequently, for all approximations considered here, the numerical scaling of the $\delta$NEGF approach corresponds to that of an ensemble of $N_\mathrm{s}$ mean-field-type propagations. \\
\begin{table}[t]
\caption{Scaling of the runtime and memory with the basis size $N_\mathrm{b}$, the number of time steps $N_\mathrm{t}$ and the number of samples $N_\mathrm{s}$ for different self-energy approximations (columns) and three basis choices (rows). The approximations considered here are the TPP, TPH(+X) and $GW$(+X). ``Standard'' refers to solving the two-time KBEs without any further approximation or reconstruction. A diagonal basis refers to a basis with interaction tensor of the form $w_{ijkl} = v_{ij} \delta_{ik}\delta_{jl}$. \label{t:scaling}}
\label{tab:scaling_memory}
\begin{ruledtabular}
\begin{tabular}{ccccc}
 & Basis & Standard  & G1--G2 & $\delta$NEGF \\
\hline\noalign{\vskip 1.5pt}
\multirow{3}{*}{Runtime}
& general\rlap{\footnotemark[1]}  & $N_\mathrm{b}^6 N_\mathrm{t}^3$  & $N_\mathrm{b}^6N_\mathrm{t}$ & $N_\mathrm{s}N_\mathrm{b}^4N_\mathrm{t}$ \\
& diagonal\rlap{\footnotemark[2]\footnotemark[3]} & $N_\mathrm{b}^5 N_\mathrm{t}^3$  & $N_\mathrm{b}^5N_\mathrm{t}$ & $N_\mathrm{s}N_\mathrm{b}^3N_\mathrm{t}$ \\
& Hubbard  & $N_\mathrm{b}^3 N_\mathrm{t}^3$  & $N_\mathrm{b}^4N_\mathrm{t}$ & $N_\mathrm{s}N_\mathrm{b}^2N_\mathrm{t}$ \\
\hline\noalign{\vskip 1.5pt}
\multirow{3}{*}{Memory}
& general\rlap{\footnotemark[1]} & $N_\mathrm{b}^4 N_\mathrm{t}^2$  & $N_\mathrm{b}^4$ & $N_\mathrm{s} N_\mathrm{b}^2$ \\
& diagonal\rlap{\footnotemark[2]} & $N_\mathrm{b}^4 N_\mathrm{t}^2$  & $N_\mathrm{b}^4$ & $N_\mathrm{s} N_\mathrm{b}^2$ \\
& Hubbard  & $N_\mathrm{b}^2 N_\mathrm{t}^2$  & $N_\mathrm{b}^4$ & $N_\mathrm{s} N_\mathrm{b}^2$ \\
\end{tabular}
\end{ruledtabular}
\footnotetext[1]{The scaling of $\delta$NEGF assumes that the interaction tensor $w$ can be factorized, e.g., by resolution of identity (RI).}
\footnotetext[2]{The scaling for $GW$ within the standard approach is $N_\mathrm{b}^3N_\mathrm{t}^3$ and $N_\mathrm{b}^2N_\mathrm{t}^2$, for runtime and memory, respectively.}
\footnotetext[3]{The runtime scaling for $\delta$RPA is $N_\mathrm{s}N_\mathrm{b}^2N_\mathrm{t}$.}
\end{table}
A detailed comparison is presented in Tab.~\ref{t:scaling}. For example, for a general basis with interaction tensor $w_{ijkl}$, the direct solution of the KBEs (denoted ``standard'') exhibits a scaling of $\mathcal{O}(N_\mathrm{b}^6N_\mathrm{t}^3)$, for the runtime, and $\mathcal{O}(N_\mathrm{b}^4N_\mathrm{t}^2)$, for the memory. Within the G1--G2 scheme, this can be reduced to $\mathcal{O}(N_\mathrm{b}^6N_\mathrm{t})$, for the runtime and $\mathcal{O}(N_\mathrm{b}^4)$, for the memory consumption. The $\delta$NEGF method further amplifies this numerical advantage, reducing the scaling to $\mathcal{O}(N_\mathrm{s}N_\mathrm{b}^4N_\mathrm{t})$ and $\mathcal{O}(N_\mathrm{s}N_\mathrm{b}^2)$, respectively. This not only removes the unfavorable cubic scaling with the number of time steps, $N_\mathrm{t}$, but also substantially mitigates the scaling with the basis size $N_\mathrm{b}$. The precise scaling depends on the choice of the basis. For example, for a diagonal basis or for a Hubbard basis, considerable additional gains are observed (see Table~\ref{t:scaling}). An additional advantage is that the ensemble of equations for the $\delta$NEGF approach only couples through the single-particle NEGF and can thus be easily parallelized.

By construction, the employed approximations based on linearizations preserve the rank of the generalized susceptibilities $\chi^{c,\mathrm{X}}$ throughout the time evolution, i.e., $N_\mathrm{r}^{c,\mathrm{X}}\equiv \mathrm{rank} \,\chi^{c,\mathrm{X}}(t)$, for all $t \geq t_0$. The same property naturally extends to propagations based on low-rank approximations, where the chosen approximate rank $N_\mathrm{s}$ remains fixed during the dynamics. Consequently, the difference between the full rank and the retained approximate rank is constant in time which, in practice, leads to a fast convergence with regard to $N_\mathrm{s}$ even for large systems and long propagation times, as will be demonstrated numerically in Sec.~\ref{s:results}.
This is a unique feature of the $\delta$NEGF approach and in stark contrast to most other low-rank compression approaches based on hierarchical structure (HODLR)~\cite{Kaye_2021,Blommel_2025,Gasperlin_2025,Blommel_2026} or quantics tensor trains (QTT)~\cite{Shinaoka_2023,Murray_2024,Sroda_2025,Inayoshi_2026,Sroda_2026} where the difference between the approximate and full rank of the dynamics can grow uncontrollably.
Nevertheless, the associated approximation error generally accumulates during the propagation, leading to increasing deviations at long propagation times.

Importantly, the use of low-rank approximations does not affect the conservation properties of the approximations within the $\delta$NEGF approach, such as the positive semi-definiteness of the associated reduced density matrices and energy conservation. However, the approximate decomposition must be chosen consistently with the symmetries of the system. In particular, symmetry constraints, such as spin symmetry, have to be incorporated explicitly into the low-rank decomposition to ensure that the corresponding physical properties are preserved during the time evolution.

In summary, our novel $\delta$NEGF approach, via linearization of the fluctuations equations, leads straightforwardly to the TPP, TPH(+X) and $GW$(+X) approximations, see Table~\ref{t:NEGF-QF_approximations}. Although the $\delta$NEGF versions of these approximations are at a lower level of self-consistency  compared to the HF-GKBA (see Fig.~\ref{fig:sc-hierarchy}), they provide several important advantages. First, the $\delta$NEGF approximations preserve the positive semi-definiteness of the reduced density matrices ($N$-representability) associated with the corresponding generalized susceptibilities, which results in improved numerical stability compared to the G1--G2 scheme. Second, low-rank approximations enable a significant reduction in computational cost, both in terms of runtime and memory consumption, since the dynamics can be obtained from the propagation of an ensemble of mean-field-type equations. Third, the two-time generalized susceptibilities can be directly reconstructed from the time-diagonal evolution of the fluctuations, eliminating the need to explicitly resolve the full two-time plane. The combination of the $\delta$NEGF framework with stochastic implementations is, therefore, expected to yield a drastic reduction in computational resources. Furthermore, as we show in Sec.~\ref{ss:observables}, the 
$\delta$NEGF framework provides easy access to all relevant single-particle and two-particle observables, but, in addition, also to important correlation functions.  In the next section, we verify the stated advantageous properties of our approach  for several examples.

\section{Numerical Results}\label{s:results}

In this section, we present five representative applications of the $\delta$NEGF method to different correlated  systems, applying the different many-body approximations we identified in Sec.~\ref{s:approx}. The systems of choice for demonstration and benchmarks are lattice models that we introduce in the following. 

\subsection{Hubbard and extended Hubbard model}\label{ss:lattice-models}
We introduce the lattice Hamiltonian
\begin{align}
     \hat{H}(t) =
      &-  \sum_{\langle \mathfrak{i},\mathfrak{j}\rangle,\sigma}J_{\mathfrak{ij}}(t) \hat{a}^\dagger_{\mathfrak{i},\sigma} \hat{a}_{\mathfrak{j},\sigma}
      + U(t) \sum_{\mathfrak{i}} \hat{n}_{\mathfrak{i},\uparrow} \hat{n}_{\mathfrak{i},\downarrow}\nonumber\\
      &+
      \frac{1}{2} \sum_{\mathfrak{i} \neq \mathfrak{j}} V_{\mathfrak{ij}}(t) \left(\hat{n}_{\mathfrak{i}}-1\right) \left(\hat{n}_{\mathfrak{j}}-1\right)\nonumber\\
      &+ \sum_\mathfrak{i} e\, \vec{r}_\mathfrak{i} \cdot \vec{\mathcal{E}}(t)\,\hat{n}_\mathfrak{i}\,,
    \label{eq:hubbard-ppp}
\end{align}
with the new index notation $\mathfrak{i},\mathfrak{j}$ representing the spatial lattice basis and spins $\sigma\in\{\uparrow,\downarrow\}$ treated separately. Here $\hat{n}_{\mathfrak{i}}= \hat{n}_{\mathfrak{i},\uparrow}+\hat{n}_{\mathfrak{i},\downarrow}$ is the local density operator, and ${\langle \mathfrak{i},\mathfrak{j}\rangle}$ indicates summation over only nearest-neighbor sites. The kinetic part contains the inter-site hopping $J_{\mathfrak{ij}}$ between neighboring sites $\mathfrak{i},\mathfrak{j}$. Generally, we consider uniform hopping, i.e., $J_{\mathfrak{ij}} \equiv J$ for all $\mathfrak{i},\mathfrak{j}$, except for some special cases, cf. Sec.~\ref{ss:spin-correlations}, where nonuniform hopping will be permitted. Further, local on-site interactions are included via the Hubbard parameter $U$. For the local Hubbard model $V_{\mathfrak{ij}}=0$, whereas for the extended Hubbard model long-range Coulomb interactions are included via the Ohno parametrization \cite{ohno_remarks_1964}
\begin{align}
        V_{\mathfrak{ij}}(t)\coloneqq U(t)\left[1+\left(\frac{U(t) R_{\mathfrak{ij}}}{k_e}\right)^2\right]^{-\frac{1}{2}}\,,\label{eq:ohno}
\end{align}
with $U$ in eV, the distance between two sites $R_{\mathfrak{ij}}$ in $\si{\angstrom}$, and $k_e$ denoting the Coulomb constant. This long-range Hamiltonian corresponds to the Pariser--Parr--Pople (PPP) model used in molecular applications~\cite{chiappe_can_2015}. The interaction parameters are time-dependent to allow for an adiabatic switch-on of the interacting ground state. The excitation of the system through an external laser pulse can be added to the Hamiltonian in two ways, either by modifying the hopping term via Peierls substitution determined by the vector potential, see~Eq.~\eqref{eq:peierls}, or via an on-site potential induced by the electric field $\mathcal{E}$.

In general the numerical demonstrations serve three distinct purposes:
\begin{enumerate}
    \item Benchmarking the $\delta$NEGF approach for the three channels introduced in Sec.~\ref{s:approx}. This will be done for small systems, where exact reference data exist and $\delta$NEGF calculations can be performed without low-rank approximations. The latter is important, as the purpose of these tests is to evaluate the capabilities and accuracy of the $\delta$NEGF method, without introducing additional statistical uncertainties.
    \item As the ultimate goal is the application of the $\delta$NEGF concept to large-scale systems, for which approximate sampling is necessary for numerical reasons, the second purpose is to explicitly evaluate the additional error introduced by stochastically sampling the initial state with a given number of samples, $N_\mathrm{s}$. This is done for medium size systems where an exact factorization is very costly but still possible.
    \item Finally, the capability of the $\delta$NEGF method to accurately describe the nonequilibrium dynamics of large-scale systems with basis sizes $N_\mathrm{b}>1000$ will be demonstrated. Since for these mesoscopic systems no other available method is able to produce reference data of equal or better quality, it is not possible to evaluate the accuracy directly. This underlines the importance of the rigorous assessment of the reliability of the $\delta$NEGF approach, as described in 1. and 2.
\end{enumerate}

Since the different channels of the $\delta$NEGF method describe distinct physical processes, we chose five benchmark systems accordingly, to test for the correct description of the respective effects.
The first application studied in Sec.~\ref{ss:cdw} is a rather small one-dimensional Hubbard charge-density-wave (CDW) system for which exact results can be obtained that allow us to rigorously benchmark the $\delta$NEGF method with the TPP approximation. In Sec.~\ref{ss:diffusion} we apply the same approximation to larger 2D Hubbard systems---a $19\times 19$ square lattice and finally, we move on to a $101 \times 101$ square lattice. There we investigate the diffusion following a confinement quench and demonstrate that the $\delta$NEGF method is very well capable to treat very large correlated systems. 

Next, in Sec.~\ref{ss:spin-correlations} we demonstrate the capability of the $\delta$NEGF method in the TPH channel to describe spin correlations for a finite $6\times 2$ Hubbard ladder, where again exact reference data are available. We study the system both, in equilibrium and nonequilibrium, after an excitation by a short laser pulse.

Finally, we consider the $\delta$NEGF approach in the $GW$ channel for two different systems. First, in Sec.~\ref{ss:conductivity} we compare the dynamic response properties of a 1D Hubbard chain in the Mott regime both, in equilibrium and after a laser excitation, to exact reference data. Second, in Sec.~\ref{ss:numerics-ppp} we give an outlook to the performance of the $\delta$NEGF approach for Coulomb systems by studying carbon-based molecules and nanostructures. To this end, we go beyond 
the local interaction limit and consider the extended Hubbard model with long-range interactions, Eq.~\eqref{eq:hubbard-ppp}.

Before moving to these examples, we summarize the most important physical observables and how they are evaluated within the new approach.

\subsection{Macroscopic observables within $\delta$NEGF}\label{ss:observables}
The evaluation of observables in the $\delta$NEGF formalism differs from other NEGF-based approaches. While in the original two-time representation two-particle correlations can be extracted from the self-energy $\Sigma$, in the time-linear G1--G2 scheme they are included in the correlated two-particle Green function $\mathcal{G}$. In contrast, in the $\delta$NEGF approach the object carrying information about correlations in the system is the single-particle fluctuation $\delta \hat G^c$ and the resulting generalized susceptibility $\chi$. In this section we will provide expressions for relevant single-time (\ref{ss:one-time}) and two-time (\ref{ss:cor-funcs}) observables within the $\delta$NEGF framework.

\subsubsection{One-time single-particle and two-particle observables}\label{ss:one-time}

The single-particle density matrix and related one-body quantities, such as the kinetic, potential, or Hartree--Fock energy, are obtained as in standard NEGF theory from the time-diagonal lesser component of the single-particle Green function. For instance, the local single-particle density is given by \cite{stefanucci_nonequilibrium_2013,balzer-book,schluenzen_jpcm_19} 
\begin{align}
    n_\mathfrak{i}(t) = -\mathrm{i}\hbar \{G^<_{\mathfrak{ii},\uparrow}(t)+G^<_{\mathfrak{ii},\downarrow}(t)\}\,.
    \label{eq:ni-def}
\end{align}
Next, we will turn to time-local two-particle observables, where we can use the fact that the correlation part of the two-particle Green function $\mathcal{G}$ can be directly calculated from the Keldysh component of the generalized susceptibilities, cf. Sec.~\ref{ss:def-ghat},
\begin{align}
  \mathcal{G}(t) = \frac{\chi^{c,\mathrm{K}}(t) - \tilde \chi^{0,c,\mathrm{K}}(t)}{2\mathrm{i}\hbar} \,.
\end{align}
This way, time-local two-particle observables can be obtained in the same way as in the G1--G2 scheme~\cite{joost_phd_2022}, but without computing and storing two-particle Green functions. For instance, the local two-particle density or double occupancy 
\begin{align}
    d_\mathfrak{i}(t) &= d^\mathrm{HF}_\mathfrak{i}(t) + d^\mathrm{corr}_\mathfrak{i}(t)\,,
\end{align}
can be separated into a single-particle Hartree--Fock contribution
\begin{align}
    d^\mathrm{HF}_\mathfrak{i}(t) &= n_{\mathfrak{i},\uparrow}(t)n_{\mathfrak{i},\downarrow}(t)\,,
\end{align}
and a correlation part
\begin{align}
    d^\mathrm{corr}_\mathfrak{i}(t) &= -\hbar^2 \mathcal{G}_{\mathfrak{i}\mathfrak{i}\mathfrak{i}\mathfrak{i}}^{\uparrow\downarrow\uparrow\downarrow}(t)\,,
\label{eq:dcorr}
\end{align}
Other important single-time two-particle observables are the pair-correlation function,
\begin{align}
    g_{ij}(t) &=-\hbar^2 \mathcal{G}_{ijij}(t)\,,
\end{align}
and the correlation energy
\begin{align}
E_{\rm corr}(t) &= -\frac{\hbar^2}{2} \sum_{ijkl} w_{ijkl}(t) \mathcal{G}_{klij}(t)\,,\label{eq:e_corr}
\end{align}
which are expressed in terms of $\mathcal{G}$ in a similar manner.

\subsubsection{Two-time observables: Physical correlation functions}\label{ss:cor-funcs}

The direct propagation of the single-particle fluctuations $\delta\hat G^c$ as the carrier of correlation effects puts the $\delta$NEGF approach in the unique position of allowing two-time, two-point response and correlation functions to be obtained directly from a single-time calculation. For any two operators $\hat{\mathcal{O}}_1$ and $\hat{\mathcal{O}}_2$ that are linear in $\hat G^c$ for the same channel, we can directly calculate the generic response \cite{giuliani2005quantum},
\begin{subequations}
    \begin{align}
\chi^\mathrm{R}_{\mathcal{O}_1\mathcal{O}_2}(t,t') &\coloneqq \frac{1}{\mathrm{i}\hbar} \Theta(t,t')\big\langle \big[ \delta\hat{\mathcal{O}}_1(t),\delta\hat{\mathcal{O}}^\dagger_2(t')\big]\big\rangle\,,\label{eq:chi_R}\\
\chi^\mathrm{A}_{\mathcal{O}_1\mathcal{O}_2}(t,t') &\coloneqq \frac{1}{\mathrm{i}\hbar} \Theta(t',t)\big\langle \big[ \delta\hat{\mathcal{O}}^\dagger_2(t'),\delta\hat{\mathcal{O}}_1(t)\big]\big\rangle\,,
\end{align}
and correlation functions
\begin{align}
\chi^>_{\mathcal{O}_1\mathcal{O}_2}(t,t') &\coloneqq \frac{1}{\mathrm{i}\hbar} \big\langle \delta\hat{\mathcal{O}}_1(t) \delta\hat{\mathcal{O}}^\dagger_2(t')\big\rangle\,,\\
\chi^<_{\mathcal{O}_1\mathcal{O}_2}(t,t') &\coloneqq \frac{1}{\mathrm{i}\hbar}\big\langle \delta\hat{\mathcal{O}}^\dagger_2(t') \delta\hat{\mathcal{O}}_1(t)\big\rangle\,,\label{eq:chi_l}
\end{align}
\end{subequations}
via the respective fluctuation operators. The explicit form of the operators for the important cases of density, spin, current, dipole and pair fluctuations is given in Tab.~\ref{t:spectra}. Other fluctuations and correlation functions can be obtained in a similar manner.

\begin{table}[t]

\begin{caption}{Explicit form of the fluctuation operators $\mathcal{O}$ for different types of correlation functions $\chi$, cf.~\eqrefr{eq:chi_R}{eq:chi_l}. Here, $\alpha,\beta \in \{x,y,z\}$ denote the Cartesian coordinates, $\sigma,\sigma'\in\{\uparrow,\downarrow\}$ represent the spin degrees of freedom and $\tau^\alpha$ denotes the three Pauli matrices. 
\label{t:spectra}} 
\end{caption}
\begin{ruledtabular}
\begin{tabular}{ccccc}
{\begin{tabular}{c} {Fluct.}\\{Type}\end{tabular} } & {Notation} 
& $\mathcal{O}_1$ & $\mathcal{O}_2$ & Fluctuation Operator \\ 
\hline\noalign{\vskip 1.5pt}
density     & $\chi_{nn}$  & $n_i$ & $n_j$ & $\delta\hat n_i = -\mathrm{i}\hbar \delta\hat G^{\overline{\mathrm{ph}}}_{ii}$  \\\noalign{\medskip}
spin     & $\chi_{SS}$  & $S^\alpha_\mathfrak{i}$ & $S^\beta_\mathfrak{j}$ & $\delta\hat S^\alpha_\mathfrak{i} = -\mathrm{i}\hbar \sum_{\sigma\sigma'} \tau^\alpha_{\sigma\sigma'}\delta\hat G^{\overline{\mathrm{ph}}}_{\mathfrak{ii},\sigma\sigma'}$\\\noalign{\medskip} 
current & $\chi_{jj}$  & $j_{ij}$ & $j_{kl}$ & $\delta\hat j_{ij} = eJ_{ij}\delta\hat G^\mathrm{\overline{ph}}_{ij} - eJ_{ji}\delta\hat G^{\overline{\mathrm{ph}}}_{ji}$ \\\noalign{\medskip} 
dipole     &  $\chi_{PP}$  & $P^\alpha$ & $P^\beta$ &  $\delta\hat P^\alpha = -\mathrm{i}\hbar e \sum_{\mathfrak{i},\sigma}\,\vec{r}^{\,\alpha}_\mathfrak{i}\, \delta\hat G^{\overline{\mathrm{ph}}}_{\mathfrak{ii},\sigma\sigma}$   \\\noalign{\medskip} 
pair & $\chi_{\kappa\kappa}$  & $\kappa_{ij}$& $\kappa_{kl}$&  $\delta\hat\kappa_{ij} = \mathrm{i}\hbar \delta\hat  G^\mathrm{pp}_{ij}$   \\
\end{tabular}
\end{ruledtabular}
\end{table}

To obtain time-resolved spectra, for all real-time components $\mathrm{X}\in\{\mathrm{R},\mathrm{A},>,<\}$, we perform a Fourier transform
\begin{align}
\chi^\mathrm{X}_{\mathcal{O}_1\mathcal{O}_2}(\omega, t) &\coloneqq \frac{1}{2\pi} \iint  f(\omega, t; \tau_1,\tau_2) \chi^\mathrm{X}_{\mathcal{O}_1\mathcal{O}_2}(\tau_1,\tau_2)\mathrm{d}\tau_1 \mathrm{d}\tau_2\,, \label{eq:fourier-trafo}
\end{align}
with a Gaussian probe window \cite{freericks_09}
\begin{subequations}
\begin{align}
f(\omega, t; \tau_1,\tau_2) &\coloneqq e^{\mathrm{i}\omega(\tau_1-\tau_2)} g(t, \tau_1) g(t, \tau_2)\,,\\ \label{eq:gauss}
g(t, \tau) &\coloneqq \frac{1}{\sqrt{2\pi}\sigma_\mathrm{pr}} e^{-\frac{(\tau-t)^2}{2\sigma^2_\mathrm{pr}}}\,,
\end{align}
\end{subequations}
where the special case of ground-state spectra follows by neglecting the $t$-dependence.

The two-point functions listed in Tab.~\ref{t:spectra} give access to many related correlation and response functions.
For instance, the density--density response is directly connected to the screened interaction, $W$, and the inverse dielectric function via \cite{hamann_prb_20,stefanucci_nonequilibrium_2013}

\begin{align}
    W_{ijkl}(\omega) &= w_{ijkl} + \sum_{pqrs} w_{ipkq} \,\chi^{\overline{\mathrm{ph}},\mathrm{R}}_{qrps}(\omega) \,w_{sjrl}\,,\\
    \varepsilon^{-1}_{ijkl}(\omega) &= \delta_{ik}\delta_{jl} + \sum_{pq} w_{ipkq} \,\chi^{\overline{\mathrm{ph}},\mathrm{R}}_{qjpl}(\omega)\,,
\end{align}
with $\chi^{\overline{\mathrm{ph}},\mathrm{R}}_{ijkl}(\omega) = \chi^{\mathrm{R}}_{n_{ik}n_{lj}}(\omega)$.

Another important quantity is the (charge) dynamic structure factor
\begin{align}
    S(q,\omega, t) &= \chi^{>}_{n_q,n_{q}}(\omega,t)\,,
\label{eq:dsf-noneq}
\end{align}
which is equivalent to the density correlation function in momentum space, where the fluctuations of the Fourier components of the density follow, in our approach, from
\begin{align}
    \delta\hat n_q(t) &= -\mathrm{i}\hbar \sum_\sigma \sum_{\mathfrak{i}} e^{-\mathrm{i}qr_\mathfrak{i}} \,\delta\hat G^{\overline{\mathrm{ph}}}_{\mathfrak{ii},\sigma\sigma}(t)\,.
\end{align}
An important quantity that follows from the current autocorrelation function $\chi_{jj}$ is the optical conductivity 
(optical response function), the real part of which is given by 
\begin{subequations}
\begin{align}
    \sigma_{\alpha\beta}(\omega,t) &\coloneqq \sigma^\mathrm{R}_{\alpha\beta}(\omega,t) = \frac{1}{\omega} \mathrm{Im}\chi^\mathrm{R}_{j^\alpha j^\beta}(\omega,t)\,,\label{eq:opt_cond_current}\\
    \delta\hat j^\alpha(t) &= \sum_\sigma \sum_{{\langle \mathfrak{ij} \rangle}_\alpha} \delta\hat j_{\mathfrak{ij},\sigma\sigma}(t)\,,
\end{align}
\end{subequations}
where $\alpha,\beta \in \{x,y,z\}$ denote the Cartesian coordinates. For finite nanostructures we may use the dipole approximation and express $\sigma$ via the dipole response function
\begin{align}
    \sigma_{\alpha\beta}(\omega,t) &\coloneqq \sigma^\mathrm{R}_{\alpha\beta}(\omega,t) = \omega\, \mathrm{Im} \chi^\mathrm{R}_{P^\alpha P^\beta}(\omega,t) \,.
\end{align}
The quality of these two-particle spectral observables obtained within $\delta$NEGF theory is discussed and compared to other  approaches in Sec.~\ref{ss:conductivity} and~\ref{ss:numerics-ppp}.

\subsection{Generation of the initial state}\label{ss:ini-state}

In the $\delta$NEGF approach the generation of the initial state includes initializing both the single-particle Green function $G^<(t_0)$, as well as the single-particle fluctuations $\{\delta G^{c,\mathrm{X},(n)}(t_0)\}_{n=1,\dots,N_\mathrm{s}}$. For simplicity, we will only discuss the generation of the ensemble of fluctuations for the lesser component as the other components follow analogously. To this end, it is advantageous to start from an uncorrelated, ideal state with $\mathcal{G}=0$, so that, following Eqs.~\eqref{eq:G2-chi_relation} and \eqref{eq:initial_sampling}, the initial value problem for the susceptibility reduces to
\begin{align}
    \tilde{\bm{\chi}}^{0,c,<}_{(ij)(kl)} (t_0) = \frac{\alpha_c}{N_\mathrm{s}} \sum_{n=1}^{N_\mathrm{s}} \delta G^{c,<,(n)}_{ij}(t_0) \delta G^{c,<,(n)*}_{kl}(t_0)\,,
\end{align}
with
\begin{align}
        \alpha_c^{-1} &\tilde{\chi}^{0,c,<}_{ijkl}(t_0) = \\\nonumber
        &\begin{cases}
            -\left\{G^<_{ik}(t_0) G^<_{jl}(t_0) - G^<_{il}(t_0) G^<_{jk}(t_0)\right\}, &c = \mathrm{pp},\\[1ex]
            G^<_{ik}(t_0) G^>_{jl}(t_0), &c = \mathrm{ph},\\[1ex]
            G^<_{il}(t_0) G^>_{jk}(t_0), &c = \overline{\mathrm{ph}}.
        \end{cases}
\end{align}
This reconstruction can be generated in various ways. For a full-rank decomposition, we employ a singular value decomposition and keep all non-zero singular values. For a system at zero temperature, this leads to a sample number of
\begin{align}
    N_\mathrm{s} =
    \begin{cases}
            \mathcal{O}(N_\mathrm{p}^2), &c = \mathrm{pp},\\[1ex]
            \mathcal{O}(N_\mathrm{p}N_\mathrm{h}), &c = \mathrm{ph},\overline{\mathrm{ph}},
        \end{cases}
\end{align}
with $N_\mathrm{p}$ and $N_\mathrm{h}$ being the number of particles and holes in the system, respectively \footnote{The same scaling holds for the greater components in the two particle-hole channels. In the particle-particle channel, however, it follows that $N_\mathrm{s}= \mathcal{O}(N_\mathrm{h}^2)$ for the greater component.}.

As explained in Sec.~\ref{s:factorization}, an efficient way to generate a low-rank reconstruction is to draw random samples from a Gaussian distribution with zero mean and covariance matrix $\alpha^{-1}_c\tilde{\bm{\chi}}^{0,c,<}(t_0)$. In practice, it is advantageous to do this in the natural orbital basis of the density matrix, where the covariance matrix simplifies to
\begin{align}
    \frac{\hbar^2}{\alpha_c}\tilde{\bm{{\chi}}}^{0,c,<}_{(ij)(kl)}(t_0) = \begin{cases}
        n_in_j \{\delta_{ik}\delta_{jl}- \delta_{il}\delta_{jk}\}, &c = \mathrm{pp}, \\[1ex]
         n_i (1-n_j) \delta_{ik}\delta_{jl}, &c = \mathrm{ph},\overline{\mathrm{ph}}.
    \end{cases}\label{eq:2nd-moment}
\end{align}
Finally, the generated initial single-particle fluctuations are  transformed back into the working basis.

It is important to highlight that the second moment of the fluctuations, Eq.~\eqref{eq:2nd-moment}, is the only initial condition to be fulfilled, besides the trivial $G^<(t_0)$. This is in stark contrast to other stochastic (fluctuation-based) approaches, such as stochastic mean-field theory~\cite{ayik_stochastic_2008, lacroix_prb14, lacroix_simplified_2016} or the truncated Wigner approximation~\cite{polkovnikov_quantum_2003}, where also the first moment and all higher moments above the second one add additional initial conditions, resulting in an over-constrained and, therefore, practically ill-posed initialization problem for these approaches.

For the initially uncorrelated setups that will be presented in Sec.~\ref{ss:cdw} and~\ref{ss:diffusion}, initializing the ideal state is sufficient. On the other hand, when starting from a correlated initial state, as will be done in Secs.~\ref{ss:spin-correlations}--\ref{ss:numerics-ppp}, the interactions are turned on via an adiabatic switching procedure, as is common for the HF-GKBA, for details see Refs.~\cite{schluenzen_jpcm_19,joost_phd_2022}.

\subsection{Decay dynamics of a charge density wave in a 1D Hubbard chain ($\delta$TPP approximation)}\label{ss:cdw}

\begin{figure}[h]
    \centering
    \includegraphics[width=\linewidth]{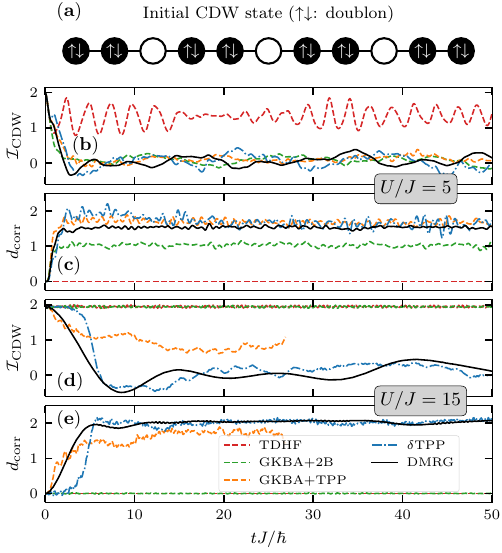}
    \caption{Benchmarking of $\delta$TPP for the melting of an 11-site CDW chain at [(b)--(c)] $U/J=5$ and [(d)--(e)] $U/J=15$. Comparison of [(b) and (d)] the density imbalance, Eq.~\eqref{eq:imbalance}  and [(c) and (e)] the correlated double occupation, Eq.~\eqref{eq:dcorr}.
    }
    \label{fig:tpp_cdw_U5_U15}
\end{figure}

\begin{figure}[h]
    \centering
    \includegraphics[width=\linewidth]{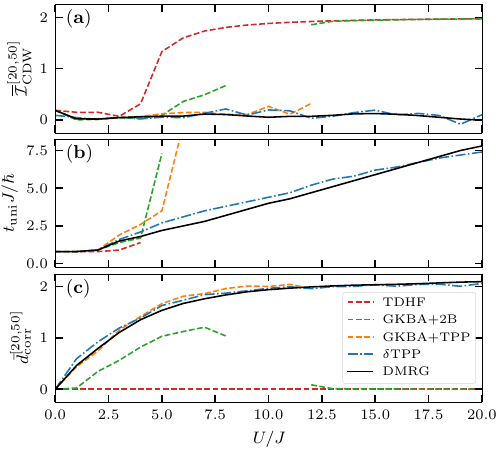}
    \caption{Further analysis of the setup of Fig.~\ref{fig:tpp_cdw_U5_U15} for a broad range of coupling strengths, $U/J=0 \dots 20$. Asymptotic values  of (a) imbalance (\ref{eq:imbalance}) and (c) correlated double occupation \eqref{eq:dcorr}, 
    averaged between $t_1J/\hbar=20$ and $t_2J/\hbar=50$. (b): first time the imbalance crosses zero. 
}
    \label{fig:tpp_cdw_U_scan}
\end{figure}

We start by studying an 11-site 1D Hubbard chain where the initial state is a charge density wave (CDW) and the sites are either doubly occupied or empty, as sketched in Fig.~\ref{fig:tpp_cdw_U5_U15}.a. We study moderate ($U/J=5$) and strong correlations ($U/J=15$) and compute the time evolution of  the density imbalance $\mathcal{I}_\mathrm{CDW}$ (a single-particle observable) and the correlated double occupancy $d_{\rm corr}$ (a two-particle observable),
\begin{align}
    \mathcal{I}_\mathrm{CDW}(t) &\coloneqq
\frac{1}{|\mathcal{O}|}\sum_{\mathfrak{i}\in \mathcal{O}} n_\mathfrak{i}(t) - \frac{1}{|\mathcal{U}|}\sum_{\mathfrak{i}\in \mathcal{U}} n_\mathfrak{i}(t),\label{eq:imbalance}
\end{align}
where the density $n_\mathfrak{i}$ was defined in Eq.~\eqref{eq:ni-def}, and 
 $\mathcal{O}$ [$\mathcal{U}$] is the set of initially occupied  [unoccupied] sites.

Consider first the exact dynamics that is obtained from DMRG simulations using the TeNPy library~\cite{TeNPy_2024}, for $U/J=5$. The imbalance (panel b) rapidly decays to zero and even becomes negative after which it oscillates around zero. At the same time, $d_{\rm corr}$ (panel c) builds up from zero and converges to a value around $1.5$. TDHF simulations completely miss these trends exhibiting oscillations of $\mathcal{I}_\mathrm{CDW}$ around $1.2$ and $d_{\rm corr}\equiv 0$. Let us now consider the remaining three curves that correspond to correlated simulations. While the behavior of the imbalance is similar for all, the dynamics of $d_{\rm corr}$ is not correctly captured by second Born simulations (GKBA+2B). On the other hand, the two $T$-matrix simulations (GKBA+TPP and $\delta$TPP) reproduce the exact results rather well. Let us now turn to the strong coupling case, panels (d) and (e). Here GKBA+2B completely fails, yielding constant values, $\mathcal{I}_\mathrm{CDW}(t)=2$ and $d_{\rm corr}(t)=0$. On the other hand, the two simulations with the TPP self-energy are more close to the DMRG dynamics. GKBA-TPP simulations are accurate for the initial time, $t \lesssim 3\hbar/ J$, but then exhibit strong deviations, in particular for the imbalance. Even worse, the GKBA simulations become unstable and break down for $t \gtrsim 27\hbar/ J$. Interestingly, our $\delta$NEGF method with the TPP self-energy behaves much better: it does not exhibit instabilities and reproduces the exact behavior of $\mathcal{I}_\mathrm{CDW}(t)$ and $d_{\rm corr}(t)$ very well. Aside from two minor deviations---the onset of the dynamics for short times is delayed and the oscillations of $\mathcal{I}_\mathrm{CDW}(t)$ are slightly out of phase---this approximation performs surprisingly well. A similar delayed onset of dynamics was observed previously when studying interaction quenches in the Hubbard model using $\delta$$GW$~\cite{joost_quenching_2026}. It thus seems to be a characteristic feature of the $\delta$NEGF approach and is connected to the reduced level of self-consistency, compared to the GKBA, as discussed in Sec.~\ref{s:approx}.

Some characteristic features of this behavior are further analyzed in Fig.~\ref{fig:tpp_cdw_U_scan} where we consider a broad range of coupling parameters, extending up to values as large as $U/J=20$. In panel (a) we present a temporal average of the imbalance $\mathcal{I}_\mathrm{CDW}$, Eq.~(\ref{eq:imbalance}), for late times where the exact curve has approached a nearly constant value close to zero, cf. Fig.~\ref{fig:tpp_cdw_U5_U15}. Similarly, in panel (c) we present the temporal average of the correlated double occupation, Eq.~(\ref{eq:dcorr}).  Obviously, time-dependent Hartree-Fock completely fails to describe the behavior. Correlated simulations using GKBA with the 2B self-energy (green) behave slightly better but break down around $U/J \approx 8$ due to an instability. Stable solutions are again possible for $U/J \gtrsim 12$ but the results are wrong, essentially reproducing TDHF. Significantly more accurate results are achieved with the TPP self-energy (orange curves), but they become unstable for $U/J \gtrsim 12$. Note that quantum fluctuations with the TPP self-energy (blue curves) are the only method that is stable in the entire coupling range and agrees impressively well with the DMRG benchmarks. Finally, in panel (b) we depict the times where the imbalance $\mathcal{I}_\mathrm{CDW}$ first crosses zero. This is a sensitive measure of the relaxation time because the system tends to uniform filling. Again, the two GKBA simulations (green and orange) reproduce the correct trends for small coupling, $U/J \lesssim 5$ but then break down. In contrast, our $\delta$TPP results are in excellent agreement with DMRG, in the entire $U$ range.

This behavior of our $\delta$TPP simulation is remarkable because the case of such strong coupling could not be well described before using NEGF methods at all. We, therefore, expect that this excellent quantitative agreement with the exact results will also persist for larger systems, which we consider next, where no benchmark data are available. 

\begin{figure*}[t]
    \centering
    \includegraphics[width=\linewidth]{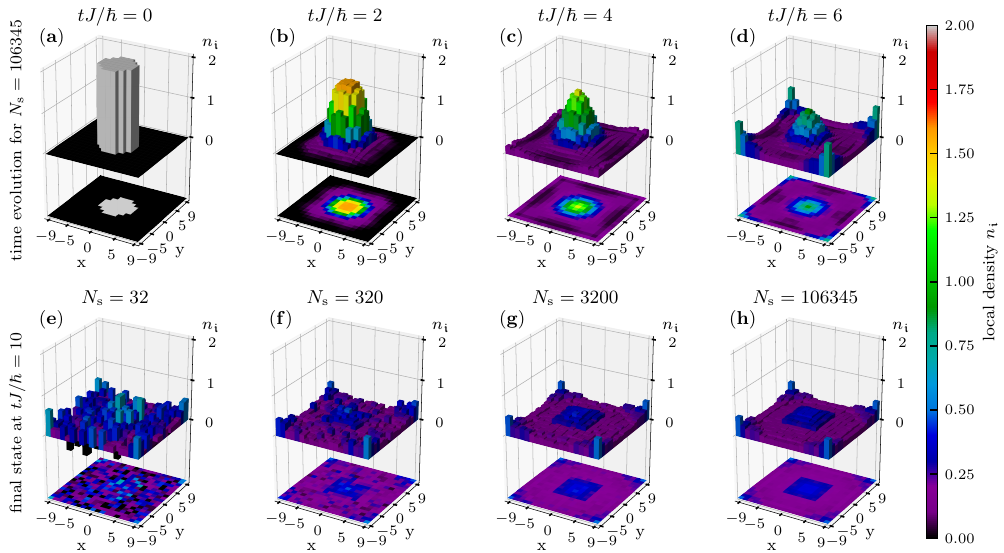}
    \caption{Diffusion of a doubly-occupied circular initial distribution containing 74 particles on a $19\times 19$ square lattice for $U/J=3$. (a)--(d): Surface plots of the local density distribution at different times for exact simulations which require $N_\mathrm{s}=106\,345$ samples. (e)--(h): Surface plots of the local density distribution at the final time $tJ/\hbar=10$, for different numbers of samples $N_\mathrm{s}$. Already for $3200$ samples [panel (g)] the results are indistinguishable from the exact result shown in panel (h).
}
    \label{fig:tpp_sampling_3D}
\end{figure*}

\begin{figure}[h]
    \centering
    \includegraphics[width=\linewidth]{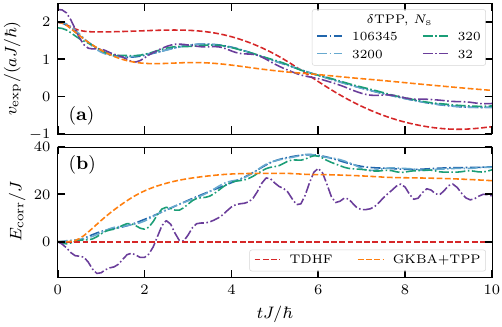}
    \caption{Further analysis of the setup of Fig.~\ref{fig:tpp_sampling_3D} for (a) the expansion velocity of the distribution, Eq.~\eqref{eq:v-exp},  and (b) the correlation energy. Results for TDHF (red) and GKBA+TPP (orange) are added for comparison.
    Notice that averaged single-particle quantities, such as the expansion velocity, converge much faster than the local density in Fig.~\ref{fig:tpp_sampling_3D} and the correlation energy in panel (b). In general, even small sample numbers, cf. curve for $N_\mathrm{s}=32$, give accurate results.}
    \label{fig:tpp_sampling_lines}
\end{figure}

\subsection{Diffusion in a large two-dimensional Fermi Hubbard system ($\delta$TPP approximation)}\label{ss:diffusion}
Our second example is the density dynamics following a confinement quench (rapid removal of confinement) for different coupling strengths $U/J$. The initial state was a doubly occupied central region sketched in Fig.~\ref{fig:tpp_sampling_3D}.(a). This setup has been studied previously in detail, both experimentally \cite{schneider_fermionic_2012} and theoretically using NEGF and DMRG simulations \cite{schluenzen_prb16,schluenzen_prb17}. In Ref.~\cite{schluenzen_prb16} it was found that the TPP approximation of Green functions theory performs very well for this setup and accurately reproduces the experimental results for the diffusion speed.
In Fig.~\ref{fig:tpp_sampling_3D} we start with a system
containing 74 particles on a $19\times 19$ square lattice, which is roughly the size previous calculations were restricted to.
This makes it the ideal system
to test the fluctuations approach and the dependence of the results on the number $N_\mathrm{s}$ of samples, since it is possible to obtain fully converged reference calculations without statistical uncertainties.

In Fig.~\ref{fig:tpp_sampling_3D}.(b)-(d) we show time-dependent $\delta$NEGF simulations with the TPP self-energy for four times using the full number of samples, $N_\mathrm{s}=106\,345$, that is required for an exact decomposition within this approximation. The density of the initially doubly occupied central part quickly decays and spreads uniformly into the entire lattice (cf. the pink color), but a peak remains well pronounced in the original circular region up to $t=6\hbar/J$.  In addition, four density peaks emerge at the four corners of the lattice. The final simulation snapshot is for $t=10\hbar/J$ and shown in panel (h). 

We now use this result for the final time, $t=10\hbar/J$,  to analyze the sensitivity of the $\delta$TPP simulations to the number of samples. Panels (e)-(h) show independent simulations where $N_\mathrm{s}$ is varied by three orders of magnitude. Except for the smallest number, $N_\mathrm{s}=32$, panel (e), all simulations reproduce the exact result rather well with the main difference being that a background of noise appears at small $N_\mathrm{s}$, see panel (f). But already a number $N_\mathrm{s}=3200$ yields results that are practically coinciding with the exact ones. Thus, the simulations indicate that it is possible to reduce the sample number $N_\mathrm{s}$ at least by a factor of 30. At the same time it is clear that the permissible reduction factor may be even larger depending on the observable that is of interest because one may expect cancellation effects when summing over parts of the lattice sites.

To verify this hypothesis, we analyze in Fig.~\ref{fig:tpp_sampling_lines} the time dependence of two relevant observables: the expansion velocity of the central cloud, cf. panel (a), where \cite{schluenzen_prb16} 
\begin{subequations}
\begin{align}
    v_{\rm exp}(t) &\coloneqq \frac{\mathrm{d}}{\mathrm{d}t}R_{\rm exp}(t)\,,\\
    R_{\rm exp}^2(t)&\coloneqq R^2(t)-R^2(0)  \,,
    \label{eq:v-exp}
    \\
    R^2(t) &\coloneqq \frac{1}{N}\sum_\mathfrak{i} n_\mathfrak{i}(t)\, |\vec{r}_\mathfrak{i} - \vec{r}_\mathrm{avg}(t)|^2\,,\\
    \vec{r}_\mathrm{avg} &\coloneqq   \frac{1}{N}\sum_\mathfrak{i} n_\mathfrak{i}(t) \,\vec{r}_\mathfrak{i}\,, 
\end{align}
\end{subequations}
with $N$ being the particle number, and the correlation energy $E_\mathrm{corr}$, Eq.~\eqref{eq:e_corr}, cf. panel (b), which in the Hubbard model is just $d_\mathrm{corr}$, Eq.~\eqref{eq:dcorr}, scaled by the interaction $U$.

Since the considered system cannot be solved exactly, evaluating the performance of the $\delta$TPP approximation is challenging. Here, we compare it to TDHF and GKBA+TPP calculations. While TDHF is expected to capture the dynamics poorly, GKBA+TPP serves as a good benchmark that is known to perform reasonably well in similar setups, as observed in Sec~\ref{ss:cdw}. For the expansion velocity the stochastically exact $\delta$NEGF data lie between the TDHF and GKBA+TPP solution indicating that $\delta$TPP provides a considerable improvement compared to a purely uncorrelated description. For the correlation energy the $\delta$TPP solution again shows a delayed growth, as already observed for the correlations in Fig.~\ref{fig:tpp_cdw_U5_U15}, before reaching an asymptotic value close to the one of GKBA+TPP. Overall, these observations suggest that the good performance of $\delta$TPP for the small 1D system established in Sec.~\ref{ss:cdw} also applies to 2D systems.

Now, we focus on $\delta$TPP simulations with $N_\mathrm{s}= 32, 320$ and $3200$ samples and compare with the exact result (blue lines). The dynamics of the correlation energy is, for all times, very well captured already with 320 samples, whereas using 32 samples only reproduces the global trend. Interestingly, the expansion velocity \eqref{eq:v-exp} is much less sensitive to the sample number and is already accurately obtained with just $32$ samples---an astonishing speedup of more than $3000$, which leads to a gain of CPU time and RAM of this order of magnitude. Based on the results of Figs.~\ref{fig:tpp_sampling_3D} and \ref{fig:tpp_sampling_lines} we conclude that the required number of samples can be adjusted to the observables of interest. While for two-particle quantities, such as the correlation energy, or space resolved single-particle quantities a gain on the order of $300$ is achieved  ($320$ samples), for global single-particle observables, such as the expansion velocity, an additional speedup factor of $10$ is achieved. At this point it is important to reiterate that the reconstruction error of the randomly sampled initial state obeys the central limit theorem, $\varepsilon_\mathrm{error}(N_\mathrm{s}) \propto 1/\sqrt{N_\mathrm{s}}$. It is therefore effectively independent of the system size, as previously demonstrated in Ref.~\cite{schroedter_cmp_22}. Therefore, the achievable speedup factors are significantly higher for larger systems, as will be demonstrated next.

\begin{figure*}[t]
    \centering
    \includegraphics[width=\linewidth]{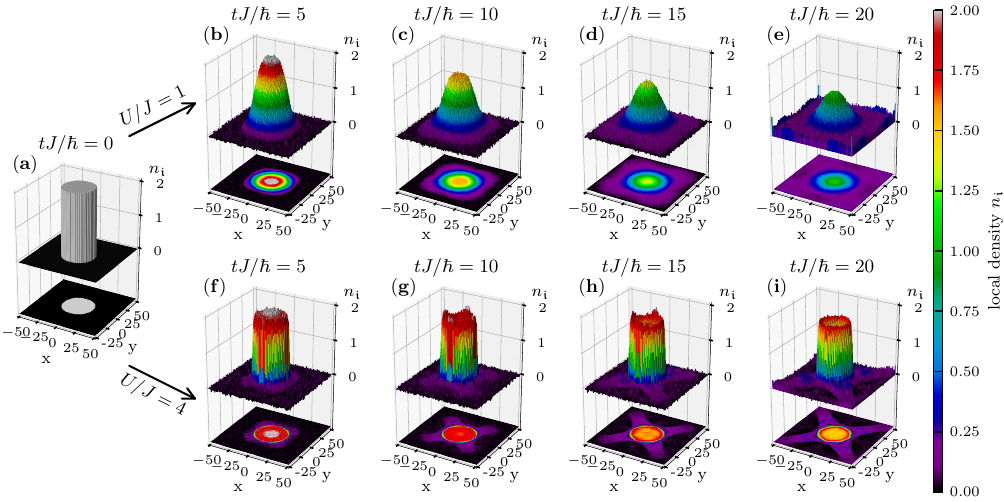}
    \caption{Diffusion of a doubly-occupied circular initial distribution containing 2514 particles on a $101\times 101$ square lattice depicted in panel (a). Surface plots of the local density distribution at different times for  $U/J=1$ [(b)--(e)] and  $U/J=4$ [(f)--(i)]. Simulations were performed using $\delta$TPP with $N_\mathrm{s}=960$ samples. 
    Notice the formation of a hollow cloud in the central part and x-shaped expansion for $U/J=4$.}
    \label{fig:tpp_large_3D}
\end{figure*}

\begin{figure}[h]
    \centering
    \includegraphics[width=\linewidth]{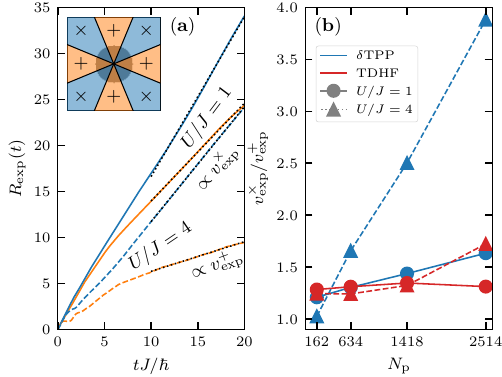}
    \caption{Analysis of the velocity anisotropy of the simulations shown in  Fig.~\ref{fig:tpp_large_3D}. (a): $\delta$TPP results for the expansion dynamics of the radius of the distribution, $d(t)$, Eq.~\eqref{eq:v-exp}, along the axes (orange) and along the diagonals (blue) for $U/J=1$ (solid, upper curves) and $U/J=4$ (dashed, lower curves), cf.~Fig.~\ref{fig:tpp_large_3D}. The fitted dashed black lines represent the asymptotic expansion velocity. (b): Ratio of the expansion velocities along the diagonals ($v^\times_{\rm exp}$) and along the axes ($v^+_{\rm exp}$) for different sizes $N_{\rm p}$ of the initial distribution. The purely diagonal expansion observed in Fig.~\ref{fig:tpp_large_3D} for $U/J=4$ is shown to be present only for large $U/J$ and when including correlations (blue triangles). This effect emerges only when $N_{\rm p} \gtrsim 160$ and monotonically increases with $N_{\rm p}$.
}
    \label{fig:tpp_large_lines}
\end{figure}

This flexibility in choosing the sample number $N_\mathrm{s}$ is a particularly attractive feature of the $\delta$NEGF method and opens the way towards simulating large correlated quantum systems. In the following, we demonstrating the capabilities of the method by considering the same setup as in Fig.~\ref{fig:tpp_sampling_3D} but for a system that is nearly two orders of magnitude larger: in Fig.~\ref{fig:tpp_large_3D} we present simulations for $2514$ particles on a $101\times 101$ square lattice. Such large systems are of high interest since the experiments with fermionic atoms, e.g. \cite{schneider_fermionic_2012} typically use several thousands of particles and previous simulations \cite{schluenzen_prb16, schluenzen_prb17} were restricted to only a few hundred particles which imposes substantial finite size effects. We performed simulations with several sample numbers and established that, in the present case, $N_\mathrm{s}=960$ is sufficient to achieve converged results for the local density dynamics.
In contrast, a $\delta$TPP calculation, using an exact decomposition of the initial state, would require around 81.6 million samples resulting, for the smaller simulation, in a speedup factor of 85 000, for both CPU time and required RAM. Despite this significant reduction of samples and computational resources needed, the amplitude of the resulting noise of the single-particle density is comparable to the case of $N_\mathrm{s}=320$ samples, for the $19\times 19$ system (speedup of 300), cf.~Fig.~\ref{fig:tpp_sampling_3D}(f) and Fig.~\ref{fig:tpp_large_3D}. This empirical observation confirms the result of the central limit theorem stating that the error induced by randomly generating the $\delta$NEGF initial state is nearly independent of the system size and mainly depends on the absolute number of samples.

In Fig.~\ref{fig:tpp_large_3D} we now show the dynamics following a confinement quench up to $t=20\hbar/J$, for two coupling strengths, $U/J=1$ and $U/J=4$. One calculation required $12\,000$ core-hours on AMD Epyc 7313 CPUs using a total of 5 TB of RAM. Consider first the case of the lower coupling strengths (top row). Here the diffusion proceeds as in the case of nearly independent particles: the initial peak is dissolved steadily, and the density spreads uniformly across the entire lattice.

More interesting is the case of moderate coupling, $U/J=4$, shown in the lower row. On the time scales shown, the central peak is only marginally decreasing, and particles are not uniformly filling the entire lattice, as in the case of $U/J=1$. Instead, particles fill only two stripes of the lattice that are connecting the four corners of the lattice. Another striking observation is the change of the shape of the central peak: while its height remains close to two, the inner part is significantly depleted, leaving ring shaped wall with a larger density. Both of these features were not observed before, apparently due to the small size of the previous simulations. Our simulations show that these feature occur only for both, sufficiently strong correlations, requiring simulations with the TPP self-energy, and sufficiently large systems.

In the following, we analyze the effect of the anisotropic expansion in more detail. To this end, we plot in the left panel of Fig.~\ref{fig:tpp_large_lines} the extension $R_\mathrm{exp}(t)$ of the particle cloud measured, respectively, along the axes ($+$) and along the diagonals ($\times$), for weak and moderate coupling strength. While for $U/J=1$ the cloud size grows fast, the growth along the diagonal is only slightly faster than along the x- and y-axes. The situation changes drastically for  $U/J=4$ (dashed lines). Here the cloud extends along the diagonal more than twice as fast than along the coordinate axes which is caused by the different expansion velocities. This dependence is explored again in panel (b) where we plot the ratio of the two velocities against different particle numbers $N_\mathrm{p}$ for both values of $U/J$ as well as for two approximations: time-dependent Hartree-Fock and $\delta$TPP, where all calculations were again performed for $N_\mathrm{s}=960$. Obviously, for $U/J=1$ the asymmetry is only weakly pronounced, in both simulations. In this case the ratio of the velocities is close to the value of $\sqrt{2}$ which corresponds to a square-shaped expansion, as it was observed in previous studies for smaller systems~\cite{schluenzen_prb16, schluenzen_prb17}. The asymmetry emerges only for $U/J=4$ and for $\delta$TPP simulations, whereas it is absent in uncorrelated simulations, cf. the TDHF curve (red triangles). Thus, the asymmetry requires both, an overcritical coupling strength in the system and a simulation with correlations included to a sufficient degree (TPP self-energy). However, there is also a third necessary condition: the system has to be large enough for the asymmetry effect to show up. Even for $U/J=4$ and $\delta$TPP simulations the asymmetry vanishes, if the system size is $N_{\rm p} \lesssim 160$. In contrast, for larger values $N_{\rm p}$ the asymmetry increases exponentially (note the log scale on the $N_{\rm p}$ axis). For $N_{\rm p}=2514$ the asymmetry is already close to $4$ and continues to increase for still larger systems.

\subsection{Spin correlations in a two-leg Fermi-Hubbard ladder ($\delta$TPH approximation)}\label{ss:spin-correlations}
After considering two examples of moderate and strong coupling where the proper self-energy was the particle-particle $T$-matrix, we now analyze the performance of the $\delta$NEGF approach with the particle-hole $T$-matrix self-energy.
To evaluate the performance of $\delta$TPH, we consider a two-leg Hubbard ladder with periodic boundary conditions. This geometry allows for a rigorous benchmark because its quasi-one-dimensional nature amplifies the particle-hole quantum fluctuations that the TPH is constructed to capture. Specifically, we examine a periodic $6\times 2$ ladder with asymmetric hopping: the hopping amplitude along the leg (x-direction) is given by $J_\mathrm{leg}/J=1$, whereas the hopping along the rungs (y-direction) is $J_\mathrm{rung}/J = 0.5$. In this system, the magnetic ordering is of particular interest which we 
 investigate  by evaluating the staggered spin-spin correlation functions, defined as 
\begin{align}
    C_\mathfrak{i}(t) \coloneqq (-1)^{d(\mathfrak{i})} \big\langle \delta\hat{S}^z_{\mathfrak{j}}(t) \delta\hat{S}^z_{\mathfrak{j}+\mathfrak{i}}(t)\big\rangle,\quad
\label{eq:spin-correlations}
\end{align}
where $\mathfrak{j}$ denotes an arbitrary site and we define the distance between the sites $\mathfrak{j}$ and $\mathfrak{j+i}$ as
\begin{align}
    d(\mathfrak{i}) \coloneqq d(\mathfrak{i}_\mathrm{leg},\mathfrak{i}_\mathrm{rung})\coloneqq\min\{\mathfrak{i}_\mathrm{leg}, 6- \mathfrak{i}_\mathrm{leg}\}+\mathfrak{i}_\mathrm{rung}\,,
\end{align}
for $\mathfrak{i}_\mathrm{leg} = 0,\dots, 5$ and $\mathfrak{i}_\mathrm{rung} = 0,1$. More specifically, we examine three quantities: 
\begin{itemize}
    \item $C_\mathrm{loc}\coloneqq C_{(0,0)}$, the local correlations;
    \item $C_\mathrm{leg}\coloneqq C_{(1,0)}$, the nearest-neighbor correlation along the leg;  
    \item $C_\mathrm{rung}\coloneqq C_{(0,1)}$, the nearest-neighbor correlation across the rung. 
\end{itemize}
 We compare the TPH approximation within the $\delta$NEGF approach to the standard TPH within the GKBA. Further, we consider the GKBA+2B approximation as a baseline reference. All approximate methods are benchmarked against exact diagonalization (ED) results using the QuSpin library \cite{QuSpin2017}. 
\begin{figure}[t]
    \centering
    \includegraphics[width=\linewidth]{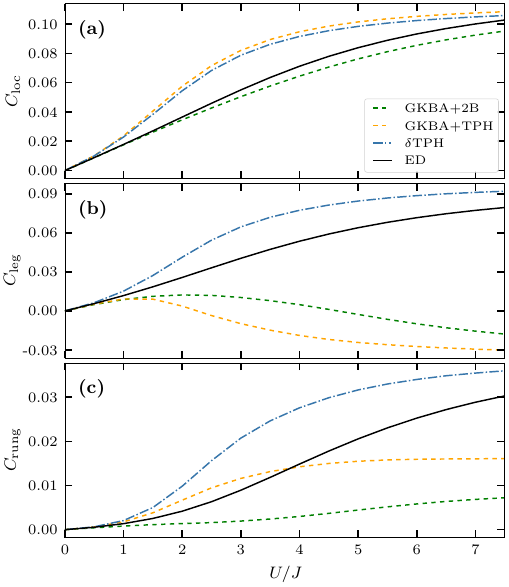}
    \caption{Ground-state staggered spin-spin correlations as a function of the interaction strength $U$ for a half-filled $6\times 2$ Hubbard ladder with periodic boundary conditions and
    asymmetric hopping, $J_\mathrm{leg}/J = 1$ and $J_\mathrm{rung}/J = 0.5$. The panels display (a) the local correlations, $C_\mathrm{loc}$, (b) the nearest-neighbor correlations along the leg, $C_\mathrm{leg}$ and (c) the nearest-neighbor correlations across the rung, $C_\mathrm{rung}$. The exact data is obtained via exact diagonalization (ED).}
    \label{fig:tph_staggered_spin_correlations_U_scan}
\end{figure}

Figure~\ref{fig:tph_staggered_spin_correlations_U_scan} displays the ground-state staggered correlations \eqref{eq:spin-correlations} for an on-site interaction ranging from the non-interacting limit, $U/J=0$, to the strongly correlated regime, $U/J = 7.5$. Consider first the lowest-order approximation: the  2B self-energy. While it accurately reproduces the exact solution for the local moment, $C_\mathrm{loc}$, only slightly underestimating the magnitude, it fundamentally breaks down for the non-local correlations, $C_\mathrm{leg}$ and $C_\mathrm{rung}$. Although the weakly interacting regime ($U/J \lesssim 1$) is described by the 2B self-energy reasonably well, the rung correlations, $C_\mathrm{rung}$, are largely underestimated. More drastically, for $U/J\gtrsim  2$, the leg correlations, $C_\mathrm{leg}$, are unphysically suppressed and systematically decrease with stronger coupling, and for $U/J \gtrsim 5$ even fall below the non-interacting value.

Let us now turn to the particle-hole $T$-matrix self-energy which we study with the standard HF-GKBA and the $\delta$NEGF method, respectively. 
Both versions of the TPH exhibit similar behavior for the local correlations $C_\mathrm{loc}$. While the $\delta$NEGF version slightly underestimates them compared to the GKBA solution, both $T$-matrix approaches overestimate $C_\mathrm{loc}$ at intermediate coupling strengths ($1 \lesssim U/J \lesssim 6$). Only for $U/J \gtrsim 7$ do both TPH variants begin to accurately reproduce the exact local correlations, clearly outperforming the 2B simulations. On the other hand, for the leg correlations, $C_\mathrm{leg}$, GKBA+TPH suffers from the same qualitative failure as 2B, suppressing the spatial ordering with increasing interaction. This suppression is even more pronounced for GKBA+TPH, which drops below the non-interacting limit already at $U/J \gtrsim 2$. However, GKBA+TPH does provide a marginal improvement over 2B for the rung correlations, $C_\mathrm{rung}$, capturing an initial enhancement for $U/J \lesssim 4$, before exhibiting an unphysical saturation. In stark contrast, our novel $\delta$TPH approximation correctly captures the qualitative physics of both non-local correlations, $C_\mathrm{leg}$ and $C_\mathrm{rung}$. While it quantitatively overestimates the exact magnitudes, it accurately predicts the simultaneous growth of robust antiferromagnetic correlations across both, the legs and the rungs, in the strongly correlated regime.

Next, we extend the analysis of spin correlations in this system to nonequilibrium. To this end, we apply a spatially uniform laser pulse polarized along the legs. The pulse is modeled via the Peierls substitution, modifying the hopping amplitude by a complex phase $J_\mathrm{leg}\rightarrow J_\mathrm{leg}e^{\mathrm{i}A(t)}$, with the vector potential given by
\begin{align}
     A(t) =A_\mathrm{L} \cos\left[\omega_\mathrm{L} (t-t_\mathrm{L})\right] e^{-\frac{(t-t_\mathrm{L})^2}{2\sigma^2}}\,,\label{eq:peierls}
\end{align}
where the laser parameters are $t_\mathrm{L} = 10\,\hbar/J$; $\hbar \omega_\mathrm{L} = 2 J$; $A_\mathrm{L} = 0.5$; and $\sigma = 2\, \hbar/J$. To isolate the intrinsic resonant frequencies, we analyze the dynamically induced fluctuations, defined as
\begin{align}
    \Delta C(t) \coloneqq C(t) - C(t_0)\,,
    \label{eq:delta-c}
\end{align}
and compute their Fourier transforms strictly for the post-pulse phase, $t > 15\,\hbar/J$.

\begin{figure*}[t!]
    \centering
    \includegraphics[width=\linewidth]{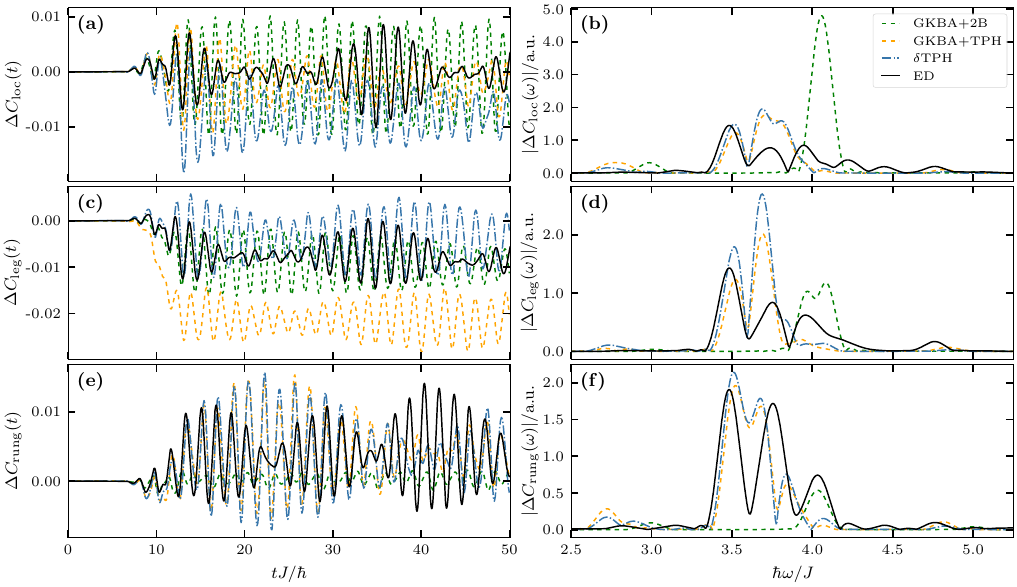}
    \caption{Nonequilibrium dynamics of the staggered spin-spin correlations for the same system as in Fig.~\ref{fig:tph_staggered_spin_correlations_U_scan} at $U/J=1$ following a laser excitation along the legs of the ladder with an amplitude $A_\mathrm{L} = 0.5$, a frequency $\hbar\omega_\mathrm{L} /J = 2 $, a width $\sigma J/\hbar = 2$ and centered at $t_\mathrm{L} J/\hbar = 10$. Left column: time evolution of the dynamically induced fluctuations in the staggered correlations, $\Delta C(t)$, Eq.~\eqref{eq:delta-c}. Right column: Fourier transforms of the staggered correlations following the laser excitation evaluated from $t_1 J/\hbar =15$ until $t_2J/\hbar= 100$. The rows correspond to the local (a, b), leg nearest-neighbor (c, d) and the rung nearest-neighbor (e, f) correlations. The exact data are computed via exact diagonalization (ED).}
    \label{fig:tph_dynamical_staggered_correlations}
\end{figure*}

The simulation results for $U/J = 1$ are displayed in Fig.~\ref{fig:tph_dynamical_staggered_correlations}. An interesting first observation is that, while for the chosen weak interaction strength, all approximations (GKBA+2B, GKBA+TPH, and $\delta$TPH) are in very good agreement with each other and with the exact ground state [cf. Fig.~\ref{fig:tph_staggered_spin_correlations_U_scan}], their predictions for the nonequilibrium response to the laser pulse differ significantly. The exact solution exhibits a multi-peak structure in the frequency domain, with three dominant excitation frequencies located at $\hbar\omega/J \approx 3.5$, $3.75$, and $4$. While the $\hbar\omega/J \approx 3.5$ peak carries most of the spectral weight globally, the highest-energy peak at $\hbar\omega/J\approx 4$ remains a distinct feature.

In stark contrast to the exact solution, the 2B dynamics are dominated by a single characteristic frequency at $\hbar\omega/J \approx 4$, with only a minor contribution near $\hbar\omega/J \approx3$. While the spectral weight of the dominant peak is significantly overestimated for the local correlations [panel (b)], it is well captured for the rung correlations [panel (f)] compared to the ED solution.  A slight peak splitting begins to emerge for 2B in the leg correlations [panel (d)], but 2B generally fails to capture the lower-energy structure of the exact spectrum. In contrast, both realizations of the TPH self-energy demonstrate very good agreement with the exact reference and successfully capture the lower-energy features of the exact spectrum ($\hbar\omega/J \approx 3.5$ and $3.75$). However, they predict that the intermediate frequency ($\hbar\omega/J \approx3.75$) carries the dominant spectral weight for the local and leg correlations, and both fail to reproduce the highest-energy peak at $\hbar\omega/J \approx4$. 

Let us now turn to an analysis in the time domain, cf. left panel of Fig.~\ref{fig:tph_dynamical_staggered_correlations}, which, in addition to the previous results, also allows us to study the short time behavior during the laser excitation. The spectral discrepancies between the approximations and the exact benchmark are, in the time domain, associated with phase drifts and amplitude variations. During the initial laser excitation phase, GKBA+TPH predicts an immediate decrease of the local correlations [panel (a)], an effect that is present also in $\delta$TPH simulations where it is somewhat overestimated. The exact solution, however, shows no significant offset during excitation. For the leg correlations [panel (c)], the exact solution decreases during the laser excitation and $\delta$TPH well captures this initial drop, whereas GKBA+TPH fails to reproduce it and also tends to much too low values of $\Delta C_\mathrm{leg}$. Finally, both TPH realizations exhibit excellent agreement for the rung correlations [panel (e)], correctly describing both, the initial excitation phase and the subsequent dominant oscillation frequencies, albeit with a slightly narrower spectral bandwidth than the exact solution [panel (f)].

Summarizing the results shown in Figs.~\ref{fig:tph_staggered_spin_correlations_U_scan} and \ref{fig:tph_dynamical_staggered_correlations},
despite its reduced self-consistency $\delta$NEGF is in good agreement with the standard GKBA results with the same self-energy and in some cases even outperforms the latter 
for the description of staggered spin-spin correlations. While $\delta$TPH quantitatively overestimates the absolute magnitude of the ground-state correlations compared to ED, it is the only one of the many-body approximations that is capable of recovering the correct qualitative physics of robust non-local antiferromagnetic ordering at strong coupling. Furthermore, the nonequilibrium dynamical response of $\delta$TPH remains highly consistent with the standard GKBA+TPH, demonstrating that the $\delta$NEGF method successfully describes ground-state correlations without severely compromising the validity of the dynamical excitation spectrum.

\subsection{Dynamical correlation and response properties of a 1D Hubbard  chain ($\delta GW$ approximation)}\label{ss:conductivity}

\begin{figure}[t]
    \centering
    \includegraphics[width=\linewidth]{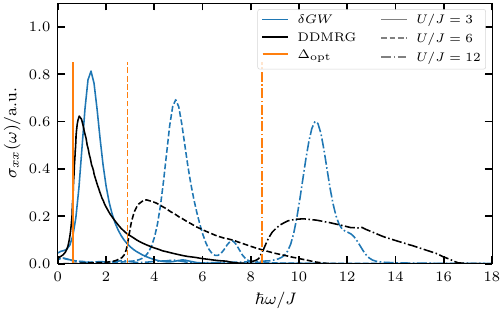}
    \caption{Real part of the optical conductivity $\sigma_{xx}(\omega)$, Eq.~\eqref{eq:opt_cond_current}, for a half-filled open 1D Hubbard chain of 128 sites in the ground state. Dynamical DMRG (DDMRG) data are taken from Jeckelmann et al. \cite{Jeckelmann2000}. The vertical orange lines indicate the optical gap given by the exact Bethe ansatz solution for the infinite Hubbard chain. The $\delta$$GW$ ground state is generated via adiabatic switching up to a time of $tJ/\hbar=200$. All data use the same numerical broadening so differences are due to the approximations. All spectra are normalized to 1. 
    }
    \label{fig:gw_opt_chain}
\end{figure}

\begin{figure*}[t]
    \centering
    \includegraphics[width=\linewidth]{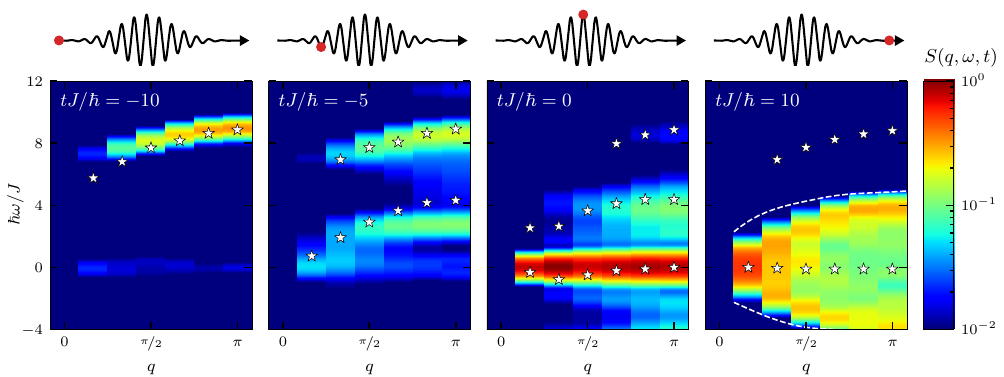}
    \caption{Time-resolved $\delta GW$ data for the (charge) dynamic structure factor of an open 12-site Hubbard chain for $U/J=8$ before (panel 1), during (panel 2 and 3) and after (panel 4) laser excitation with photon energy $\sim 4.4\,J$. The initial state is generated via adiabatic switching up to a time of $tJ/\hbar=200$. Stars indicate exact reference data by Wang et al.\cite{Wang2017}}
    \label{fig:gw_laser_chain}
\end{figure*}

In this section, we demonstrate one of the principal advantages of our $\delta$NEGF method, namely the direct and efficient access to time-resolved dynamical correlation and response properties. Conventionally, equilibrium dynamical response functions are often obtained by explicitly simulating the response of the system to an external perturbation, such as a short kick or a monochromatic drive~\cite{kwong_prl_00}. This strategy has several well-known drawbacks. In the monochromatic case, a large number of separate calculations is required to sample a broad frequency or wavenumber range. In time-dependent nonequilibrium settings, the procedure becomes even more costly, since the perturbation must be applied at different times in order to recover temporal resolution. Moreover, this route naturally yields only the response functions, i.e., the retarded
real-time component. In equilibrium, the corresponding correlation functions, i.e., the lesser and greater real-time components, may still be reconstructed through the fluctuation-dissipation theorem, e.g., Ref.~\cite{giuliani2005quantum}, but in nonequilibrium no such relation is generally available.

By contrast, within the $\delta$NEGF framework, two-time response functions and the associated correlation functions can be obtained directly from the single-particle fluctuations, as discussed in detail in Sec.~\ref{ss:cor-funcs} and in Refs.~\cite{schroedter_23, schroedter_pssb23}. To illustrate these capabilities, we first consider in Fig.~\ref{fig:gw_opt_chain} the real part of the optical conductivity as defined in Eq.~\eqref{eq:opt_cond_current}, for $\omega>0$.
We study a half-filled open one-dimensional Hubbard chain with $128$ sites in the ground state, focusing on moderate to strong interaction strengths: $U/J=3, 6$ and $12$.

As an exact reference, we compare with the DDMRG results of Jeckelmann et al.~\cite{Jeckelmann2000}, which exhibit a pronounced peak slightly above the optical gap together with a broad tail extending to higher energies. With increasing $U$, this peak structure becomes progressively less distinct. In the strong-coupling regime, the spectral weight is centered approximately around $U$ and broadened over an energy window of about $8J$, as is clearly visible for $U/J=12$.

The $\delta GW$ ground state is prepared by adiabatic switching starting from a slightly spin-symmetry-broken initial state, resulting in an antiferromagnetically correlated spin-density-wave (SDW) state. Owing to the sizable spin-density dynamics during the switching procedure, the final state is not a perfectly converged ground state. For $U/J=12$, this is visible in the form of weak additional spectral contributions around $\hbar\omega/J\approx 0$ and $\hbar\omega/J\approx 4-6$, indicating a slight residual excitation.
The optical conductivity, Eq.~\eqref{eq:opt_cond_current}, is then obtained in the ground state from the two-time current-current response, cf. Tab.~\ref{t:spectra}, via 
the Fourier transform \eqref{eq:fourier-trafo} of $\chi^\mathrm{R}_{jj}$,
using a Gaussian probe function \eqref{eq:gauss} with $\sigma_\mathrm{pr} = 3 \hbar J^{-1}$. To facilitate comparison with the DDMRG data, the optical conductivity spectra have been convoluted with a Lorentzian of width (HWHM) $\hbar\eta/J = 0.1$ and normalized to unity.

Overall, the $\delta$$GW$ approach reproduces the qualitative opening of the Mott gap well, even at large interaction strengths. Quantitatively, however, the gap is overestimated, particularly for stronger interactions, while the spectral broadening is underestimated, leading to a main peak that remains too sharp. One notable feature that is captured correctly is the emergence of a secondary peak around $\hbar\omega/J\approx 7$ for $U/J=6$ and around $\hbar\omega/J\approx 12.5$ for $U/J=12$, both in good agreement with the exact solution.

In summary, these findings are consistent with earlier observations that the two-particle spectra obtained within $\delta$$GW$ are comparable in quality to RPA-like results~\cite{kwong_prl_00, schroedter_23}. For the present system, this conclusion is further supported by the Gutzwiller+RPA data of Seibold et al.~\cite{Seibold2001,Seibold2003}, which show a performance similar to that of our $\delta$$GW$ spectra. 
Finally, it is important to emphasize that, in the $\delta$NEGF formalism, the response functions obtained at the $\delta GW$ level are not merely postprocessed observables. They are propagated quantities that re-enter the equations of motion self-consistently. This is in marked contrast to conventional perturbation--response schemes, where response properties are extracted from the reaction of an underlying time-local propagation to an external perturbation. In such approaches, the response kernel may contain correlation effects that are not present at the same level in the quantities entering the actual equations of motion~\cite{kwong_prl_00}, as in TDHF~\cite{McLACHLAN_1964}, TDDFT~\cite{Casida_1995,Petersilka_1996}, or static and adiabatic $GW$/BSE-based response schemes~\cite{Sangalli_2019,Marek2025}. In $\delta$NEGF, by contrast, the same fluctuation objects determine both the correlated dynamics and the response functions. We return to this point in Sec.~\ref{ss:numerics-ppp}.

Next, in Fig.~\ref{fig:gw_laser_chain} we again turn to the nonequilibrium dynamics following a short laser pulse, cf.~Eq.~\eqref{eq:peierls} with parameters $A_\mathrm{L} = 2.4$, $\hbar\omega_\mathrm{L} = 4.4J$ and $\sigma_\mathrm{L} = 3\hbar/J$, which is sketched on the top of the figure. The system studied is an open 12-site Hubbard chain at strong coupling, $U/J=8$.
In the main row of Fig.~\ref{fig:gw_laser_chain} 
we present results for the dynamic structure factor, Eq.~\eqref{eq:dsf-noneq}, 
at different times (indicated by the red dots) before, during and after the pulse. For the Gaussian envelope, cf.~Eq.~\eqref{eq:gauss}, we choose $\sigma_\mathrm{pr} = 2 \hbar J^{-1}$. As emphasized above, one of the central strengths of $\delta$NEGF is the direct availability of two-time response functions and their associated correlation functions. In the present nonequilibrium example, the full time-, momentum-, and frequency-resolved charge dynamical structure factor is obtained from a single $\delta$$GW$ simulation. 

The $\delta$$GW$ ground state is again an antiferromagnetic SDW state prepared in the same way as for Fig.~\ref{fig:gw_opt_chain}. In the left panel, at $tJ/\hbar= -10$, weak contributions around $\hbar\omega/J\approx0$ indicate that the state is not a perfectly pure ground state, but only very weakly excited. Note, however, that these features are visible only on the logarithmic scale employed here. As a reference, we indicate by white stars the peak positions of the different branches of the charge dynamical structure factor obtained from exact diagonalization (ED) by Wang et al.~\cite{Wang2017}.

Before the arrival of the pulse (panel 1), the $\delta$$GW$ spectra reproduce the equilibrium branch structure reasonably well, although, as already seen in Fig.~\ref{fig:gw_opt_chain}, the Mott gap is somewhat overestimated. During the laser pulse (panels 2 and 3), transient sideband features appear, which can be interpreted as virtual Floquet-like replicas of the equilibrium excitations. As the pulse evolves (panel 3), these sidebands flatten, consistent with a renormalization of the effective bandwidth, while at the same time a narrow, nearly gapless low-energy band of charge excitations builds up. Importantly, these low-energy features persist even after the pulse has ended (panel 4), indicating that they correspond to real photoinduced electronic states rather than purely virtual dressed states. The width of this low-energy band agrees well with the ED reference and is consistent with the white dashed  lines shown in panel 4 to guide the eye.

The main deficiency of the $\delta GW$ description is found in the bleaching of the gapped Mott excitation branch. In the ED data of Wang et al., this branch is strongly suppressed by roughly an order of magnitude but remains clearly visible throughout the driven dynamics. In the $\delta GW$ results, by contrast, the corresponding branch is over-bleached and nearly disappears completely. Despite this shortcoming, the overall nonequilibrium spectral evolution is captured well at a qualitative level.

Most importantly, the present example highlights the practical scope of the method: whereas exact diagonalization is limited here to a 12-site Hubbard chain, the same type of analysis within the $\delta$NEGF framework can be carried out for far larger and structurally much more complex systems containing thousands of basis functions, as demonstrated in the next section.

\subsection{Dynamical correlation and response properties of carbon nanostructures with long range interaction ($\delta$RPA approximation)}\label{ss:numerics-ppp}

So far, all calculations have been performed for the standard local Hubbard model. This is a natural starting point for assessing the performance of the two $T$-matrix approximations, which are designed to describe short-range particle-particle and particle-hole scattering processes. For the $GW$/RPA class of approximations, however, the physically relevant regime is one with dynamical screening and long-range Coulomb interactions.
To this end, we present an outlook of applying the $\delta$NEGF approach to Coulomb systems in the following.
We, therefore, now turn to the extended Hubbard model, Eq.~(\ref{eq:hubbard-ppp}) with parameters $J=2.34\,\mathrm{eV}$ and $U=3.54J$, which extends the Hubbard Hamiltonian by including nonlocal Coulomb matrix elements.

\begin{figure}[t]
    \centering
    \includegraphics[width=\linewidth]{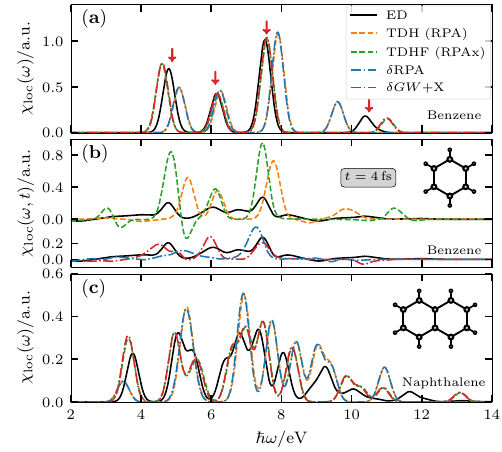}
    \caption{Local retarded density response \eqref{eq:chi-r} of (a) Benzene and (c) Naphthalene described in the extended Hubbard model \eqref{eq:hubbard-ppp}. $\delta$RPA and $\delta GW$+X results are compared to TDH and TDHF data obtained from time-dependent calculations following an infinitesimal kick and to ED data obtained using the QuSpin library~\cite{QuSpin2017}. The red arrows indicate the peak positions obtained from a kick calculation using $\delta GW$+X.
    (b) Laser induced change in the local retarded density response, Eq.~\eqref{eq:chi-r}, with laser parameters given in the text. For the sake of clarity, the lines have been shifted and the ED result is shown twice.}
    \label{fig:ppp-small}
\end{figure}

This modification of the interaction tensor increases the numerical cost of $\delta$NEGF simulations for most approximations, changing the scaling from $\mathcal{O}(N_\mathrm{s}N_\mathrm{b}^2N_\mathrm{t})$ to $\mathcal{O}(N_\mathrm{s}N_\mathrm{b}^3N_\mathrm{t})$, with one notable exception:  $\delta$RPA, which contains no explicit exchange terms, retains the same favorable scaling, even in the presence of long-range interactions. Exchange effects are nevertheless incorporated implicitly through the generalized susceptibility $\chi$ entering the collision integral in Eq.~\eqref{eq:i-def-chi-k}. This makes $\delta$RPA a particularly attractive candidate for large-scale simulations with nonlocal interactions. In Fig.~\ref{fig:ppp-small} we, therefore, assess its performance by comparing the local retarded density response function
\begin{align}
    \chi_{\rm loc}(\omega, t) &= \sum_i \chi^\mathrm{R}_{n_in_i}(\omega, t)
    \label{eq:chi-r}\,,
\end{align}
for benzene and naphthalene, to exact diagonalization (ED), as well as to time-dependent Hartree (TDH) and time-dependent Hartree--Fock (TDHF) calculations. The latter two correspond to the direct random-phase approximation (RPA) and the random-phase approximation with Hartree--Fock exchange kernel (RPAx), respectively~\cite{Hellgren2018}.

Note that, as we employ the extended Hubbard model here, this comparison should not be interpreted as an attempt to obtain the most accurate optical spectra of benzene or naphthalene in an \textit{ab initio} sense. In particular, the ED solution of the extended Hubbard model deviates from literature \textit{ab initio} excitation energies~\cite{Marek2025}. The relevant question here is instead whether the $\delta$NEGF approximations reproduce the ED reference within the model.

For benzene, panel (a) shows that the ED ground-state spectrum contains four dominant peaks at approximately $4.8$, $6.2$, $7.5$, and $10.4\,$eV. The $\delta$RPA spectrum gives the correct overall structure but slightly overestimates the first three excitation energies, with deviations up to about $0.4\,$eV, while the fourth peak is underestimated by roughly $0.7\,$eV. The $\delta GW$+X approximation improves the description of the intermediate-energy part of the spectrum: the second and third peaks are reproduced very well, while the first peak is underestimated by only about $0.2\,$eV. The fourth peak is instead shifted upward by about $0.6\,$eV. Notably, the $\delta$RPA curve coincides with TDH/RPA, whereas $\delta GW$+X coincides with TDHF/RPAx, as discussed in Sec.~\ref{ss:conductivity}.

The same overall picture emerges for naphthalene in panel (c). The $\delta$RPA approximation again gives a reasonable account of the main spectral features, but misses an important splitting of the structures around $5.3\,$eV and $7\,$eV. This splitting is captured by $\delta GW$+X, consistent with the known need for exchange effects beyond RPA in the description of these excitations~\cite{schindlmayr_prl_98,kwong_prl_00,reining_jcp_09}. As for the case of benzene, the equivalence between $\delta$RPA and TDH/RPA, and between $\delta GW$+X and TDHF/RPAx, respectively, is fulfilled in the linear-response spectra. 

\begin{figure*}[t]
    \centering
    \includegraphics[width=\linewidth]{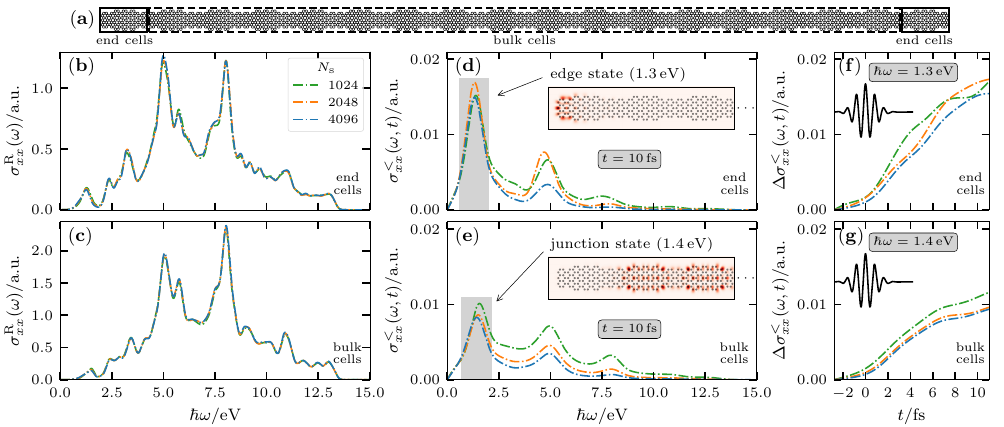}
    \caption{(a): Sketch of a 7--9-AGNR heterostructure containing 18 unit cells ($\,1728$ sites). (b, c): Ground state retarded longitudinal optical conductivity, Eq.~\eqref{eq:sigma-R,<}, for three numbers of samples, for an end and bulk cell, respectively. (d, e): Laser-induced lesser longitudinal optical conductivity, Eq.~\eqref{eq:sigma-R,<}, $10$ fs after the pulse maximum, for three different numbers of samples, for an end and bulk cell, respectively. (f, g): Laser-induced spectral weight of the edge and junction state, respectively, Eq.~\eqref{eq:delta-sigma-<}.}
    \label{fig:ppp_large_797}
\end{figure*}

This agreement in the equilibrium spectra should, however, be interpreted with care, since the TDH and TDHF response functions are obtained from the explicit time-dependent response to a sudden perturbation (``kick''). As a consequence, these quantities are output observables only and may contain higher-order correlation contributions beyond those that are actually fed back into the equations of motion, see e.g., Refs.~\cite{kwong_prl_00, bonitz_qkt}, for a discussion. Therefore, they are not an accurate representation of the physical effects that are captured by these methods when describing nonlinear nonequilibrium dynamics. In contrast, within the $\delta$NEGF approach the correlation functions are generated self-consistently and enter directly into the dynamical evolution. 

This difference becomes particularly relevant in Fig.~\ref{fig:ppp-small}b, where we show the local retarded density response function $4\,$fs after an interaction with a short laser pulse
\begin{align}
    \mathcal{E}_x(t) = \mathcal{E}_\mathrm{L} \cos\left[\omega_\mathrm{L} (t-t_\mathrm{L})\right] e^{-\frac{(t-t_\mathrm{L})^2}{2\sigma_\mathrm{L}^2}}\,,
\end{align}
with parameters $\mathcal{E}_\mathrm{L}=1J/(ea)$, $\sigma=1.7\,$fs. Additionally, we choose $\omega_\mathrm{L} = 3.8\,$eV for ED, TDHF, and $\delta GW$+X, while $\omega_\mathrm{L} = 4.0\,$eV is used for TDH and $\delta$RPA to account for the slightly different excitation energies. 

The pump mainly addresses the third peak through a two-photon process, $2\omega_\mathrm{L} \approx 7.6–8.0\,$eV. The ED result shows pronounced bleaching of the equilibrium absorption peaks, accompanied by spectral broadening and the emergence of new photoinduced structures. Both $\delta$RPA and $\delta GW$+X reproduce these qualitative nonequilibrium features, with $\delta GW$+X giving the more accurate description. By contrast, TDH and TDHF strongly underestimate the laser-induced redistribution of spectral weight and show only weak changes relative to the equilibrium spectrum. This demonstrates that the agreement of TDH/TDHF with $\delta$RPA / $\delta GW$+X in the equilibrium density-response spectra does not carry over to the correlated nonequilibrium dynamics, where the self-consistent $\delta$NEGF treatment is essential.

Of course, the common way of calculating the density response via a small kick perturbation can also be employed with the $\delta$NEGF approach to include additional vertex corrections in the output observable. This is demonstrated for $\delta GW$+X in Fig.~\ref{fig:ppp-small}a, where the red arrows indicate the improved excitation energies obtained by this method which show the best agreement with the ED data.

Finally, in Fig.~\ref{fig:ppp_large_797} we present an application of the $\delta$NEGF method to a large 7-9-AGNR nanoribbon heterostructure containing $1728$ sites (18 unit cells) which is described within the extended Hubbard model \eqref{eq:hubbard-ppp}. Similar systems can be synthesized reliably and have been studied both experimentally and numerically before~\cite{rizzo_topological_2018,joost_prr_25}. A particular feature of these systems are the correlated, topologically protected edge states~\cite{joost_19_nanolett}. We compute the optical conductivity, $\sigma^{\rm R}_{xx}$ in the ground state and compare it to the conductivity that is induced by a short laser pulse
\begin{subequations}
    \begin{align}
        \sigma^{\mathrm{R},<}_{xx}(\omega,t) &= \omega\, \mathrm{Im} \chi^{\mathrm{R},<}_{P^x P^x}(\omega,t) \,,\label{eq:sigma-R,<}\\
        \Delta \sigma^{<}_{xx}(\omega, t) &\coloneqq \int_{{\omega}-\Delta \omega}^{{\omega}+\Delta \omega}[\sigma^{<}_{xx}(\omega', t) - \sigma^{<}_{xx}(\omega', t_0)]\mathrm{d}\omega',
    \label{eq:delta-sigma-<}
\end{align}
\end{subequations}

with $\Delta \omega = 0.75\,\mathrm{eV}$.

A central issue addressed in Fig.~\ref{fig:ppp_large_797} is the convergence with respect to the number of stochastic samples. The fully converged reference requires around 1.5 million samples. Already at $N_\mathrm{s}=1024$, the ground-state spectra are very well converged. The spectra of the bulk [panel (c)] and end [panel (b)] cells are overall very similar, with the most notable differences appearing in the low-energy regime. In the end cell, one finds an in-gap particle-hole edge state at approximately $1.3\,\mathrm{eV}$, whereas in the bulk, a corresponding in-gap particle-hole junction state appears at about $1.4\,\mathrm{eV}$.

A snapshot taken $10\,\mathrm{fs}$ after the laser pulse [panels (d) and (e)] shows that the particle-hole occupation is concentrated predominantly in these two in-gap states, an effect that is even more pronounced at the end cell. The nonequilibrium results are somewhat less well converged than the ground-state spectra. In particular, the bulk-cell data, averaged over 16 cells, exhibit a significantly smoother convergence behavior than the end-cell data, which are averaged over only 2 cells. For the latter, the $N_\mathrm{s}=2048$ result deviates more strongly from the $N_\mathrm{s}=4096$ reference than the $N_\mathrm{s}=1024$ result, in some frequency regions. Nevertheless, all sample numbers considered yield the same qualitative physical picture.

The integrated spectral weight of the main excitonic peaks [panels (f) and (g)] reveals a delayed build-up of the corresponding populations, after the laser pulse. This behavior suggests a two-step mechanism in which the pulse first excites hot carriers, which subsequently scatter into bound electron-hole states. The excitonic occupation appears to be essentially converged for times later than about $10\, \mathrm{fs}$. This agrees well with the results obtained previously, for smaller graphene nanostructures \cite{joost_prr_25}, where an exciton formation time of roughly $10$--$15\, \mathrm{fs}$ after short pulse excitation was  observed.

Once again, convergence is substantially smoother for the bulk than for the end region. This further supports the conclusion already reached for the diffusion setup in the Hubbard model in Sec.~\ref{ss:diffusion} [cf. Fig.~\ref{fig:tpp_sampling_lines}]: strongly averaged observables converge much faster with the number of samples than more localized quantities. Even so, the calculation with only $1024$ samples already reproduces the correct trend.

Finally, we comment on the computational effort. The largest calculation, for $N_\mathrm{s}=4096$, required around $8000$ core-hours on AMD Epyc 7313 CPUs using a total of 1.2~TB of RAM. Even though the present system is still only of moderate size, since the $\delta$NEGF implementation is extremely well suited for parallel execution, substantially larger systems should be accessible on appropriately sized HPC architectures.

\section{Summary and Outlook}\label{s:discussion}

In this paper we presented a novel approach to the ultrafast dynamics of correlated quantum systems: a nonequilibrium Green functions-based quantum fluctuations theory, denoted $\delta$NEGF. This was motivated by severe restrictions of previous NEGF-based simulations: full NEGF simulations scale cubically with the simulation time, which severely limits their range of applicability.
Recent concepts that overcome this bottleneck and achieve linear  runtime scaling, such as the G1--G2 scheme \cite{schluenzen_prl_20, joost_prb_20, joost_prb_22}---a time-local reformulation of the HF-GKBA---face novel problems:  numerical instabilities that are caused by the HF-GKBA, difficulties in handling strong correlations, as well as memory bottlenecks, due to the need to store the correlated part $\mathcal{G}$ of the two-particle Green function. As we have demonstrated in this paper, focusing on fluctuations is a promising way to eliminate $\mathcal{G}$ without compromising on computing speed or accuracy. This way $\delta$NEGF overcomes the basis-size bottleneck of the G1--G2 scheme while retaining dynamical self-energy effects, thereby opening correlated real-time simulations of spatially inhomogeneous systems beyond the reach of conventional two-time NEGF and G1--G2 implementations.

Let us put our theory into the context of other fluctuations  approaches. The mathematical investigation of fluctuations was introduced by Einstein \cite{Einstein1905}, Smoluchowski \cite{Smoluchowski1906}, Langevin \cite{Langevin1908} and others in order to describe the random impacts of fluid molecules on a macroparticle (Brownian motion) and to describe the transport properties of the latter without resolving the dynamics of the former. Similarly, the influence of thermal noise in macroscopic systems, such as conductors, has been effectively described by stochastic methods in terms of the Nyquist theorem \cite{Nyquist1928,Johnson1928}. More generally, the analysis of the behavior of a many-particle system exposed to external influences (such as a thermal bath) has led to the theory of open systems, as formulated, e.g., by Lindblad \cite{Lindblad1976,gorini_1976}.

Common to all these problems is that the fluctuations are caused by external influences (``bath'') on a physical system transforming the latter into an ``open system'' \cite{Weiss1993,BreuerPetruccione2002}. In contrast, our theory concentrates on isolated systems where the fluctuations are intrinsic and caused by correlations between the particles [even though our theory can be straightforwardly generalized to open systems as well].
The idea to map correlation effects on density fluctuations (density inhomogeneities) has been developed for fluids in the frame classical density functional theory (DFT) by Evans \cite{evans1979} and, for quantum systems, in the frame of orbital-free DFT, e.g., \cite{huang2010wt,witt2018ofdft} and stochastic DFT \cite{baer_prb_18}.
An independent concept to represent correlations in terms of fluctuations has been introduced, for classical kinetic theory of plasmas, by Klimontovich \cite{klimontovich_method_1958, dufty_jpcs_05,bonitz_cpp_24} who also introduced approximations for the fluctuations, such as the approximation of second moments and the polarization approximation that were discussed in Sec.~\ref{s:approx}. Similar approximations for the fluctuations were used by DuBois to investigate plasma turbulence  \cite{dubois_phys-fl_76}.
However, these approximations were used for analytical derivations only and not for computational analysis, and primarily focused on classical systems. In contrast, our theory is of intrinsically nonequilibrium nature and formulated in the framework of nonequilibrium  Green functions.

We also mention an independent direction of research  that is based on quantum fluctuations. Examples are stochastic mean field theory \cite{ayik_stochastic_2008, lacroix_prb14, lacroix_simplified_2016} and the truncated Wigner approximation \cite{polkovnikov_quantum_2003,rousse_prb_23}. The present theory has conceptual similarities with these methods but  differs substantially:
Our approach is based on the fact that two-particle correlation functions can be expressed as a tensor decomposition in terms of single-particle quantum fluctuations, which was first discovered for the equilibrium case by Garrod and Percus~\cite{garrod_reduction_1964} and found later application, for example, in Refs. \cite{Valdemoro2000,Mazziotti2004,mazziotti_reduced-density-matrix_2007}. In nonequilibrium, being able to express the dynamics of a full two-particle object by propagating single-particle fluctuations has numerous positive implications: First, it preserves time-linearity and, at the same time, guarantees positivity of the generalized susceptibilities and, thus,  stable solutions, without the need for any regularization. Second, the computational cost is reduced to that of a finite number $N_\mathrm{s}$ of Hartree-Fock propagations that can be performed in parallel. Third, a rank-reduced initial state can be generated very efficiently via random sampling which greatly reduces the numerical cost and allows to perform nonequilibrium simulations for nonuniform systems, as large as $10^4$ lattice sites.
Since the introduced approximations ($\delta$TPP, $\delta$TPH(+X), $\delta GW$(+X), $\delta$RPA) by construction preserve the rank of the initial state, this efficient representation remains valid also for long propagation times. Fourth, it allows to calculate two-time response and correlation functions directly from a single-time calculation without the need to propagate observables in the two-time plane explicitly. This, in combination with the numerical advantages mentioned above, enables the calculation of time-resolved resonant inelastic X-ray scattering (tr-RIXS) spectra for large systems~\cite{Chen_2019,Mitrano2020}.

At the same time, the reduced level of self-consistency of the $\delta$NEGF approach, compared to the HF-GKBA and the rank-conserving property of its approximations, can lead to the underestimation of short-time relaxation behavior, as has been observed in Ref.~\cite{joost_quenching_2026}. One purpose of this paper was, therefore, to present thorough benchmarks of the accuracy of the $\delta$NEGF approach for various lattice setups and to evaluate the convergence behavior with respect to the sample number $N_\mathrm{s}$. In general, the $\delta$NEGF approach was found to be in good agreement with the HF-GKBA for the same self-energy approximations.

Moreover, in certain scenarios $\delta$NEGF even outperformed the HF-GKBA, for instance when treating strong coupling up to $U/J=20$, which was previously far out of reach with NEGF-based simulations. This fact is not surprising, since it has already been observed in the past that a reduced level of self-consistency can lead to an improved description of certain effects. For example, the HF-GKBA resolves the overdamping problem exhibited by the fully self-consistent solution of the KBEs~\cite{von_friesen_successes_2009,puig_von_friesen_kadanoff-baym_2010,schluenzen_prb17,schluenzen_prb17_comment}.
Regarding the low-rank approximation of the two-particle fluctuations via random sampling of the initial state, the number of samples $N_\mathrm{s}$ required for an accurate representation depends on the desired observables with global observables converging much faster than local ones. Further, we observed that, for a given number of samples, the local noise is independent of the system size.

Combined with the rank-preserving property of the $\delta$NEGF approach this allows for a highly efficient representation of large systems of up to $N_\mathrm{b}\sim 10^4$ basis functions using only $N_\mathrm{s} = 10^3$--$10^4$ samples resulting in speed-up and compression factors of $\sim10^5$ compared to equivalent G1--G2 calculations.

This opens the way to treat even larger lattice model systems with more particles in arbitrary geometries, which is especially relevant for the modeling of strongly correlated and multiorbital materials using downfolding Hamiltonians~\cite{Aryasetiawan_2004,Aryasetiawan_2006,Tubman_2005}. A particular strength of the $\delta$NEGF approach is expected to lie in its ability to describe large heterogeneous systems with, e.g., defects~\cite{Jie_2019}, or heterostructures~\cite{joost_19_nanolett}, as well as interfaces between materials and molecules~\cite{Baumgaertner_2026}, which all require large basis sets.

However, the $\delta$NEGF approach is not restricted to lattice models. Real materials are directly accessible by using Kohn-Sham orbitals from a DFT simulation as done, e.g., in the Yambo code \cite{marini_yambo_2009,Sangalli_2019}. The $\delta$NEGF idea would significantly extend such a combination to more advanced self-energies. Further, the $\delta$NEGF approach can be directly applied to atoms or molecules by using basis sets from quantum chemistry as input~\cite{perfetto_cheers_2018,pavlyukh_pss_23}. This would enable the simulation of attosecond Auger~\cite{Covito_2018}, shake-up~\cite{Mansson_2021,pavlyukh_interacting_2022}, and charge migration~\cite{perfetto_ultrafast_2018,Perfetto_2019,Perfetto_2020} processes in large nanostructures and biologically relevant molecules with thousands of active orbitals, which is out of reach for current state-of-the-art methods~\cite{Nisoli_2017,Palacios_2020,Calegari_2023,Petropoulos_2024}.

Among the open problems of the $\delta$NEGF approach we mention the modeling of uniform systems with plane wave basis sets. Here, as is the case for the G1--G2 scheme~\cite{joost_prb_20}, the scaling with the basis size is less favorable than for lattice models and requires additional optimizations by exploiting symmetries of the system. Further open questions include the development of factorizable fluctuation formulations for more advanced self-energy approximations, such as the dynamically screened ladder approximation (DSL)~\cite{joost_prb_22}, which would particularly strongly benefit from the guaranteed positivity of the $\delta$NEGF approach. Another important direction is the extension of the $\delta$NEGF theory to open quantum systems. For the G1--G2 scheme, various embedding strategies have already been developed to describe, for example, Lindblad dynamics~\cite{blommel_arxiv_2026}, electron--phonon~\cite{karlsson_fast_2021,pavlyukh_time-linear_2022,pavlyukh_time-linear_2022-1} and electron--photon coupling~\cite{Tuovinen_2024}, ion--surface interactions~\cite{balzer_prb_23}, and coupling to external leads~\cite{tuovinen_prl_23,pavlyukh_prb_25}. Adapting these embedding schemes to the $\delta$NEGF framework would substantially broaden its range of applications and, in particular, enable time-dependent transport simulations of large, heterogeneous, and strongly correlated nanostructures. Finally, while $\delta$NEGF provides direct access to two-particle spectra, which can be expressed in terms of products of single-particle fluctuations, the extraction of correlated single-particle spectral functions remains an open challenge since the method derives from the HF-GKBA. A promising route is to combine the $\delta$NEGF approach with the recently developed real-time Dyson expansion scheme, which allows for the calculation of high-quality spectral functions~\cite{Reeves_PRL_2024,Reeves_JCTC_2025,Blommel_PRB_2026}.

As a concluding remark we note that, overcoming the high numerical cost, has been a major direction of research in the field of NEGF theory in recent years. Next to the time-linear formulation that was introduced in Ref.~\cite{schluenzen_prl_20} and its fluctuations extension presented here, there exist other promising approaches. They are based on the concept of applying a low-rank compression of the two-time structure, namely quantics tensor trains (QTT)~\cite{Shinaoka_2023,Murray_2024,Sroda_2025,Inayoshi_2026,Sroda_2026} and an hierarchical off-diagonal low-rank representation (HODLR)~\cite{Kaye_2021,Blommel_2025,Gasperlin_2025}. The latter has recently been published within the open-source software package H-NESSi~\cite{Blommel_2026}. Both concepts---the $\delta$NEGF approach, on one side, and the two-time compression techniques, on the other---are highly complementary. The former excels for heterogeneous, finite systems, while extensions to exploit symmetries for homogeneous systems are still under active development. In contrast, the latter performs best for large translationally invariant systems where the self-energy becomes diagonal, while for large finite systems the compression techniques will have to be extended to include also the basis dimensions. Future work toward a universal framework that overcomes the problem of high numerical cost equally across all possible use cases may therefore require the combination of both concepts.

\begin{acknowledgments}
We acknowledge valuable comments by Tim Kalsberger and Christopher Makait. This work has been supported by the Deutsche Forschungsgemeinschaft (DFG, German Research Foundation) via projects no. 440395346 and 464370560. This research was supported in part through high-performance computing resources available at the Kiel University Computing Centre. 
\end{acknowledgments}

\appendix
\section{Tensor algebra}\label{a:tensor_algebra}
The set of rank-4 tensors $\CC^{N_\mathrm{b}\times N_\mathrm{b}\times N_\mathrm{b}\times N_\mathrm{b}}$ naturally forms a vector space under the usual operations of entry-wise addition and scalar multiplication. However, there is no canonical choice of multiplication or taking adjoints on this space. To this end, we adapt the notation of Ref.~\cite{Pavlyukh_PRB_2021} and identify any rank-4 tensor $A \in \mathbb{C}^{N_\mathrm{b}\times N_\mathrm{b}\times N_\mathrm{b}\times N_\mathrm{b}}$ in the single-particle space of rank-4 tensors with a different matrix $\bm{A}^c\in \mathbb{C}^{N_\mathrm{b}^2\times N_\mathrm{b}^2}$ for each of the three channels in the two-particle space (indicated by bold symbols and superscript ``$c$''):
\begin{align}
   \bm{A}^c_{(ij)(kl)} \coloneqq \begin{cases}
        A_{ijkl}, & c = \mathrm{pp},\\
        A_{ilkj} , & c = \mathrm{ph},\\
        A_{iljk}, & c= {\overline{\mathrm{ph}}}.
    \end{cases}
\end{align}
Then, for each of the channels, multiplication and adjoints can be defined as usual, i.e., products of matrices in the two-particle spaces are given by
\begin{align}
    (\bm{A}^c \bm{B}^c)_{(ij)(kl)} = \begin{cases}
        \sum_{mn} A_{ijmn} B_{mnkl} , & c = \mathrm{pp},\\[1ex]
        \sum_{mn} A_{imnj} B_{nlkm} , & c = \mathrm{ph},\\[1ex]
        \sum_{mn} A_{imjn}B_{nlmk}, & c = \overline{\mathrm{ph}},
    \end{cases}
\end{align}
and adjoints
\begin{align}
    (\bm A^c)^\dagger_{(ij)(kl)} = \bm{A}^{c*}_{(kl)(ij)}= \begin{cases}
        A^*_{klij}, & c = \mathrm{pp}, \\[1ex]
        A^*_{kjil}, & c= \mathrm{ph},\\[1ex]
        A^*_{kjli}, & c = \overline{\mathrm{ph}}.
    \end{cases} 
\end{align}
Further, traces are given by 
\begin{align}
    \mathrm{Tr}\bm{A}^c \coloneqq \begin{cases}
        \sum_{ij} A_{ijij},& c = \mathrm{pp},\mathrm{ph}\\[1ex]
        \sum_{ij} A_{ijji},& c = \overline{\mathrm{ph}}.
    \end{cases}
\end{align}
In addition, we define partial traces as contractions of matrices in two-particle space to matrices in the single-particle space:
\begin{align}
        \big(\mathrm{Tr}_2\bm{A}^c\big)_{ij} \coloneqq \begin{cases}
        \sum_{k} A_{ikjk},& c = \mathrm{pp},\mathrm{ph}\\[1ex]
        \sum_{k} A_{ikkj},& c = \overline{\mathrm{ph}},
    \end{cases}
\end{align}
and analogously the other partial trace $\mathrm{Tr}_1$.

Given two matrices in the single-particle space, $C,D\in \mathbb{C}^{N_\mathrm{b}\times N_\mathrm{b}}$, we can associate a matrix in the different channels of the two-particle space as
\begin{align}
    (C\otimes_c D)_{(ij)(kl)}\coloneqq C_{ik}D_{jl}.
\end{align}
Lastly, for any tensor that has already been defined for a specific channel, i.e., $A^c$, we drop one channel superscript ``$c$'' and write $\bm A^c \equiv  (\bm{A}^c)^c$.

\section{Symmetries and energy conservation in the $\delta$NEGF approach} \label{a:symmetries-energy_conservation}
In Sec.~\ref{sss:conservation_laws} we briefly noted that the approximations developed in this article, for which we presented simulation results, are energy conserving for the considered systems. Here we give more details on the analysis of total energy conservation in the present $\delta$NEGF formalism.

\subsection{Symmetries of the generalized susceptibilities}\label{as:chi-symmetry}
Conservation laws are directly linked to symmetries (Noether's theorems). In 
NEGF theory conserving approximations are linked to symmetries of the two-particle two-time Green function $G^{(2)}$ or of the self-energy
\cite{baym_conservation_1961,kadanoff_quantum_1962,stefanucci_nonequilibrium_2013}: 
\begin{align}
    G^{(2)}_{ijkl}(z_1,z_2,z^+_1,z^+_2)= G^{(2)}_{jilk}(z_2,z_1,z^+_2,z^+_1).
\end{align}
While most conservation laws only require the fulfillment of this symmetry condition, energy conservation additionally requires that the approximate two-particle Green function obeys the same boundary conditions as the exact one. Alternatively, when considering only the time diagonal, analogous exchange symmetries of the approximate three-particle NEGF suffice to guarantee energy conservation \cite{bonitz_qkt}.

While the $\delta$NEGF approach is based on NEGF theory and its common approximations, to speed up simulations and to access large correlated systems, the theory eliminates both, the self-energy and the two-particle Green function, in favor of the generalized susceptibilities $\chi^c$, where the definition of $\chi^c$ is aligned to the respective channel [cf. Eq.~\eqref{eq:def-chi-2times}]. Although the generalized susceptibilities share a number of common properties, such as the positive semi-definiteness of the associated reduced density matrices, they differ, among other things, in their symmetries. For example, in the particle-particle channel and particle-hole channels one finds, respectively:
\begin{subequations}
\begin{align}
    \chi^{\mathrm{pp},\gtrless}_{ijkl}(t,t') &= \chi^{\mathrm{pp},\gtrless}_{jilk}(t,t')\,,\label{eq:pp-exchange-symmetry}\\
    \chi^{c,>}_{ijkl}(t,t') &= \chi^{c,<}_{jilk}(t',t)\,,\quad  c = \mathrm{ph}, \overline{\mathrm{ph}}.
\end{align}\end{subequations}

In order to recover the same symmetries as of $\mathcal{G}$, in all channels, we introduced the Keldysh component of the susceptibility, Eq.~\eqref{eq:def-K_component-chi} which obeys
\begin{align}
    \chi^{c,\mathrm{K}}_{ijkl}(t) &= \chi^{c,\mathrm{K}}_{jilk}(t)\,,
\end{align}
in all channels simultaneously, on the time diagonal.

Let us now consider energy conservation. We will split the discussion into several parts for the different channels  and discuss the conservation properties of each of the considered approximations. In particular, we will show the properties for a general basis and then investigate interactions of the form for which each approximation has been applied.
\subsection{Energy conservation in the particle-particle channel}\label{as:econs-pp}
We start by considering the particle-particle channel. Here, we do not consider the symmetrized expression given by $\chi^{c,\mathrm{K}}$ for the investigation of energy conservation, as the greater and lesser components $\chi^{\mathrm{pp},\gtrless}$ already satisfy the necessary exchange symmetry \eqref{eq:pp-exchange-symmetry}. In the particle-particle channel, this symmetrization is only necessary in order to restore the particle-hole symmetry.

The total energy is given by the sum of the single-particle, $E^{(1)}$, and two-particle energies, $E^{(2)}$:
\begin{align}
    E(t) &= E^{(1)}(t)+E^{(2)}(t),\\
   E^{(1)}(t) &\coloneqq -\mathrm{i}\hbar \mathrm{Tr}\big[h(t) G^<(t)\big], \label{eq:1p-energy}\\
   E^{(2)}(t)&\coloneqq-\frac{h^2}{2}\mathrm{Tr}\big[\bm{w}^\mathrm{pp}(t) \bm{G}^{(2),\mathrm{pp},<}(t)\big]. \label{eq:2p-energy}
\end{align}
As we investigate the energy conservation of the discussed approximations, we assume that the single-particle Hamiltonian and the pair interaction are independent of time, i.e.,  $h(t) \equiv h$ and $w(t) \equiv  w$.

For the dynamics of the single-particle energy, $E^{(1)}$, we obtain:
\begin{align}
    \partial_t E^{(1)}(t) 
    &=-\mathrm{Tr}\big[h\big\{\big[h,G^<(t)\big] + \mathrm{Tr}_2\big[\bm{\chi}^{\mathrm{pp},<}(t),\bm{w}^\mathrm{pp}\big]\big\} \big]\nonumber\\
    &= -\mathrm{Tr}\big[h \mathrm{Tr}_2\big[\bm{\chi}^{\mathrm{pp},<}(t),\bm{w}^\mathrm{pp}\big]\big],
\end{align}
where we used the cyclic property of the trace to see that $\mathrm{Tr}[h[h,G^<(t)]]\equiv 0$.

Next, we consider the two-particle energy and find 
\begin{align}
    \partial_t E^{(2)}(t) =& \frac{1}{2}\mathrm{Tr}\big[\bm{w}^\mathrm{pp}\big\{ \bm{\mathfrak{h}}^\mathrm{pp}(t) \bm{\chi}^{\mathrm{pp},<}(t)-  \bm{\chi}^{\mathrm{pp},<}(t)\bm{\mathfrak{h}}^{\mathrm{pp},\dagger}(t)\big\} \big]\nonumber\\
    =& \mathrm{Tr}\big[h^\mathrm{HF}(t) \mathrm{Tr}_2\big[\bm{\chi}^{\mathrm{pp},<},\bm{w}^\mathrm{pp}\big](t)\big]\,,
\end{align}
where we again used the cyclic property of the trace to verify that the contribution due to $\mathfrak{h}^{\mathrm{cor},\mathrm{pp}}$ vanishes:
\begin{align}
   &\mathrm{Tr}\big[\bm{w}^\mathrm{pp}\bm{\chi}^{0,\mathrm{pp},\Delta}(t)\bm{w}^\mathrm{pp}\bm{\chi}^{\mathrm{pp},<}(t)\big]\nonumber\\&\quad=\mathrm{Tr}\big[ \bm{w}^\mathrm{pp}\bm{\chi}^{\mathrm{pp},<}(t)\bm{w}^\mathrm{pp}\bm{\chi}^{0,\mathrm{pp},\Delta}(t) \big]\,.
\end{align}
Further, again using the cyclic property and also the exchange property \eqref{eq:pp-exchange-symmetry} it follows
\begin{align}
  & \frac{1}{2}\mathrm{Tr}\big[ \bm{w}^\mathrm{pp}\bm{\mathfrak{h}}^{\mathrm{HF},\mathrm{pp}}(t)\bm{\chi}^{\mathrm{pp},<}(t)\big]\nonumber
   \\ &\quad = \mathrm{Tr}\big[h^\mathrm{HF}(t)\mathrm{Tr}_2\big[\bm{\chi}^{\mathrm{pp},<} (t)\bm{w}^\mathrm{pp}\big]\big].
\end{align}
Hence, it holds for the total energy within the LA:
\begin{align}
    \partial_t E(t) = \mathrm{Tr}\big[\Sigma^\mathrm{HF}(t)\mathrm{Tr}_2\big[\bm{\chi}^{\mathrm{pp},<}(t),\bm{w}^\mathrm{pp}\big]\big]\,. \label{eq:energy-dynamics_pp}
\end{align}
As the term on the r.h.s. does not vanish for a general basis with pair interaction $w_{ijkl}$, the LA is not an energy conserving approximation. This violation arises from the inclusion of the effective Hartree-Fock Hamiltonian $h^\mathrm{HF}$ in the equation of motion for the generalized susceptibilities, whereas the single-particle equation only includes the ideal single-particle Hamiltonian $h$. Consequently, a modification of the LA, arising from considering only the ideal Hamiltonian instead of the Hartree-Fock Hamiltonian, yields a conserving approximation.

The term on the r.h.s. of Eq.~\eqref{eq:energy-dynamics_pp}, however, vanishes for special pair interactions. For example, for spatially local pair interactions (explicitly considering the spin dependence of the interaction denoted with $\sigma, \tau$) given by
\begin{align}
    w_{(\mathfrak{i},\sigma)(\mathfrak{j},\tau)(\mathfrak{k},\sigma')(\mathfrak{l},\tau')}= U_\mathfrak{i} \delta_{\mathfrak{ij}}\delta_{\mathfrak{ik}}\delta_{\mathfrak{jl}}\delta_{\sigma\sigma'}\delta_{\tau\tau'},
\end{align}
the LA yields an energy conserving approximation. Consequently, the $\delta$TPP approximation is conserving for the standard Hubbard model with its purely local interactions. Additionally, any pair interaction $w_{ijkl}$ that is unitarily equivalent to a local interaction via a unitary transformation of the single-particle basis, for example, the Fourier transform, yields an equivalent conserving pair interaction for the $\delta$TPP approximation.
\subsection{Energy conservation in the particle-hole channels}\label{as:econs-ph}

Next, we consider the particle-hole channels.
For completeness, we will, at this point, give the expression for the equations of motion for the generalized susceptibilities in their lowest-order approximations that we have discussed, the approximation of second moments \eqref{eq:2M}:
\begin{align}
    \mathrm{i}\hbar \partial_t \bm{\chi}^{c,\gtrless}(t) =& \big\{\bm{\mathfrak{h}}^{0,c}(t)+\bm{\mathfrak{\bm{h}}}^{\mathrm{cor},c}(t) \big\}\bm{\chi}^{c,\gtrless}(t)\nonumber\\
    &-\bm{\chi}^{c,\gtrless}(t)\big\{\bm{\mathfrak{h}}^{0,c}(t)+\bm{\mathfrak{\bm{h}}}^{\mathrm{cor},c}(t) \big\}^\dagger,
\end{align}
where we introduced the effective channel-dependent two-particle Hamiltonian 
\begin{align}
    \bm{\mathfrak{h}}^{0,c}(t) &\coloneqq
        h^\mathrm{0,c}(t)\otimes_c \mathbbm{1}-\mathbbm{1}\otimes_c h^{0,c,\mathrm{T}}(t) \,.
    \label{eq:2p-0,c-Hamiltonian}
\end{align}

Unlike the particle-particle channel where the generalized susceptibility essentially corresponds to the two-particle Green function, the particle-hole susceptibilities have to be added to expressions depending on single-particle Green functions in order to reconstruct the full two-particle Green function. As exchange properties are essential for energy conservation, we now have to consider the Keldysh component $\chi^{c,\mathrm{K}}$ as it has the necessary symmetries, i.e, $\chi^{c,\mathrm{K}}_{ijkl}(t) = \chi^{c,\mathrm{K}}_{jilk}(t)$. The time-diagonal two-particle Green function, $G^{(2),<}(t)= G^{(2),\mathrm{pp},<}(t)$ (we omit the superscript ``$\mathrm{pp}$'' in the following), can then be expressed in terms of $\chi^{c,\mathrm{K}}$ as
\begin{align}
    G^{(2), <}_{ijkl}(t) = G^{(2),0, c,<}_{ijkl}(t)+
    X^c_{ijkl}(t)+\frac{1}{2\mathrm{i}\hbar}\chi^{c,\mathrm{K}}_{ijkl}, \label{eq:X+chi}
\end{align}
where we introduced the mean-field part of the two-particle Green function
\begin{align}
    G^{(2),0,c,<}_{ijkl}(t) \coloneqq \begin{cases}
        -G^<_{il}(t) G^<_{jk}(t), & c = \mathrm{ph}, \\[1ex]
        G^<_{ik}(t) G^<_{jl}(t) , & c = \overline{\mathrm{ph}},
    \end{cases}
\end{align}
and an additional contribution of the following form
\begin{align}
    \bm{X}^c(t) \coloneqq -\frac{1}{2\alpha_c} \big\{ G^<(t)\otimes_c \mathbbm{1}+\mathbbm{1}\otimes_c G^{<,\mathrm{T}}(t)\big\}.
\end{align}
It is important to highlight that a similar contribution would be present without the symmetrization as well and would be of the form $\alpha_c^{-1} \mathbbm{1}\otimes G^{<,\mathrm{T}}(t)$.

Inserting expression \eqref{eq:X+chi} into the two-particle energy contribution \eqref{eq:2p-energy} yields
\begin{align}
    E^{(2)}(t) = &-\frac{\mathrm{i}\hbar}{2}\mathrm{Tr}\big[\Sigma^{0,c}(t) G^<(t)\big]- \frac{\hbar^2}{2}\mathrm{Tr}\big[\bm{w}^c\bm{X}^c(t)\big]\nonumber\\&+\frac{\mathrm{i}\hbar}{4}\mathrm{Tr}\big[\bm{w}^c \bm{\chi}^{c,\mathrm{K}}(t)\big]\,,
\end{align}
where, due to the symmetries of the pair interaction,
\begin{align}
    - \frac{\hbar^2}{2}\mathrm{Tr}\big[\bm{w}^c\bm{X}^c(t)\big]  
    &= -\mathrm{i}\hbar \mathrm{Tr}\big[w^c_\mathrm{loc} G^<(t)\big]\,,
\end{align}
Again, we underline that this energy contribution is always present, even without symmetrization.

Furthermore, we have
\begin{align}
    \partial_t \mathrm{Tr}\big[\Sigma^{0,c}(t), G^<(t)\big]=2\mathrm{Tr}\big[ \Sigma^{0,c}(t), \partial_t G^<(t)\big]
\end{align}
Thus, it follows for the change in the two-particle energy:
\begin{align}
    \partial_t E(t) = &-\mathrm{i}\hbar\mathrm{Tr}\big[\big(h^c+w^c_\mathrm{loc}\big) \partial_tG^<(t)\big]\nonumber\\&+\frac{\mathrm{i}\hbar}{4}\mathrm{Tr}\big[\bm{w}^c\partial_t \bm{\chi}^{c,\mathrm{K}}(t)\big].
\end{align}
The first term on the r.h.s. essentially corresponds to the change in the effective single-particle energy, while the second term on the r.h.s. describes the change in the remaining part of the two-particle energy. Due to the structure of the equation, we can proceed analogously as for the particle-particle channel. Here, we again start by considering the lowest-level approximation: the approximation of second moments \eqref{eq:2M}. It follows that this approximation is not generally conserving, as the contributions involving $w^c_\mathrm{loc}$ are not compensated,  i.e.,
\begin{align}
    \partial_t E(t) = \frac{1}{2}\mathrm{Tr}\big[w^c_\mathrm{loc}\mathrm{Tr}_2\big[\bm{\chi}^{c,\mathrm{K}}(t),\bm{w}^{c}\big]\big]\,. \label{eq:energy-dynamics_ph}
\end{align}
It is possible to improve this deficiency of the approximation by also applying the same symmetrization to the equation of motion for the fluctuations, so that $w_\mathrm{loc}^c$ is also included in the equation of motion for the generalized susceptibilities, thus leading to an approximation that is conserving for arbitrary interactions. 

For a diagonal interaction of the form $w_{ijkl}= v_{ij}\delta_{ik}\delta_{jl}$, it follows that 
\begin{align}
    \big[w^{\overline{\mathrm{ph}}}_\mathrm{loc}, G^<\big]= \big[w^{\overline{\mathrm{ph}}}_\mathrm{loc}, \delta\hat{G}^{\overline{\mathrm{ph}}}\big]=0.
\end{align}
Thus, it follows that $\delta$RPA is a conserving approximation for these types of interactions, for example, a standard Coulomb interaction.

Let us next consider the more advanced approximations within the particle-hole channels. For $\delta$TPH and $\delta GW$, it follows
\begin{align}
    \partial_t E(t) = \frac{1}{2}\mathrm{Tr}\big[\big\{h^\mathrm{HF}(t)-h^c(t)+w^c_\mathrm{loc}\big\}\mathrm{Tr}_2\big[\bm{\chi}^{c,\mathrm{K}}(t),\bm{w}^{c}\big]\big]\,. \label{eq:energy-dynamics_ph_2}
\end{align}
This contribution again vanishes for purely local (Hubbard-type) interactions. 

Finally, we consider the  $\delta$TPH+X and $\delta GW$+X approximations, both arising from the full PA, Eq.~\eqref{eq:pa}. Here, Hartree-Fock contributions are included for, both the effective single-particle Hamiltonian, $h^\mathrm{HF}$, and the effective field induced by the fluctuations, $\delta \hat{\Sigma}^\mathrm{HF}$. Within the standard NEGF theory, these approximations are only conserving for local interactions.
For the $\delta$NEGF approach, we also find that the total energy is not conserved:
\begin{align}
     \partial_t E(t) =&\frac{1}{2} \mathrm{Tr}\big[\big\{h^\mathrm{HF}(t)-h^c(t)+w^c_\mathrm{loc}\big\}\mathrm{Tr}_2\big[\bm{\chi}^{c,\mathrm{K}}(t),\bm{w}^{c}\big]\big]\nonumber\\
     &+\frac{\mathrm{i}\hbar}{4}\mathrm{Tr}\big[\bm{w}^{c}\big\{\bm{\chi}^{0,c,\Delta}(t)\{\bm{w}^{c,-}-\bm{w}^{c}\}\bm{\chi}^{c,\mathrm{K}}(t)\nonumber\\&-\bm{\chi}^{c,\mathrm{K}}(t) \{\bm{w}^{c,-}-\bm{w}^{c}\}\bm{\chi}^{0,c,\Delta}(t)\big\}\big]\,.
     \label{eq:energy-dynamics_ph_3}
\end{align}
Remarkably, within the $\delta$NEGF framework, these approximations are conserving for broader classes of interactions: $\delta GW$+X is conserving for diagonal interactions of the form $w_{ijkl}= v_{ij}\delta_{ik}\delta_{jl}$, whereas $\delta$TPH+X is conserving for interactions of the form $w_{ijkl} = v_{ij}\delta_{il}\delta_{jk}$. In this case, the additional contributions cancel.

In summary, it follows that all of the considered approximations are energy conserving, following symmetrization, for local (Hubbard-type) interactions and $\delta$RPA and $\delta GW$+X also for diagonal (Coulomb-type) interactions. 
\bibliography{library,mb-ref,mb-ref-1, literature}

\end{document}